\documentclass[a4paper,11pt]{article}
\pdfoutput=1

\usepackage{jcappub}
\usepackage{booktabs}
\usepackage{subcaption}
\usepackage[normalem]{ulem}

\title{Pressure Parametrization of Dark Energy: First and Second-Order Constraints with Latest Cosmological Data}

\author[a,b]{Hanyu Cheng,}
\author[b]{Eleonora Di Valentino,}
\author[b]{Luis A.\ Escamilla,}
\author[c]{ \newline Anjan A.\ Sen,}
\author[d, e]{and Luca Visinelli}

\affiliation[a]{Tsung-Dao Lee Institute \& School of Physics and Astronomy, Shanghai Jiao Tong University, Shanghai 201210, China}
\affiliation[b]{School of Mathematical and Physical Sciences, University of Sheffield, Hounsfield Road, Sheffield S3 7RH, United Kingdom}
\affiliation[c]{Centre for Theoretical Physics, Jamia Millia Islamia, New Delhi-110025, India}
\affiliation[d]{Dipartimento di Fisica ``E.R.\ Caianiello'', Universit\`a degli Studi di Salerno,\\ Via Giovanni Paolo II, 132 - 84084 Fisciano (SA), Italy}
\affiliation[e]{Istituto Nazionale di Fisica Nucleare - Gruppo Collegato di Salerno - Sezione di Napoli,\\ Via Giovanni Paolo II, 132 - 84084 Fisciano (SA), Italy}

\emailAdd{hcheng19@sheffield.ac.uk}
\emailAdd{e.divalentino@sheffield.ac.uk}
\emailAdd{luis.escamilla@icf.unam.mx}
\emailAdd{anjan.ctp@jmi.ac.in}
\emailAdd{lvisinelli@unisa.it}

\date{\today}

\abstract{
We explore an extension of the $\Lambda$CDM model in which the pressure $p$ of the dark energy (DE) fluid evolves with the expansion of the Universe, expressed as a function of the scale factor $a$. The corresponding energy density $\rho$ is derived from the continuity equation, resulting in a dynamical equation-of-state parameter $w \equiv p/\rho$ during the late-time expansion of the Universe. The pressure is modeled using a Taylor expansion around the present epoch ($a = 1$), introducing deviations from a cosmological constant within the dynamical dark energy (DDE) framework. At first order, a single new parameter $\Omega_1$ captures linear deviations, while a second-order parameter, $\Omega_2$, accounts for quadratic evolution in the pressure. We constrain the first- and second-order DDE models using multiple observational datasets and compare their performance against $\Lambda$CDM and the CPL parameterization. A joint analysis of \textit{Planck} CMB, DESI, and DESY5 data yields the strongest evidence for DDE, with a $2.7\sigma$ deviation in the first-order model and over $4\sigma$ in the second-order model—providing strong statistical support for a departure from a cosmological constant. The reconstructed DE evolution in the second-order case reveals a distinctive non-monotonic behavior in both energy density and $w_{\rm DE}(a)$, including clear phantom-crossing phenomena. Notably, the late-time evolution of $w_{\rm DE}(a)$ remains consistent across datasets and shows strong agreement with the CPL parameterization, underscoring the robustness of the pressure-based approach.
}

\begin{document}
\maketitle
\flushbottom

\section{Introduction}
\label{sec:Introduction}

Theoretical models, combined with high-precision cosmological observations, have significantly refined our understanding of the universe’s fundamental dynamics. One of the most remarkable discoveries of the past century is that the expansion of the universe is not decelerating, as originally expected due to gravitational attraction, but is instead accelerating. The first direct evidence for this phenomenon came from observations of high-redshift Type Ia supernovae~\cite{SupernovaSearchTeam:1998fmf, SupernovaCosmologyProject:1998vns}, which act as standard candles due to their well-calibrated intrinsic luminosity. This unexpected result pointed to the existence of a dark energy (DE) component driving the accelerated expansion on cosmological scales. Subsequent measurements of the cosmic microwave background (CMB) anisotropies~\cite{WMAP:2012fli, Planck:2018nkj} provided strong support for this framework, offering precise constraints on the universe’s energy content and indicating that DE constitutes approximately 70\% of the total energy density. Large-scale structure surveys, such as the Sloan Digital Sky Survey (SDSS)~\cite{SDSS:2003eyi, SDSS:2003lnz, SDSS:2005xqv} and the 2-Degree Field Galaxy Redshift Survey (2dF)~\cite{2dFGRSTeam:2002tzq}, further corroborated this picture by mapping the spatial distribution of galaxies and detecting baryon acoustic oscillation (BAO) signatures. Together with Type Ia supernovae, BAO measurements historically favored the standard cosmological model $\Lambda$CDM at low redshift~\cite{Efstathiou:2021ocp, Krishnan:2021dyb, Keeley:2022ojz, Gariazzo:2024sil}, where $\Lambda$ is the cosmological constant and CDM stands for Cold Dark Matter.

However, this concordance has recently been called into question. New results from the Dark Energy Spectroscopic Instrument (DESI)~\cite{DESI:2024mwx, DESI:2024aqx, DESI:2024jis, DESI:2025zgx, DESI:2025fii} and the Dark Energy Survey (DES) Y5 supernovae sample~\cite{DES:2024jxu} challenge the previous consistency, suggesting that low-redshift BAO and SNe data may no longer support the standard $\Lambda$CDM model.
These findings point to a potential dynamical nature of DE and open the door to alternative scenarios. 
Despite its dominant role in the cosmic energy budget, the fundamental nature of DE remains unknown. Whether it corresponds to a true cosmological constant, a dynamical scalar field evolving over cosmic time, or modifications to general relativity on large scales is still an open question. Addressing this mystery is a central goal of current and upcoming observational efforts, including next-generation CMB experiments, galaxy redshift surveys, and gravitational wave observatories, all aiming to probe the physics driving cosmic acceleration. Validated data from DESI already improve upon earlier releases, offering tantalizing hints of new physical mechanisms operating at late times, which may also play a role in resolving the persistent Hubble tension~\cite{Giare:2024smz, Wang:2024hks, Cortes:2024lgw, Colgain:2024xqj, Yin:2024hba, Sabogal:2025mkp}.

In fact, aside from the potential dynamics of DE, there are also the long-standing cosmological tensions that arise when analyzing the $\Lambda$CDM model using different combinations of data sets~\cite{Abdalla:2022yfr,DiValentino:2020vvd,DiValentino:2020srs,DiValentino:2020zio,Perivolaropoulos:2021jda,DiValentino:2025sru}. The most persistent and statistically significant among them is the Hubble tension~\cite{Verde:2019ivm, DiValentino:2020zio,DiValentino:2021izs,Perivolaropoulos:2021jda,Schoneberg:2021qvd,Shah:2021onj,Abdalla:2022yfr,DiValentino:2022fjm,Kamionkowski:2022pkx,Giare:2023xoc,Hu:2023jqc,Verde:2023lmm,DiValentino:2024yew,Perivolaropoulos:2024yxv,DiValentino:2025sru}. The latest values of the Hubble constant $H_0$ inferred from CMB measurements~\cite{Planck:2018vyg,SPT-3G:2022hvq,ACT:2025fju}, assuming the $\Lambda$CDM model, yield a Hubble constant in more than $5\sigma$ tension with direct astrophysical observations~\cite{Freedman:2020dne,Birrer:2020tax,Wu:2021jyk,Anderson:2023aga,Scolnic:2023mrv,Jones:2022mvo,Anand:2021sum,Freedman:2021ahq,Uddin:2023iob,Huang:2023frr,Pesce:2020xfe,Kourkchi:2020iyz,Schombert:2020pxm,Blakeslee:2021rqi,deJaeger:2022lit,Murakami:2023xuy,Breuval:2024lsv,Freedman:2024eph,Riess:2024vfa,Vogl:2024bum,Gao:2024kkx,Scolnic:2024hbh,Said:2024pwm,Boubel:2024cqw,Scolnic:2024oth,Li:2025ife,Jensen:2025aai}.

The mounting evidence for a time-evolving DE component, together with the persistent cosmological tensions observed when applying the $\Lambda$CDM model across various datasets, motivates the exploration of alternative scenarios in which novel properties of the DE fluid emerge. Recently, the second data release (DR2) from DESI hints at a dynamical nature of DE~\cite{DESI:2025zgx}. An assessment using the Chevallier–Polarski–Linder (CPL) parametrization~\cite{Chevallier:2000qy, Linder:2002et} (also referred to as the $w_0$–$w_a$ parametrization) reveals a deviation from a cosmological constant at more than $2\sigma$ confidence level, when the data are combined with CMB and various Type Ia supernova datasets~\cite{DESI:2025zgx}. This result remains robust under alternative choices of DDE parametrizations and dataset combinations~\cite{DESI:2024mwx,Cortes:2024lgw,Shlivko:2024llw,Luongo:2024fww,Yin:2024hba,Gialamas:2024lyw,Dinda:2024kjf,Najafi:2024qzm,Wang:2024dka,Tada:2024znt,Carloni:2024zpl,Chan-GyungPark:2024mlx,DESI:2024kob,Ramadan:2024kmn,Notari:2024rti,Orchard:2024bve,Hernandez-Almada:2024ost, Malekjani:2024bgi, Giare:2024gpk,Reboucas:2024smm,Giare:2024ocw,Chan-GyungPark:2024brx,Menci:2024hop,Li:2024qus,Li:2024hrv,Notari:2024zmi,Gao:2024ily,Fikri:2024klc,Jiang:2024xnu,Zheng:2024qzi,Gomez-Valent:2024ejh, RoyChoudhury:2024wri, Lewis:2024cqj, Wolf:2025jlc, Shajib:2025tpd, Giare:2025pzu, Chaussidon:2025npr, Kessler:2025kju, Pang:2025lvh, RoyChoudhury:2025dhe, Scherer:2025esj, Li:2025cxn, Yang:2025mws, Lin:2025gne, Cheng:2025hug, An:2025vfz}.

In this work, we adopt a pressure parametrization framework to describe a DDE, following the approach proposed in Ref.~\cite{Sen:2007gk}. Instead of parametrizing the DE equation of state $w$, this method considers a Taylor expansion of the DE pressure around the cosmological constant behavior, $p = -p_0 + (1 - a)p_1 + \ldots$, where constant pressure corresponds to $\Lambda$CDM. Deviations from this baseline are modeled through the introduction of additional K-essence fields, analogous to the assisted inflation scenario, where higher-order terms in the expansion require multiple scalar fields. While the original model was tested against supernovae, BAO, and X-ray cluster gas fraction data available at the time, our goal is to revisit and constrain this framework using the latest cosmological datasets, including recent BAO measurements from DESI and SDSS, PantheonPlus and DESY5, as well as updated CMB observations.

The paper is organized as follows. The modeling of the pressure parametrization is outlined in Section~\ref{sec:DDE}. The methodology and datasets used are discussed in Section~\ref{sec:methods}. The results are presented in Section~\ref{sec:results}, followed by a discussion in Section~\ref{sec:discussion}. Conclusions are summarised in Section~\ref{sec:conclusions}. We work in units where $\hbar = c = 1$.

\section{Dynamical Dark Energy Parametrization}
\label{sec:DDE}

In the pressure parametrization approach of Ref.~\cite{Sen:2007gk}, the term that parametrizes the evolution of the DE pressure is expanded in a Taylor series around the present time, as
\begin{equation}
    \label{eq:Taylorpressure}
    p = -p_0 + \sum_{n \geq 1} \frac{1}{n!} (1 - a)^n p_n\,,
\end{equation}
where $a$ denotes the scale factor and the series is truncated at the desired order in the resolution. Note that the zeroth-order term in the expansion is explicitly negative, since we set $p_0 > 0$. In the absence of interaction with other fluids, the DE component follows the continuity equation,
\begin{equation}
    \label{eq:continuity}
    \frac{{\rm d}\rho}{{\rm d}t} + 3H(p + \rho) = 0\,,    
\end{equation}
so that eq.~\eqref{eq:Taylorpressure} results in an expression for the energy density as $\rho = \rho(a)$.

We first consider the truncation of the Taylor series in eq.~\eqref{eq:Taylorpressure} for the pressure field at first order. For a single DE fluid, eq.~\eqref{eq:continuity} leads to the following expression for the energy density:
\begin{equation}
    \label{eq:rho1}
    \rho = \rho_{\rm DE,0} - \frac{3}{4} (1 - a) p_1\,,
\end{equation}
where $\rho_{\rm DE,0}$ is the DE energy density today and $p_1$ is a DE parameter. Note that for $a = 1$, the expression recovers $\rho = \rho_{\rm DE,0}$. The corresponding pressure term from eq.~\eqref{eq:Taylorpressure} at first order reads:
\begin{equation}
    p = -\rho_\mathrm{DE,0} + \left(\frac{3}{4} - a\right) p_1\,.
\end{equation}
We introduce a reformulation of the pressure parameters in terms of the dimensionless quantities:
\begin{equation}
    \label{eq:Omega}
    \Omega_\mathrm{DE,0} \equiv \frac{\rho_\mathrm{DE,0}}{\rho_\mathrm{crit}}\,,\quad \Omega_1 \equiv \frac{3}{4}\frac{p_1}{\rho_\mathrm{crit}}\,,
\end{equation}
where $\rho_\mathrm{crit} = 3H_0^2/(8\pi G)$ is the critical energy density today. This parameterization ensures that the present-day DE abundance $\Omega_\mathrm{DE,0}$ has a direct correspondence with the standard cosmological model in the limiting case where $p_1 = 0$. Under these assumptions, the equation-of-state parameter $w_{\rm DE} \equiv p/\rho$ reduces to
\begin{equation}
    \label{eq:neweos1}
    w_{\rm DE} = -1 + \frac{1}{3}\frac{\Omega_1\,a}{\Omega_1\,(1 - a) - \Omega_\mathrm{DE,0}}\,.
\end{equation}

We choose to expand the dark energy pressure $p(a)$ directly, rather than the equation of state parameter $w_{\rm DE}$, for both theoretical and practical reasons. From a theoretical standpoint, the pressure appears explicitly in the Einstein field equations and directly governs the acceleration of the Universe. In particular, the spatial component of the Einstein equation shows that the total pressure determines the dynamics of the cosmic expansion at the background level. Since cold DM is pressureless, it does not contribute to this equation; instead, the dynamics at late times is driven by DE. This means that the DE pressure $p_{\rm DE}(a)$ can be directly reconstructed from observables such as the angular diameter distance or the luminosity distance, without assuming any value for the matter density. As a result, any statistically significant deviation of $p_{\rm DE}(a)$ from a constant directly indicates dynamical dark energy, independent of uncertainties in the matter sector. From a practical modeling perspective, expanding $p(a)$ ensures that both $\rho_{\rm DE}(a)$ and $w_{\rm DE}(a)$ are derived consistently from the energy-momentum conservation equation in eq.~\eqref{eq:continuity}, preserving the self-consistency of the cosmological evolution. In contrast, parameterizing $w_{\rm DE}(a)$ implicitly assumes that the ratio of pressure to density is the most fundamental quantity to model. This choice can bias the interpretation of the results, as the derived cosmological dynamics are forced to conform to a specific, and potentially restrictive, functional form for $w_{\mathrm{DE}}(a)$.

Our expansion of the pressure can be theoretically motivated by scalar field models of dark energy, such as quintessence, where the pressure evolves smoothly with time in terms of the field $\phi$ as $p = \dot\phi^2/2 - V(\phi)$, where $V(\phi)$ is the effective potential. A Taylor expansion in $p(a)$ around $a = a_0$ thus corresponds to a late-time expansion of the scalar field dynamics. More generally, $p(a)$ arises directly in the effective stress-energy tensor in covariant theories of modified gravity, such as Horndeski theories, making it a more fundamental and stable quantity to parametrize than the derived function $w_{\rm DE}(a)$. Moreover, a concrete Lagrangian realization of our pressure expansion was presented in
~\cite{Sen:2007gk}, where a multi-field $k$-essence model is constructed, yielding the desired pressure expansion as a sum of effective field contributions. Each term in the Taylor expansion of $p(a)$ corresponds to a distinct $k$-essence component with its own Lagrangian. Specifically, eq.~(9) of Ref.~\cite{Sen:2007gk} provides the form of the action that reproduces a given pressure expansion to arbitrary order, starting from a constant Lagrangian at zeroth order, and building higher-order corrections with appropriately constructed kinetic and potential terms. This construction offers a consistent field-theoretic embedding of our phenomenological model.

\begin{figure*}[t!]
    \centering
    \begin{subfigure}[t]{0.49\textwidth}
        \centering
        \includegraphics[width=0.99\linewidth]{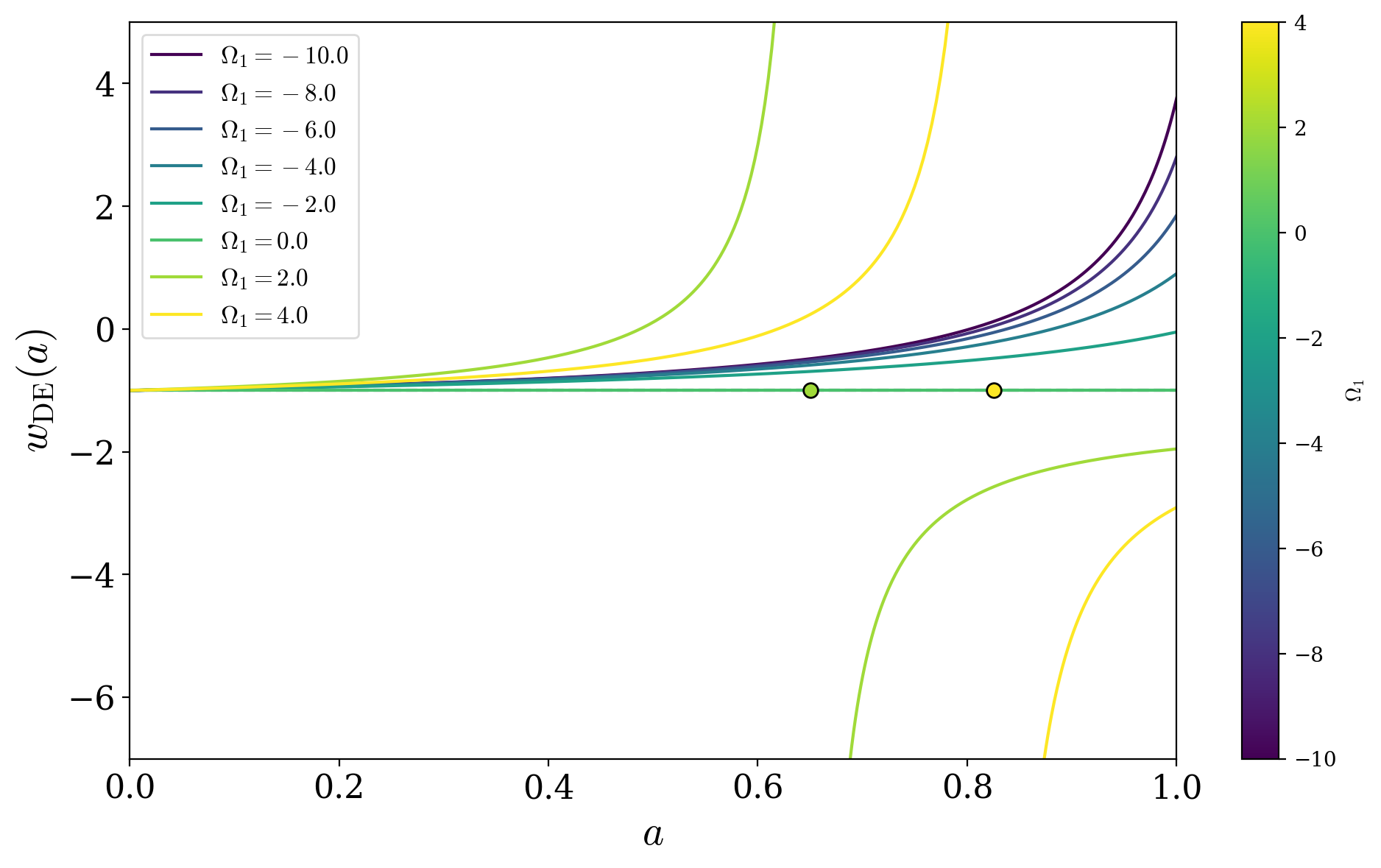}
    \end{subfigure}
    \begin{subfigure}[t]{0.49\textwidth}
        \centering
        \includegraphics[width=0.99\linewidth]{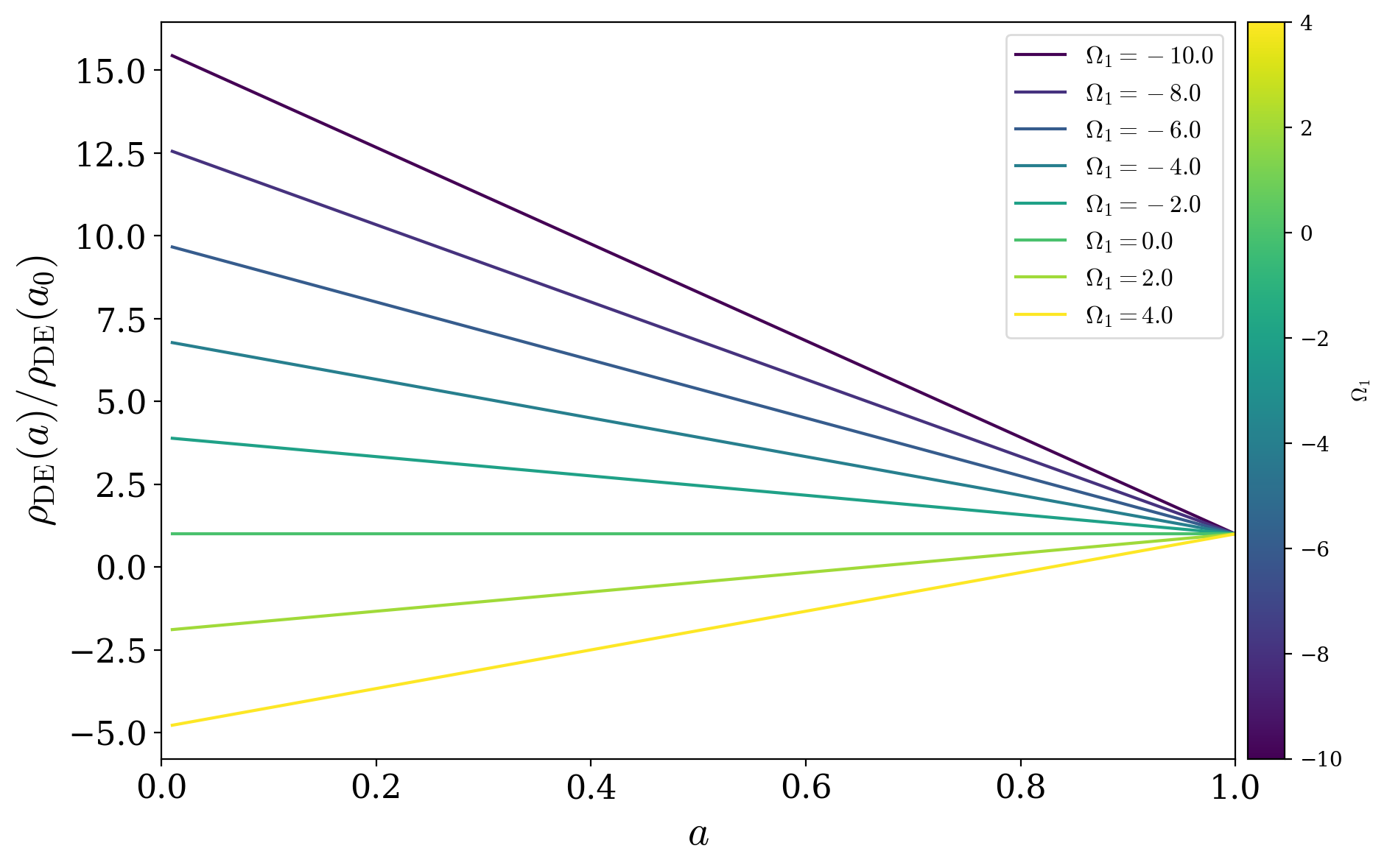}
    \end{subfigure}
    \includegraphics[width=0.5\linewidth]{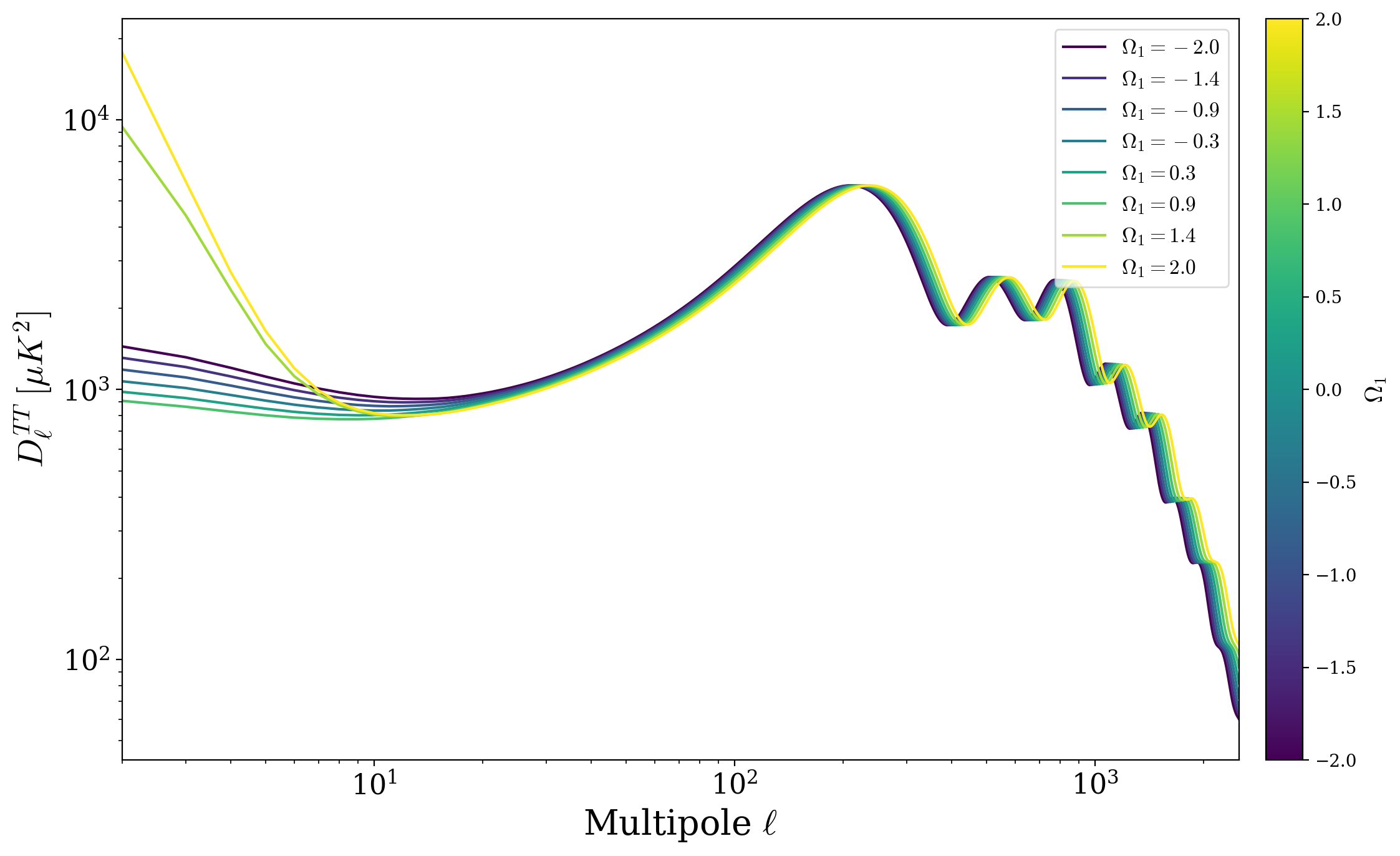}
    \caption{{\bf Expansion to first order.} \textit{Top panel:} The equation of state $w_{\rm DE}$ from eq.~\eqref{eq:neweos} (left) and the energy density ratio $\rho_{\rm DE}(a)/\rho_{\rm DE}(a_0)$ (right), shown as functions of the scale factor $a$. For illustration, the present-day DE contribution is fixed at $\Omega_{\mathrm{DE,0}} = 0.7$. \textit{Bottom panel:} The impact of varying the modified DE parameter $\Omega_1$ on the CMB TT power spectrum.}
    \label{fig:w_de_first}
\end{figure*}

The top-left panel in figure~\ref{fig:w_de_first} shows the evolution of the DE equation of state, $w_{\rm DE}$, from eq.~\eqref{eq:neweos1} under a first-order expansion, highlighting the effect of varying the modified DE parameter $\Omega_{1}$. For illustrative purposes, the present-day DE density parameter is fixed at $\Omega_{\mathrm{DE,0}} = 0.7$. 
Here, $\Omega_{\mathrm{DE,0}}$ is defined as $\Omega_{\mathrm{DE,0}} = 1 - \Omega_{\mathrm{m}} - \Omega_{\mathrm{k}} - \Omega_{\mathrm{r}}$, where $\Omega_{\mathrm{m}}$ is the matter density parameter, $\Omega_{\mathrm{k}}$ is the curvature density parameter, and $\Omega_{\mathrm{r}}$ is the radiation density parameter. The top-right panel illustrates the impact of varying $\Omega_{1}$ on the energy density ratio $\rho_{\rm DE}(a)/\rho_{\rm DE}(a_0)$. The bottom panel in the figure shows the impact of varying $\Omega_1$ at first order on the CMB TT power spectrum. The most significant effects are observed at low multipoles, primarily due to modifications in the late-time Integrated Sachs–Wolfe (ISW) effect. This behavior is particularly relevant in light of eq.~\eqref{eq:neweos1}, which predicts the presence of a pole in the equation of state at  
\begin{equation}
    a_{\rm pole} \equiv 1 - \frac{\Omega_{\rm DE,0}}{\Omega_1}\,.
\end{equation}
When $\Omega_1 > \Omega_{\rm DE,0}$, this pole appears at a scale factor $a_{\rm pole}$ between the early Universe and the present epoch, modifying the evolution of the gravitational potentials and enhancing the ISW contribution to the CMB anisotropies\footnote{ In some regions of the parameter space, the pressure expansion may lead to a divergence in $w_{\rm DE}(a)$ at early times, typically due to the denominator in eq.~\eqref{eq:neweos1} crossing zero. To ensure that the model remains numerically stable we enforce a regularization condition: when a divergence is encountered, i.e.\ $a = a_{\rm pole}$, we set $w_{\rm DE}(a_{\rm pole}) = 0$ if $\Omega_1 > \Omega_{\rm DE}$. This prescription ensures that $w_{\rm DE}(a)$ remains finite across the entire domain relevant for our analysis and avoids unphysical poles without introducing artificial discontinuities in the observable quantities. We have verified that this procedure has negligible impact on the background or perturbation evolution within the range probed by current data.}.

We now turn to the truncation of the Taylor series at second order in the pressure parametrization. This leads to the following expression for the energy density:
\begin{equation}
    \label{eq:rho2}
    \rho = \rho_{\rm DE,0} - \frac{3}{4}\left(p_1 + p_2\right)(1 - a) + \frac{3}{10}\,p_2\,(1 - a^2)\,,
\end{equation}
where, in addition to the parameter $p_1$ from eq.~\eqref{eq:rho1}, an extra DE parameter $p_2$ appears.
The corresponding pressure term from eq.~\eqref{eq:Taylorpressure}, truncated at second order, reads:
\begin{equation}
    p = -\rho_\mathrm{DE,0} + \left(\frac{3}{4} - a\right)p_1 + \left(\frac{9}{20} - a + \frac{1}{2}a^2\right)p_2\,.
\end{equation}

\begin{figure*}[t!]
    \centering
    \begin{subfigure}[t]{0.49\textwidth}
        \centering
        \includegraphics[width=0.99\linewidth]{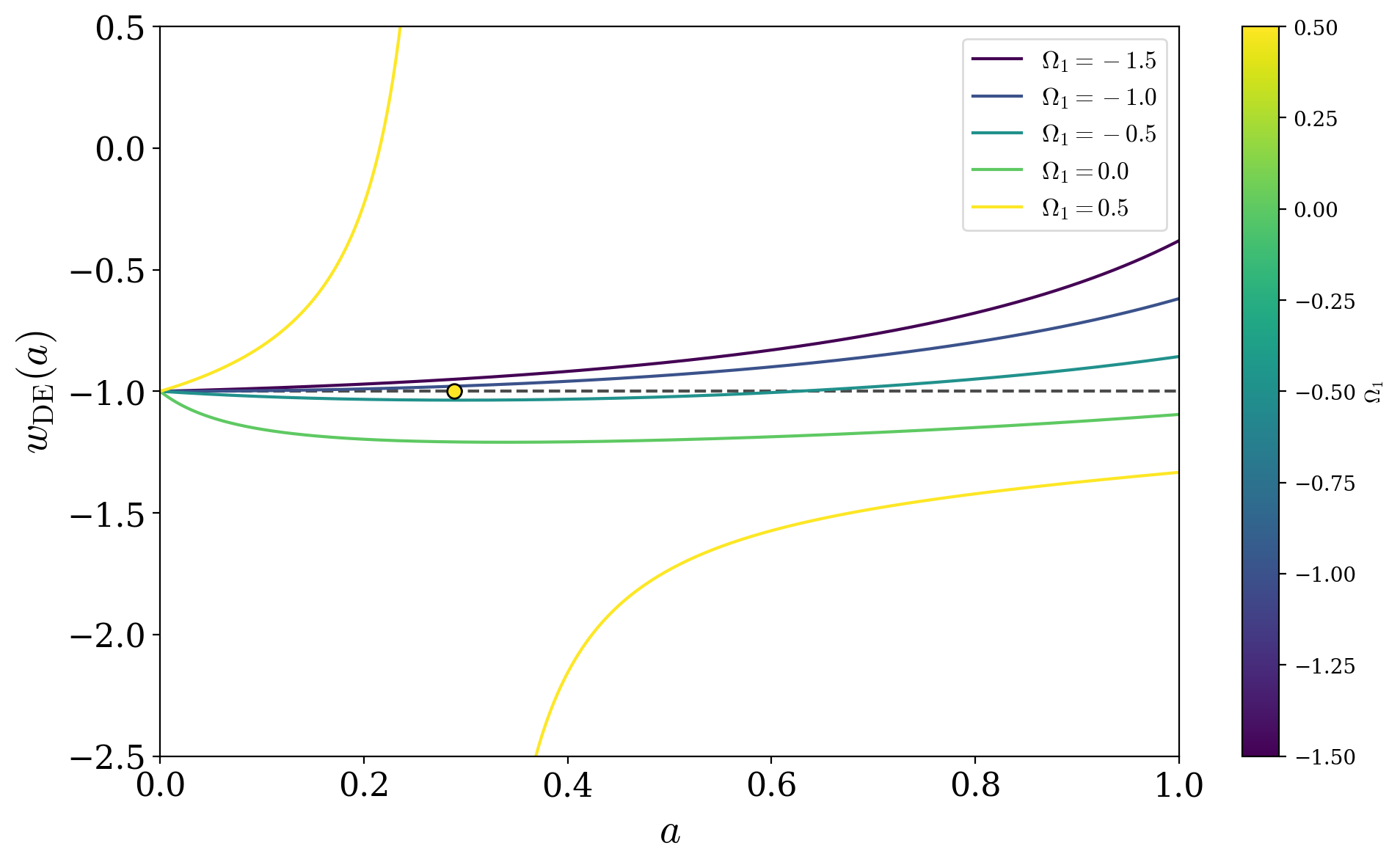}
    \end{subfigure}
    \begin{subfigure}[t]{0.49\textwidth}
        \centering
        \includegraphics[width=0.99\linewidth]{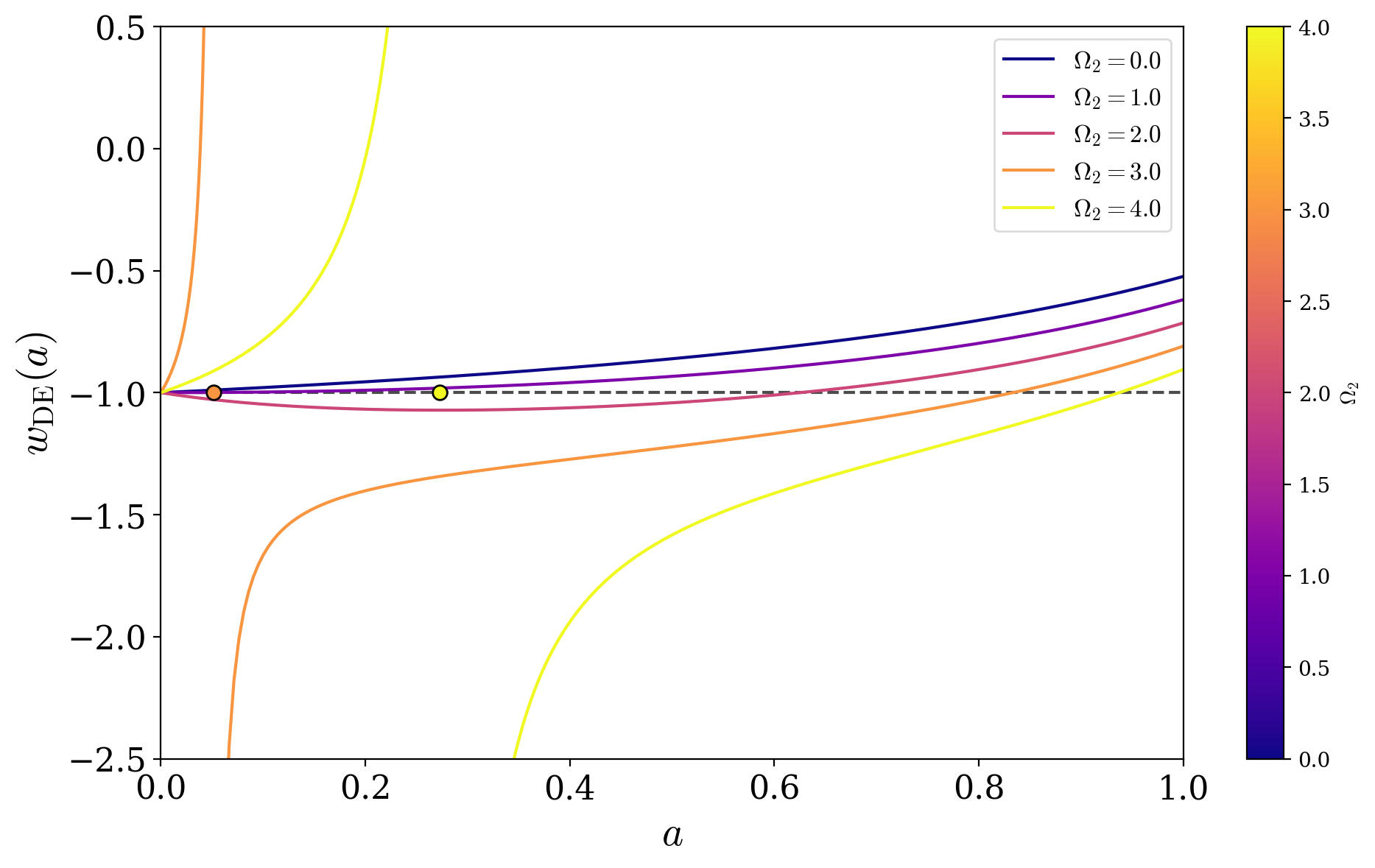}
    \end{subfigure}
    \begin{subfigure}[t]{0.49\textwidth}
        \centering
        \includegraphics[width=0.99\linewidth]{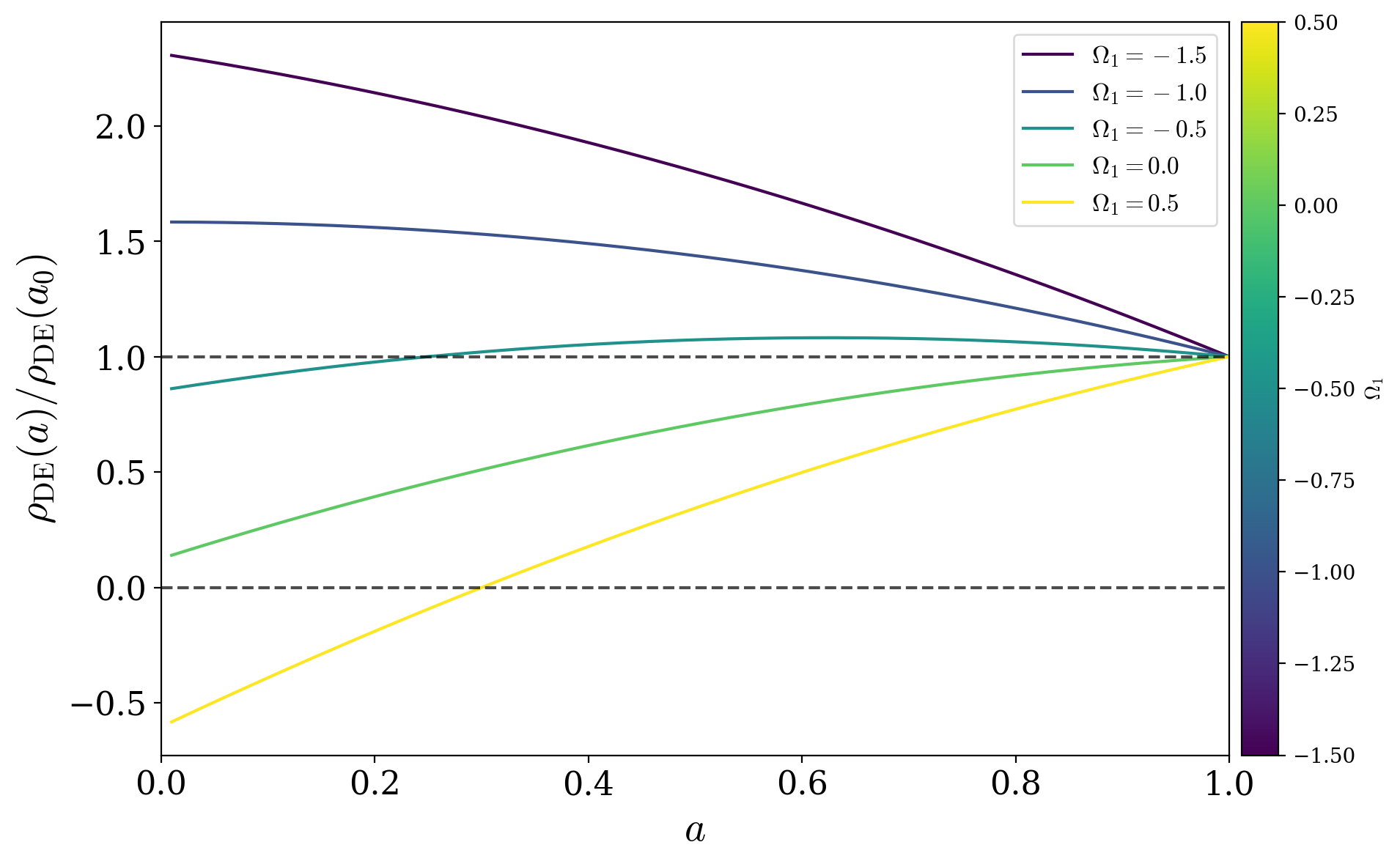}
    \end{subfigure}
    \begin{subfigure}[t]{0.49\textwidth}
        \centering
        \includegraphics[width=0.99\linewidth]{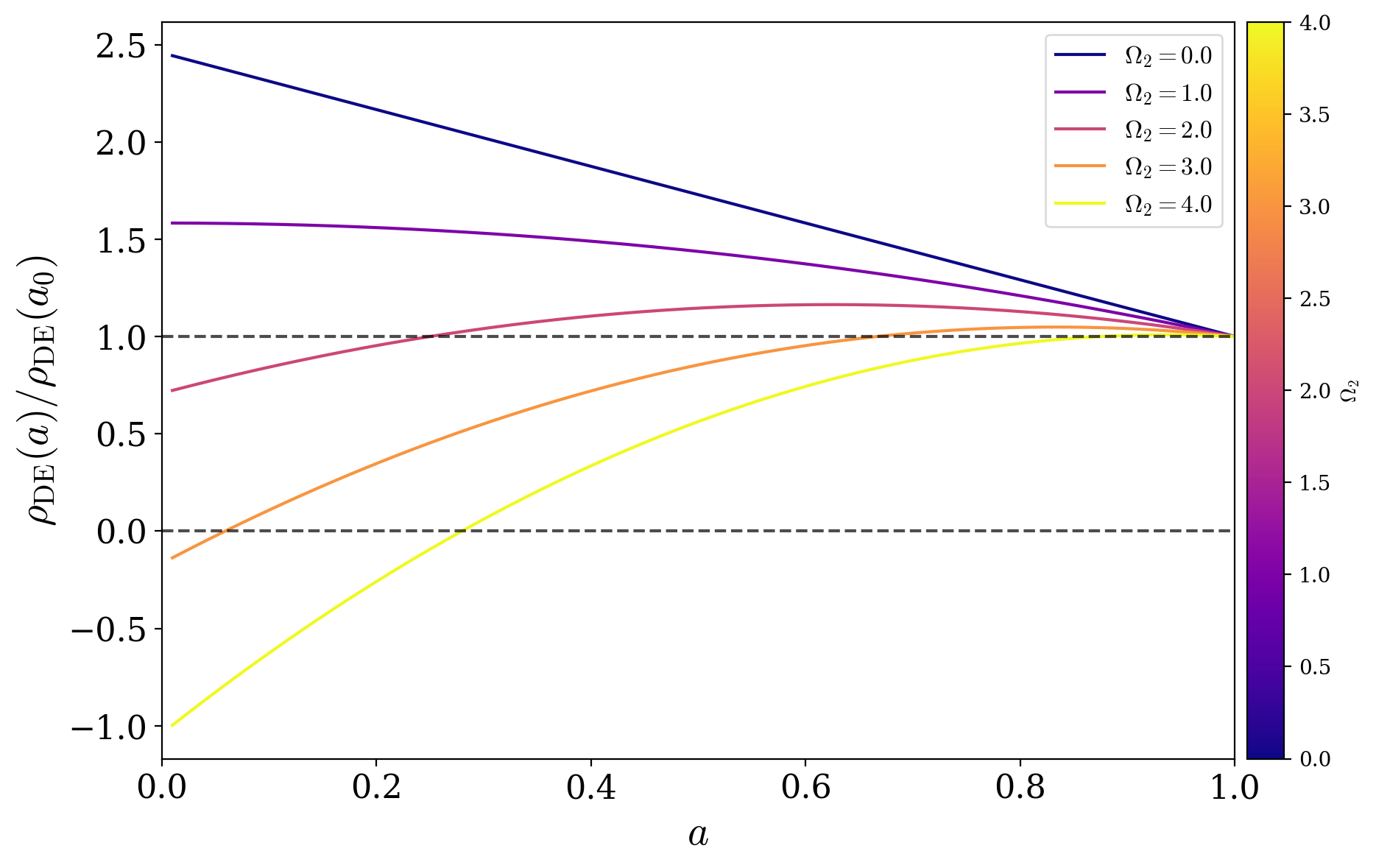}
    \end{subfigure}
    \begin{subfigure}[t]{0.49\textwidth}
        \centering
        \includegraphics[width=0.99\linewidth]{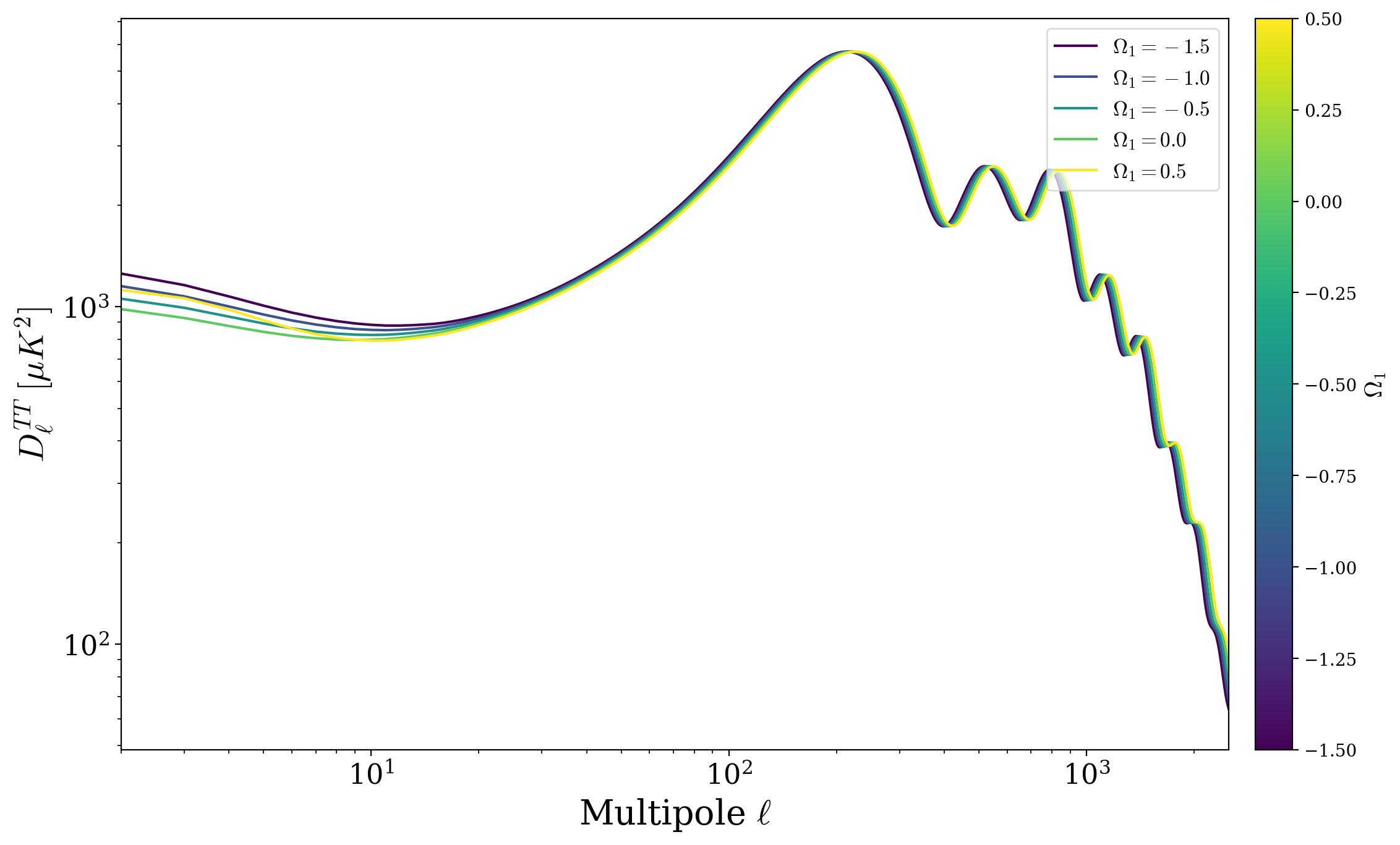}
    \end{subfigure}
    \begin{subfigure}[t]{0.49\textwidth}
        \centering
        \includegraphics[width=0.99\linewidth]{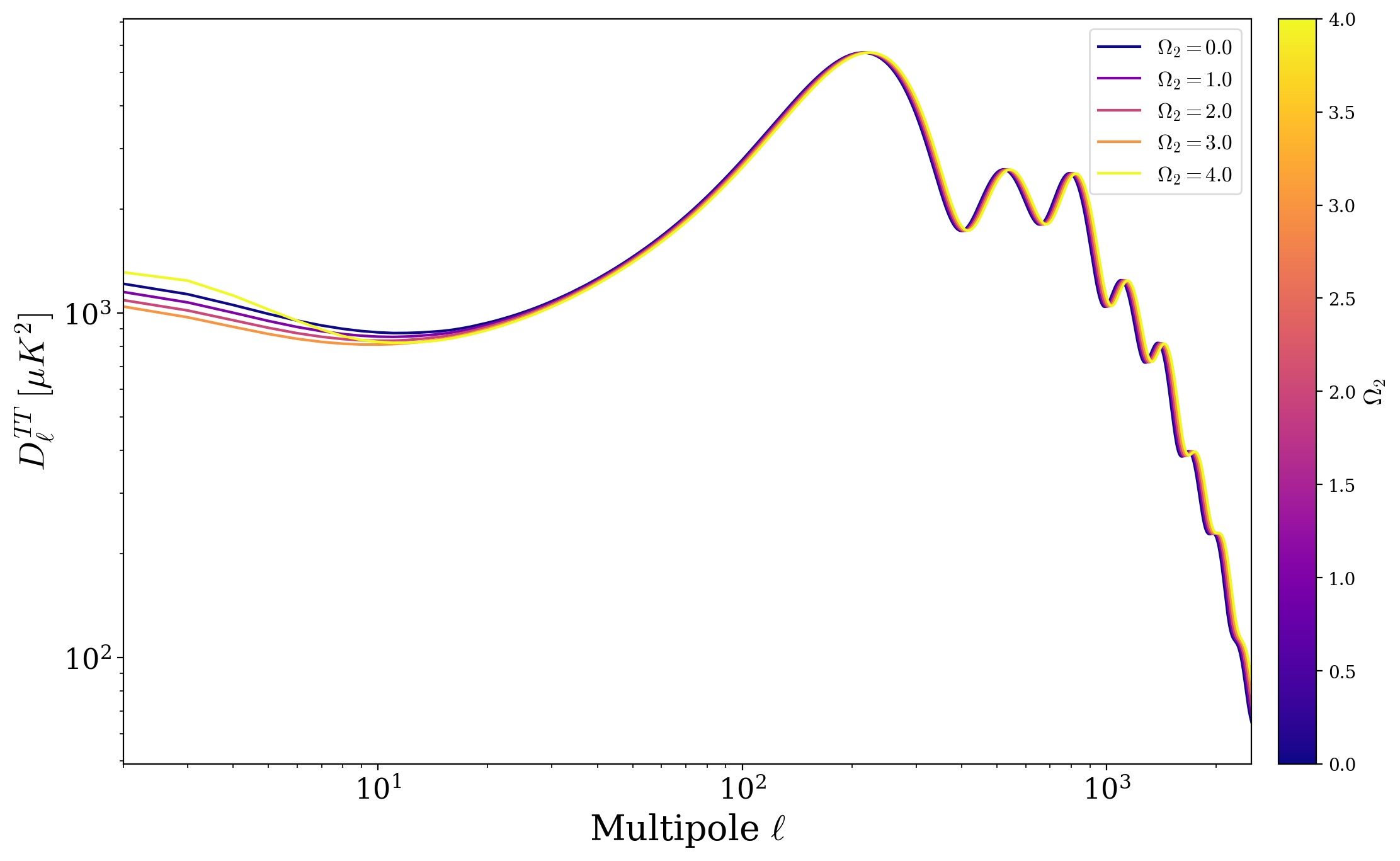}
    \end{subfigure}
    \caption{{\bf Expansion to second order.} \textit{Top panel:} The equation of state $w_{\rm DE}$ from eq.~\eqref{eq:eos} as a function of the scale factor $a$. \textit{Middle panel:} The energy density ratio $\rho_{\rm DE}(a)/\rho_{\rm DE}(a_0)$ from eq.~\eqref{eq:rho_a} as a function of the scale factor $a$. For both top and middle panels, the left side shows fixed $\Omega_2 = 1$ with varying $\Omega_1$ as indicated in the legend, while the right side shows fixed $\Omega_1 = -1$ with varying $\Omega_2$. The present-day DE contribution is fixed at $\Omega_{\mathrm{DE,0}} = 0.7$ for illustration. \textit{Bottom panel:} The impact on the CMB TT power spectrum when varying $\Omega_1$ with fixed $\Omega_2 = 1$ (left) and when varying $\Omega_2$ with fixed $\Omega_1 = -1$ (right), where $\Omega_{\mathrm{DE,0}}$ is defined as $\Omega_{\mathrm{DE,0}} = 1 - \Omega_{\mathrm{m}} - \Omega_{\mathrm{k}} - \Omega_{\mathrm{r}}$.} 
    \label{fig:w_de_second}
\end{figure*}

Similarly to eq.~\eqref{eq:Omega}, we introduce the dimensionless quantities $\Omega_{1,2} \equiv \frac{3}{4}\frac{p_{1,2}}{\rho_\mathrm{crit}}$, so that the energy density of a single DE component, truncated at second order in the Taylor expansion, takes the form:
\begin{equation}
\label{eq:rho_a}
    \rho = \rho_\mathrm{DE,0} \left( 1 + (a - 1)\left(\frac{\Omega_1}{\Omega_\mathrm{DE,0}} + \frac{\Omega_2}{\Omega_\mathrm{DE,0}}\right) + \frac{2}{5}(1 - a^2)\frac{\Omega_2}{\Omega_\mathrm{DE,0}} \right).
\end{equation}
Similarly, the pressure associated with a single DE component at second order is given by:
\begin{equation}
    p = \rho_\mathrm{DE,0} \left[ -1 + \left(1 - \frac{4}{3}a\right)\frac{\Omega_1}{\Omega_\mathrm{DE,0}} + \left(\frac{3}{5} - \frac{4}{3}a + \frac{2}{3}a^2\right)\frac{\Omega_2}{\Omega_\mathrm{DE,0}} \right].
\end{equation}
From this, the equation-of-state parameter can be expressed as:
\begin{equation}
    \label{eq:eos}
    w_{\rm DE} = -1 + \frac{1}{3} \frac{\left(\Omega_1 + \left(1 - \frac{4}{5}a\right)\Omega_2\right)a}{\left[\Omega_1 + \frac{3}{5}\left(1 - \frac{2}{3}a\right)\Omega_2\right](1 - a) - \Omega_\mathrm{DE,0}}\,.
\end{equation}
The expression above diverges for values of the scale factor equal to
\begin{equation}
    a_\mathrm{pole} = \frac{5}{4} \, \frac{(\Omega_{\mathrm{1}}+\Omega_{\mathrm{2}}) \pm \sqrt{(\Omega_{\mathrm{1}}+\Omega_{\mathrm{2}})^2 - \frac{8}{5} \Omega_{\mathrm{2}} \left(\Omega_{\mathrm{1}}+\frac{3}{5} \Omega_{\mathrm{2}} - \Omega_{\mathrm{DE,0}}\right)}}{\Omega_{\mathrm{2}}}\,,
\label{eq:simplified_formula}
\end{equation}
where the expression is valid for $\Omega_2 \neq 0$. If $\Omega_2 = 0$, the equation-of-state parameter $w_{\rm DE}$ in eq.~\eqref{eq:eos} reduces to
\begin{equation}
    \label{eq:neweos}
    w_{\rm DE} = -1 + \frac{1}{3}\frac{\Omega_1\,a}{\Omega_1\,(1 - a) - \Omega_\mathrm{DE,0}}\,.
\end{equation}
The top panels of figure~\ref{fig:w_de_second} show the evolution of the DE equation of state $w_{\rm DE}$ in eq.~\eqref{eq:neweos} under a second-order expansion, highlighting the effects of different values of the modified DE parameters $\Omega_1$ and $\Omega_2$. For illustrative purposes, the present-day DE density parameter is fixed at $\Omega_{\mathrm{DE,0}} = 0.7$. The middle panels illustrate the impact of varying the modified DE parameters $\Omega_1$ and $\Omega_2$, respectively, on the energy density ratio $\rho_{\rm DE}(a)/\rho_{\rm DE}(a_0)$.

The top and middle panels in figure~\ref{fig:w_de_second} demonstrate that our second-order expansion model captures diverse DE behaviors. Quintessence ($-1/3 > w_{\rm DE} > -1$) scenarios are illustrated by the curves with $\Omega_1 = -1.5, -1.0$ in the left panels, and $\Omega_2 = 0.0, 1.0$ in the right panels. Phantom DE ($w_{\rm DE} < -1$) is represented by the curve with $\Omega_1 = 0.0$ in the left panel. The phantom crossing behavior, where $w_{\rm DE}$ traverses the boundary $w = -1$, is shown by the curve with $\Omega_1 = -0.5$ in the left panel and the curves with $\Omega_2 = 2.0, 3.0, 4.0$ in the right panel. Additionally, singularities of $w_{\rm DE}$~\cite{Ozulker:2022slu}(as shown in eq.~\eqref{eq:simplified_formula}), depicted by the $\Omega_1 = 0.5$ curve in the left panel and $\Omega_2 = 3.0, 4.0$ curves in the right panel, which occur when DE density transitions from negative to positive values. Notably, in the top right panel, the orange ($\Omega_2 = 3.0$) and yellow ($\Omega_2 = 4.0$) curves exhibit singularities in the early universe, followed by phantom crossing behavior in the late universe. 

The bottom panels in figure~\ref{fig:w_de_second} illustrate how variations in DE parameters affect the Cosmic Microwave Background (CMB) TT power spectrum, primarily at large angular scales (low-$\ell$ multipoles) dominated by the late-time Integrated Sachs-Wolfe (ISW) effect. Specifically, the yellow curve ($\Omega_1 = 0.5$ in the left panel and $\Omega_2 = 4.0$ in the right panel) shows enhanced power at low multipoles compared to the other curves. Unlike the orange curve in the right panel ($\Omega_2 = 3.0$), whose singularity occurs earlier, the yellow curve's later singularity leads to stronger ISW contributions because it influences the universe at a time when DE plays a more dominant role in the cosmic energy budget.

In the following, we compare the results for the DDE parametrization truncated at first or second order in the expansion of the pressure term (see eq.~\eqref{eq:Taylorpressure}) with the $\Lambda$CDM model, as well as with the dynamical $w_0$–$w_a$ (CPL) parametrization~\cite{Chevallier:2000qy, Linder:2002et}, which arises from Taylor expanding a general $w(a)$ to linear order as
\begin{equation}
    \label{eq:CPL}
    w_{\rm DE}(a) = w_0 + w_a\,(1 - a)\,.
\end{equation}
The parameters $w_0$ and $w_a$ are crucial for characterizing the evolution of the equation-of-state parameter $w(a)$ in cosmological models, and are treated as constants in the CPL model. The $\Lambda$CDM model is recovered for $w_0 = -1$ and $w_a = 0$. In the DDE framework, the parameter $w_0$ is defined as the value of the equation-of-state parameter evaluated at the current scale factor ($a = 1$), and is explicitly given by:
\begin{equation}
    \label{eq:w0_definition}
    w_0 = -1 - \frac{\Omega_1 + \frac{1}{5}\Omega_2}{3\Omega_{\mathrm{DE,0}}}.
\end{equation}
Additionally, the parameter $w_a$, which quantifies the rate of change of the equation-of-state parameter with respect to the scale factor evaluated at $a = 1$, is given by:
\begin{equation}
    \label{eq:wa_definition}
    w_a \equiv -\left.\frac{{\rm d}w(a)}{{\rm d}a}\right|_{a = 1} = \frac{\Omega_{\mathrm{DE,0}}\left(\Omega_1 - \frac{3}{5}\Omega_2\right) - \left(\Omega_1 + \frac{1}{5}\Omega_2\right)^2}{3\Omega_{\mathrm{DE,0}}^2}\,.
\end{equation}

\section{Methodology and Data}
\label{sec:methods}

To perform parameter inference, we utilize {\tt Cobaya}~\cite{Torrado:2020dgo}, a Markov Chain Monte Carlo (MCMC) sampler designed for cosmological models, in conjunction with a modified version of {\tt CAMB}~\cite{Lewis:1999bs}, a Boltzmann solver adapted to incorporate the DDE parametrization. To model dark energy perturbations, we use the default parameterized post-Friedmann (PPF) method implemented in {\tt CAMB}~\cite{Lewis:1999bs}, which allows $w(a)$ to cross the phantom divide ($w = -1$) without introducing divergences in the perturbation equations~\cite{Hu:2007pj,Fang:2008sn}. To assess the convergence of our chains, we use the Gelman–Rubin statistic $R - 1$~\cite{gelman_inference_1992}, adopting $R - 1 < 0.01$ as the threshold for convergence. The MCMC results are analyzed and visualized using {\tt getdist}~\cite{Lewis:2019xzd}.

To systematically compare the DDE models under consideration, we use the following datasets:
\begin{itemize}
    \item \textbf{CMB:} Cosmic Microwave Background (CMB) measurements from the \textit{Planck} 2018 legacy data release, including the high-$\ell$ Plik TT, TE, and EE likelihoods, the low-$\ell$ TT-only Commander likelihood, and the low-$\ell$ EE-only SimAll likelihood~\cite{Planck:2018nkj,Planck:2019nip,Planck:2018lbu}. Additionally, we include lensing measurements from the Atacama Cosmology Telescope (ACT) DR6, specifically the \texttt{actplanck baseline} likelihood~\cite{ACT:2023kun,ACT:2023dou}. This combined dataset is collectively referred to as \textbf{CMB}.
    
    \item \textbf{BAO:} Baryon Acoustic Oscillation (BAO) and Redshift-Space Distortion (RSD) measurements from the completed SDSS-IV eBOSS survey~\cite{eBOSS:2020yzd}, which include both isotropic and anisotropic distance measurements and expansion rates across a broad redshift range, incorporating Lyman-$\alpha$ BAO data. This dataset is referred to as \textbf{SDSS}. We also use BAO data from the first three years of observations by the Dark Energy Spectroscopic Instrument (DESI DR2)~\cite{DESI:2025zgx,DESI:2025fii,DESI:2025qqy}, referred to as \textbf{DESI}.
    
    \item \textbf{Type Ia Supernovae:} Distance modulus measurements of Type Ia Supernovae (SNe Ia) from the PantheonPlus sample~\cite{Scolnic:2021amr,Brout:2022vxf}, which includes 1701 light curves from 1550 distinct SNe Ia spanning redshifts $z \in [0.001, 2.26]$. This dataset is referred to as \textbf{PantheonPlus}. In addition, we include the complete five-year dataset from the Dark Energy Survey (DES), containing 1635 SNe Ia with redshifts in the range $0.1 < z < 1.13$~\cite{DES:2024hip,DES:2024jxu,DES:2024upw}, referred to as \textbf{DESY5}.
\end{itemize}

In our analysis, we adopt flat and uniform priors as summarized in table~\ref{tab:prior_table}. The extended cosmological models build upon the standard six-parameter $\Lambda$CDM framework, which includes the baryon density $\Omega_{\mathrm{b}} h^2$, cold dark matter density $\Omega_{\mathrm{c}} h^2$, optical depth $\tau$, the amplitude and spectral index of scalar fluctuations $\ln(10^{10} A_{\mathrm{s}})$ and $n_{\mathrm{s}}$, and the angular size of the sound horizon at last scattering $\theta_{\mathrm{s}}$. 
Beyond these, the CPL model introduces two additional parameters, $w_0$ and $w_a$, as defined in eq.~\eqref{eq:CPL}. For the DDE model, we introduce one additional parameter, $\Omega_1$, in the first-order expansion, denoted as DDE1 in table~\ref{tab:prior_table}. The second-order expansion includes both $\Omega_1$ and $\Omega_2$, denoted as DDE2 in table~\ref{tab:prior_table}.

\begin{table}[htbp]
\centering
\begin{tabular}{ccc}
\hline
Model & Parameter & Prior \\
\hline\hline
$\Lambda$CDM & $\Omega_{\mathrm{b}} h^2$ & $[0.005 , 0.1]$ \\
$\Lambda$CDM & $\Omega_{\mathrm{c}} h^2$ & $[0.001 , 0.99]$ \\
$\Lambda$CDM & $\tau$ & $[0.01 , 0.8]$ \\
$\Lambda$CDM & $100\,\theta_{\mathrm{s}}$ & $[0.5 , 10]$ \\
$\Lambda$CDM & $\ln(10^{10} A_{\mathrm{s}})$ & $[1.61 , 3.91]$ \\
$\Lambda$CDM & $n_{\mathrm{s}}$ & $[0.8 , 1.2]$ \\
\hline
CPL & $w_0$ & $[-3, 1]$\\
CPL & $w_a$ & $[-3, 2]$\\
\hline
DDE1 & $\Omega_1$ & $[-2.0 , 2.0]$ \\
DDE2 & $\Omega_1$,$\Omega_2$ & $[-10.0 , 10.0]$ \\
\hline\hline
\end{tabular}
\caption{Flat prior distributions imposed on the cosmological parameters used in our analysis. The CPL and DDE models include the standard $\Lambda$CDM parameters, along with the additional parameters specific to each extended model.}
\label{tab:prior_table}
\end{table}

To quantify the statistical performance of the DDE pressure parametrizations relative to the standard $\Lambda$CDM scenario or the CPL parametrization, we evaluate the differences in the minimum chi-square values:
\begin{equation}
    \label{eq:delta_chi_square}
    \Delta \chi^2_{\mathrm{min},\Lambda \mathrm{CDM/CPL}} = \chi^2_{\mathrm{min},1\mathrm{st}/2\mathrm{nd}} - \chi^2_{\mathrm{min},\Lambda \mathrm{CDM/CPL}}.
\end{equation}
A negative value of the difference in eq.~\eqref{eq:delta_chi_square} indicates that the data favor the DDE parametrization over the baseline model, either $\Lambda$CDM or CPL. The more negative the value, the stronger the preference for the DDE scenario. Since $\Lambda$CDM is nested within both the first-order DDE expansion (corresponding to $\Omega_{1}=0$) and the second-order DDE expansion (corresponding to $\Omega_{1}=0$ and $\Omega_{2}=0$), Wilks' theorem~\cite{Wilks:1938dza} implies that $\Delta\chi^{2}_{\mathrm{min}}$ should follow a $\chi^{2}$ distribution with $k$ degrees of freedom under the assumption that the null hypothesis ($\Lambda$CDM model) holds and the errors are Gaussian and correctly estimated, where $k=1$ for the first-order and $k=2$ for the second-order expansion.
To express $\Delta\chi^{2}_{\mathrm{min}}$ in more familiar terms, we convert it to a frequentist significance $N\sigma$ for a one-dimensional Gaussian distribution using: 
\begin{equation}
    \mathrm{CDF}_{\chi^{2}}\!\left(
        \Delta\chi^{2}_{\mathrm{min}} \mid k
    \right)
    = \frac{1}{\sqrt{2\pi}}
      \int_{-N}^{N} \! e^{-t^{2}/2}\,\mathrm{d}t ,
\label{eq:Nsigma_conversion}
\end{equation}
where the left-hand side denotes the cumulative distribution function of the $\chi^{2}$ distribution with $k$ degrees of freedom. Solving eq.~\eqref{eq:Nsigma_conversion} for $N$ yields the Gaussian-equivalent significance with which the DDE parameterization is preferred over the $\Lambda$CDM model.

Furthermore, we perform Bayesian model comparisons by computing the logarithm of the Bayesian evidence using \texttt{MCEvidence}~\cite{Heavens:2017afc}, accessed through the {\tt Cobaya} wrapper provided in the \texttt{wgcosmo} repository~\cite{giare2025wgcosmo}. According to the Bayes' theorem, the posterior probability distribution $P(\Theta|D,\mathcal{M}_{i})$ for a model $\mathcal{M}_{i}$ with parameters $\Theta$, given data $D$, is expressed as:
\begin{equation}
    P(\Theta|D, \mathcal{M}_{i}) = \frac{\mathcal{L}(D|\Theta, \mathcal{M}_{i}) \pi(\Theta|\mathcal{M}_{i})}{\mathcal{E}(D|\mathcal{M}_{i})},
    \label{eq:bayes_theorem}
\end{equation}
where $\mathcal{L}(D|\Theta,\mathcal{M}_{i})$ denotes the likelihood, $\pi(\Theta|\mathcal{M}_{i})$ the prior, and $\mathcal{E}(D|\mathcal{M}_{i})$ the Bayesian evidence, defined by:
\begin{equation}
    \mathcal{E}(D|\mathcal{M}_{i}) = \int_{\mathcal{M}_{i}} \mathcal{L}(D|\Theta, \mathcal{M}_{i}) \pi(\Theta|\mathcal{M}_{i}) \, {\rm d}\Theta\,.
    \label{eq:bayesian_evidence}
\end{equation}

The relative posterior probabilities between two competing models $\mathcal{M}_{i}$ and $\mathcal{M}_{j}$ are given by the ratio:
\begin{equation}
\frac{P(\mathcal{M}_{i}|D)}{P(\mathcal{M}_{j}|D)} = \mathcal{Z}_{ij} \, \frac{P(\mathcal{M}_{i})}{P(\mathcal{M}_{j})},
\label{eq:model_posterior_ratio}
\end{equation}
where the Bayes factor $\mathcal{Z}_{ij}$ is defined as:
\begin{equation}
\mathcal{Z}_{ij} = \frac{\mathcal{E}(D|\mathcal{M}_{i})}{\mathcal{E}(D|\mathcal{M}_{j})} \equiv \frac{\mathcal{Z}_{i}}{\mathcal{Z}_{j}}.
\label{eq:bayes_factor}
\end{equation}
We therefore define the relative log-Bayesian evidence comparing the DDE parametrization to the $\Lambda$CDM or CPL scenarios as:
\begin{equation}
    \Delta \ln \mathcal{Z}_{ij} \equiv \ln \mathcal{Z}_i - \ln \mathcal{Z}_j\,,
    \label{eq:relative_log_bayesian_evidence}
\end{equation}
where $i \in \{1\mathrm{st}, 2\mathrm{nd} \}$ denotes the DDE model truncated at either first or second order, and $j \in \{\Lambda\mathrm{CDM}, \mathrm{CPL} \}$ represents the baseline model.
Positive values of $\Delta \ln \mathcal{Z}_{ij}$ indicate a preference for the DDE model, with the strength of evidence interpreted according to the revised Jeffreys' scale~\cite{Kass:1995loi}: the evidence is considered \textit{inconclusive} if $0 \leq |\Delta \ln \mathcal{Z}_{ij}| \leq 1$, \textit{weak} if $1 < |\Delta \ln \mathcal{Z}_{ij}| \leq 2.5$, \textit{moderate} if $2.5 < |\Delta \ln \mathcal{Z}_{ij}| \leq 5$, \textit{strong} if $5 < |\Delta \ln \mathcal{Z}_{ij}| \leq 10$, and \textit{very strong} if $|\Delta \ln \mathcal{Z}_{ij}| > 10$, in favor of the preferred model.

\section{Results}
\label{sec:results}
In this section, we present observational constraints on our DDE extensions at first and second order and evaluate their model preference statistics compared to $\Lambda\mathrm{CDM}$ and the CPL parametrization.

\subsection{First-Order Expansion Results}
\begin{figure}[htbp]
    \centering
    \includegraphics[width=0.9\linewidth]{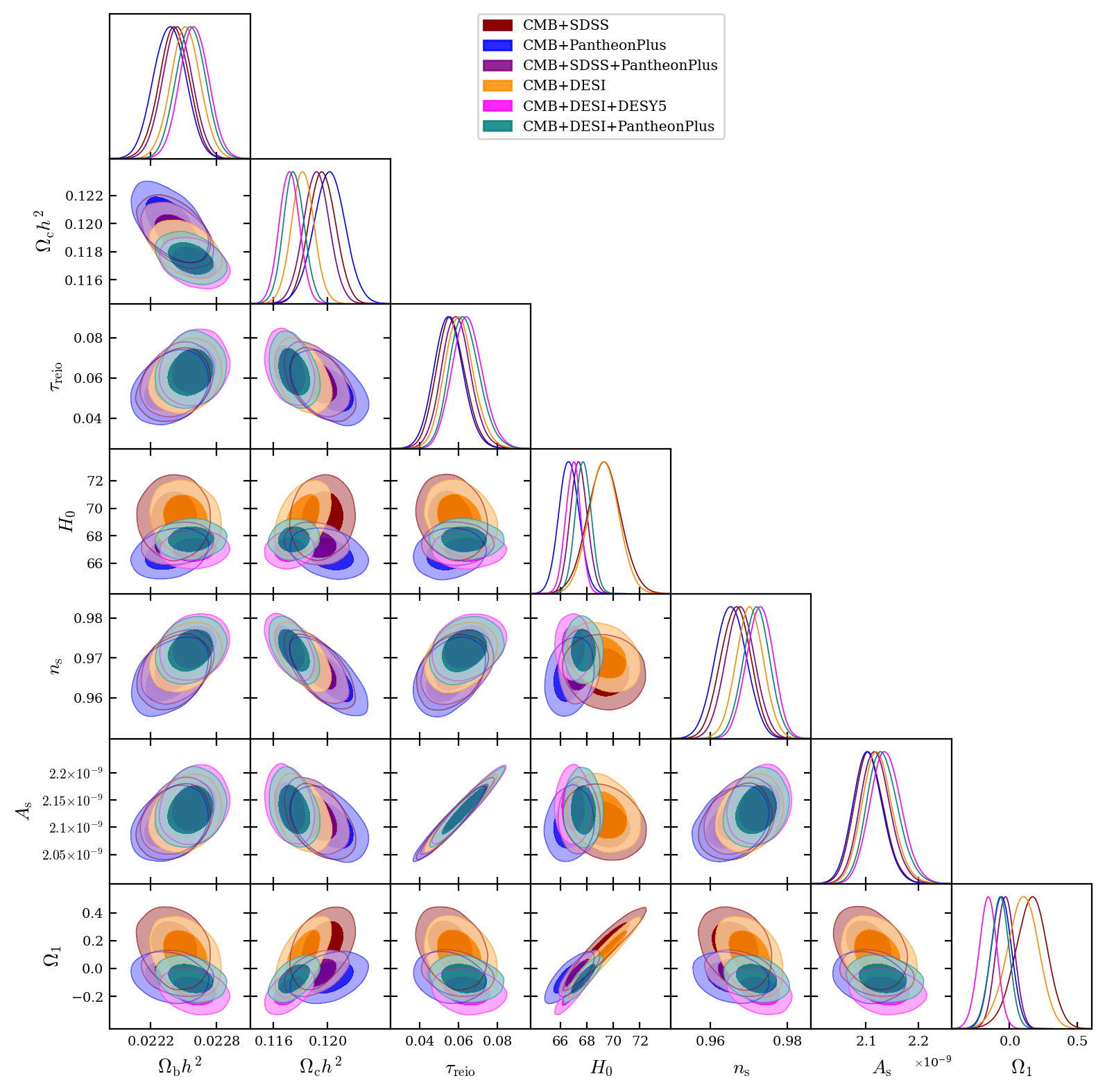}
    \caption{{\bf Expansion to first order.} One-dimensional posterior distributions and two-dimensional marginalized contours for the key parameters of the model. Results are obtained by considering different dataset combinations: CMB+SDSS, CMB+PantheonPlus, CMB+SDSS+PantheonPlus, CMB+DESI, CMB+DESI+DESY5, and CMB+DESI+PantheonPlus.}
    \label{fig:Post_distribution_O1}
\end{figure}

For the first-order expansion, we allow $\Omega_{\mathrm{1}}$ to vary according to the priors set in table~\ref{tab:prior_table}. The marginalized contours and posterior distributions are shown in figure~\ref{fig:Post_distribution_O1}, with the parameter constraints at 68\% confidence level (CL) summarized in table~\ref{tab:CosmoParams_first}.
The first-order expansion model is characterized by the parameter $\Omega_1$, which quantifies deviations from the cosmological constant model (where $\Lambda$CDM corresponds to $\Omega_1 = 0$). Our results reveal that constraints on $\Omega_1$ and other cosmological parameters vary depending on the dataset combination:
\begin{itemize}
\item With CMB data alone, parameters are poorly constrained, showing a strong preference for a high $H_0 = 83^{+10}_{-7}$\,km/s/Mpc and a low matter density $\Omega_{\mathrm{m}} = 0.217^{+0.019}_{-0.074}$. For $\Omega_1$, we find $\Omega_1 = 0.89^{+0.62}_{-0.14}$, indicating a significant deviation from zero, though with large uncertainty. Both these parameters trend in the right direction to alleviate the cosmological tensions on $H_0$ and $S_8$.
\item The addition of SDSS data breaks parameter degeneracies, substantially strengthening the constraints, with $H_0 = 69.3 \pm 1.3$\,km/s/Mpc and $\Omega_{\mathrm{m}} = 0.297 \pm 0.011$, closer to the values obtained under $\Lambda$CDM. The constraint on $\Omega_1$ tightens to $\Omega_1 = 0.16^{+0.12}_{-0.11}$, showing a mild preference for positive values but consistent with zero at approximately $1.5\sigma$.
\item The CMB+PantheonPlus combination yields $\Omega_1 = -0.062 \pm 0.075$, consistent with zero at less than $1\sigma$, and $H_0 = 66.68 \pm 0.76$\,km/s/Mpc, notably lower than with other datasets. A lower $H_0$ corresponds to a higher $\Omega_{\mathrm{m}} = 0.3221 \pm 0.0084$, reinstating the cosmological tensions.
\item For CMB+SDSS+PantheonPlus, we find $\Omega_1 = -0.032 \pm 0.068$, showing excellent consistency with $\Lambda$CDM. Consequently, both the Hubble constant, $H_0 = 67.40 \pm 0.64$\,km/s/Mpc, and the matter density, $\Omega_{\mathrm{m}} = 0.3132 \pm 0.0065$, agree well with the Planck 2018 $\Lambda$CDM results~\cite{Planck:2018vyg}.
\item Replacing SDSS with the more recent DESI DR2 data in CMB+DESI yields $\Omega_1 = 0.095^{+0.12}_{-0.11}$, showing a slight preference for positive values but still consistent with zero at less than $1\sigma$. The parameters $H_0 = 69.3 \pm 1.1$\,km/s/Mpc and $\Omega_{\mathrm{m}} = 0.2946 \pm 0.0090$ are comparable to those from CMB+SDSS.
\item The most significant deviation from $\Lambda$CDM appears with CMB+DESI+DESY5, giving $\Omega_1 = -0.162 \pm 0.067$, indicating a preference for negative $\Omega_1$ at more than $2\sigma$. The $\Delta \chi^2_{\mathrm{min},\Lambda \mathrm{CDM}}$ value is $-7.35$, corresponding to a preference for the DDE model over $\Lambda$CDM at the $2.7\sigma$ level, as calculated using eq.~\eqref{eq:Nsigma_conversion}. This represents the strongest indication against $\Lambda$CDM among our dataset combinations, suggesting that the addition of DESY5 data may be particularly sensitive to deviations from the cosmological constant model.
\item Finally, CMB+DESI+PantheonPlus yields $\Omega_1 = -0.072 \pm 0.068$, consistent with zero at just over $1\sigma$, and gives $H_0 = 67.74 \pm 0.61$\,km/s/Mpc and $\Omega_{\mathrm{m}} = 0.3068 \pm 0.0056$.
\end{itemize}

We now present the goodness of fit and Bayesian evidence of the first-order expansion compared to the standard $\Lambda$CDM and CPL ($w_0$--$w_a$) models. When comparing to $\Lambda$CDM, table~\ref{tab:CosmoParams_first} shows the first-order model performs better in terms of the maximum-likelihood fit ($\Delta \chi^2_{\mathrm{min},\Lambda \mathrm{CDM}}$) for the CMB-only and CMB+DESI+DESY5  datasets, while for all other cases it is comparable to $\Lambda$CDM. Nevertheless, having one additional parameter means any improvement in $\chi^2$ must be sufficiently large to yield a positive $\Delta \ln \mathcal{Z}_{\Lambda \mathrm{CDM}}$. In practice, most combinations in table~\ref{tab:CosmoParams_first}  aa show $\Delta \ln \mathcal{Z}_{\Lambda \mathrm{CDM}} < 0$, indicating that the evidence does not favor adding this extra parameter.  A notable example is the CMB+DESI+DESY5 dataset, where $\Delta \chi^2_{\mathrm{min},\Lambda \mathrm{CDM}} = -7.35$ suggests a better fit, but $\Delta \ln \mathcal{Z}_{\Lambda \mathrm{CDM}} = 0.0$ reveals no clear Bayesian preference for the first-order model over $\Lambda$CDM. Similarly, in the CMB-only case, the resulting $\Delta \ln \mathcal{Z}_{\Lambda \mathrm{CDM}}$ is less than 1, indicating that two models are statistically indistinguishable in terms of Bayesian evidence. 
When comparing to the CPL parameterization, table~\ref{tab:CosmoParams_first} reveals that across all data combinations, the CPL parameterization consistently provides a better fit compared to the first-order expansion model (as evidenced by positive $\Delta \chi^2_{\mathrm{min},\mathrm{CPL}}$ values). However, the Bayesian evidence typically favors the simpler first-order expansion model unless the improvement in fit is sufficiently large to overcome the penalty due to the increased model complexity. As shown in the bottom part of table~\ref{tab:CosmoParams_first}, in cases where $\Delta \chi^2_{\mathrm{min},\mathrm{CPL}}$ is much higher, the Bayesian evidence shifts to favor the CPL parameterization over the first-order expansion model.

\begin{table}[htbp]
\scriptsize
\centering
\resizebox{\textwidth}{!}{%
\begin{tabular}{l|ccccccc}
\toprule
\textbf{Parameters} 
& \textbf{CMB} 
& \textbf{CMB} 
& \textbf{CMB} 
& \textbf{CMB+SDSS}
& \textbf{CMB} 
& \textbf{CMB} 
& \textbf{CMB+DESI} \\
&
& \textbf{+SDSS} 
& \textbf{+PantheonPlus} 
& \textbf{+PantheonPlus}
& \textbf{+DESI} 
& \textbf{+DESI+DESY5} 
& \textbf{+PantheonPlus} \\
\midrule
$\Omega_{\mathrm{b}}\,h^2$ 
& $0.02243 \pm 0.00015$ 
& $0.02242 \pm 0.00014$ 
& $0.02238 \pm 0.00015$ 
& $0.02245 \pm 0.00014$
& $0.02252 \pm 0.00013$ 
& $0.02260 \pm 0.00013$ 
& $0.02257 \pm 0.00013$ \\[6pt]

$\Omega_{\mathrm{c}}\,h^2$ 
& $0.1193 \pm 0.0012$ 
& $0.11960 \pm 0.00098$ 
& $0.1201 \pm 0.0012$ 
& $0.11918 \pm 0.00096$
& $0.11817 \pm 0.00084$ 
& $0.11719 \pm 0.00075$ 
& $0.11752 \pm 0.00076$ \\[6pt]

$100\,\theta_{\mathrm{MC}}$ 
& $1.04101 \pm 0.00031$ 
& $1.04099 \pm 0.00030$ 
& $1.04092 \pm 0.00031$ 
& $1.04105 \pm 0.00029$
& $1.04118 \pm 0.00028$ 
& $1.04130 \pm 0.00028$ 
& $1.04125 \pm 0.00028$ \\[6pt]

$\tau_{\mathrm{reio}}$ 
& $0.0526 \pm 0.0076$ 
& $0.0560 \pm 0.0074$ 
& $0.0549 \pm 0.0075$ 
& $0.0588^{+0.0069}_{-0.0077}$
& $0.0604^{+0.0068}_{-0.0079}$ 
& $0.0646 \pm 0.0078$ 
& $0.0632^{+0.0069}_{-0.0083}$ \\[6pt]

$n_{\mathrm{s}}$ 
& $0.9671 \pm 0.0043$ 
& $0.9665 \pm 0.0039$ 
& $0.9653 \pm 0.0041$ 
& $0.9677 \pm 0.0038$
& $0.9702 \pm 0.0036$ 
& $0.9728 \pm 0.0035$ 
& $0.9719 \pm 0.0035$ \\[6pt]

$\ln\bigl(10^{10}\,A_{\mathrm{s}}\bigr)$ 
& $3.038 \pm 0.014$ 
& $3.048 \pm 0.013$ 
& $3.047 \pm 0.014$ 
& $3.053^{+0.013}_{-0.014}$
& $3.055^{+0.013}_{-0.014}$ 
& $3.063 \pm 0.014$ 
& $3.060^{+0.013}_{-0.015}$ \\[6pt]

$\Omega_{1}$ 
& $0.89^{+0.62}_{-0.14}$ 
& $0.16^{+0.12}_{-0.11}$ 
& $-0.062 \pm 0.075$ 
& $-0.032 \pm 0.068$
& $0.095^{+0.12}_{-0.11}$ 
& $-0.162 \pm 0.067$ 
& $-0.072 \pm 0.068$ \\[6pt]
\midrule

$H_0$ [km/s/Mpc] 
& $83^{+10}_{-7}$ 
& $69.3 \pm 1.3$ 
& $66.68 \pm 0.76$ 
& $67.40 \pm 0.64$
& $69.3 \pm 1.1$ 
& $66.98 \pm 0.56$ 
& $67.74 \pm 0.61$ \\[6pt]

$\sigma_{8}$ 
& $0.939^{+0.11}_{-0.053}$ 
& $0.828 \pm 0.013$ 
& $0.8067 \pm 0.0088$ 
& $0.8097 \pm 0.0082$
& $0.820 \pm 0.013$ 
& $0.7957 \pm 0.0083$ 
& $0.8039 \pm 0.0085$ \\[6pt]

$S_{8}$ 
& $0.783^{+0.024}_{-0.043}$ 
& $0.8236 \pm 0.0095$ 
& $0.836 \pm 0.011$ 
& $0.8272 \pm 0.0093$
& $0.8128 \pm 0.0077$ 
& $0.8128 \pm 0.0078$ 
& $0.8128 \pm 0.0078$ \\[6pt]

$\Omega_{\mathrm{m}}$ 
& $0.217^{+0.019}_{-0.074}$ 
& $0.297 \pm 0.011$ 
& $0.3221 \pm 0.0084$ 
& $0.3132 \pm 0.0065$
& $0.2946 \pm 0.0090$ 
& $0.3131 \pm 0.0053$ 
& $0.3068 \pm 0.0056$ \\[6pt]
\midrule

$\Delta \chi^2_{\mathrm{min},\Lambda \mathrm{CDM}}$ 
& $-4.13$ 
& $0.23$ 
& $0.34$ 
& $0.28$
& $0.7$ 
& $-7.35$ 
& $-0.33$ \\[6pt]

$\Delta \ln \mathcal{Z}_{\Lambda \mathrm{CDM}}$ 
& $0.57$ 
& $-1.98$ 
& $-3.0$ 
& $-3.09$
& $-2.5$ 
& $0.0$ 
& $-2.68$ \\[6pt]

$\Delta \chi^2_{\mathrm{min},\mathrm{CPL}}$ 
& $0.96$ 
& $2.68$ 
& $1.85$ 
& $4.45$
& $8.25$ 
& $12.1$ 
& $7.2$ \\[6pt]

$\Delta \ln \mathcal{Z}_{\mathrm{CPL}}$ 
& $0.55$ 
& $1.21$ 
& $2.05$ 
& $0.93$
& $-1.94$ 
& $-2.71$ 
& $-0.1$ \\
\bottomrule
\end{tabular}
}
\caption{{\bf Expansion to first order.} Parameter constraints at $68\%$ CL from different datasets for the first-order expansion. $\Delta \chi^2_{\mathrm{min}}$ and $\Delta \ln \mathcal{Z}$ are defined as $\Delta \chi^2_{\mathrm{min},\Lambda \mathrm{CDM}/\mathrm{CPL}} = \chi^2_{\mathrm{min},1\mathrm{st}} - \chi^2_{\mathrm{min},\Lambda \mathrm{CDM}/\mathrm{CPL}}$, and $\Delta \ln \mathcal{Z}_{\Lambda \mathrm{CDM}/\mathrm{CPL}} \equiv \ln \mathcal{Z}_{\mathrm{1st}} - \ln \mathcal{Z}_{\Lambda \mathrm{CDM}/\mathrm{CPL}}$. Negative values of $\Delta \chi^2_{\mathrm{min},\Lambda \mathrm{CDM}/\mathrm{CPL}}$ favor the Pressure Parameterization DE model over the standard $\Lambda$CDM/CPL scenario, while positive values of $\Delta \ln \mathcal{Z}_{\Lambda \mathrm{CDM}/\mathrm{CPL}}$ indicate a preference for the DDE model.}
\label{tab:CosmoParams_first}
\end{table}

\subsection{Second-Order Expansion Results}

\begin{figure}[htbp]
    \centering
    \includegraphics[width=0.9\linewidth]{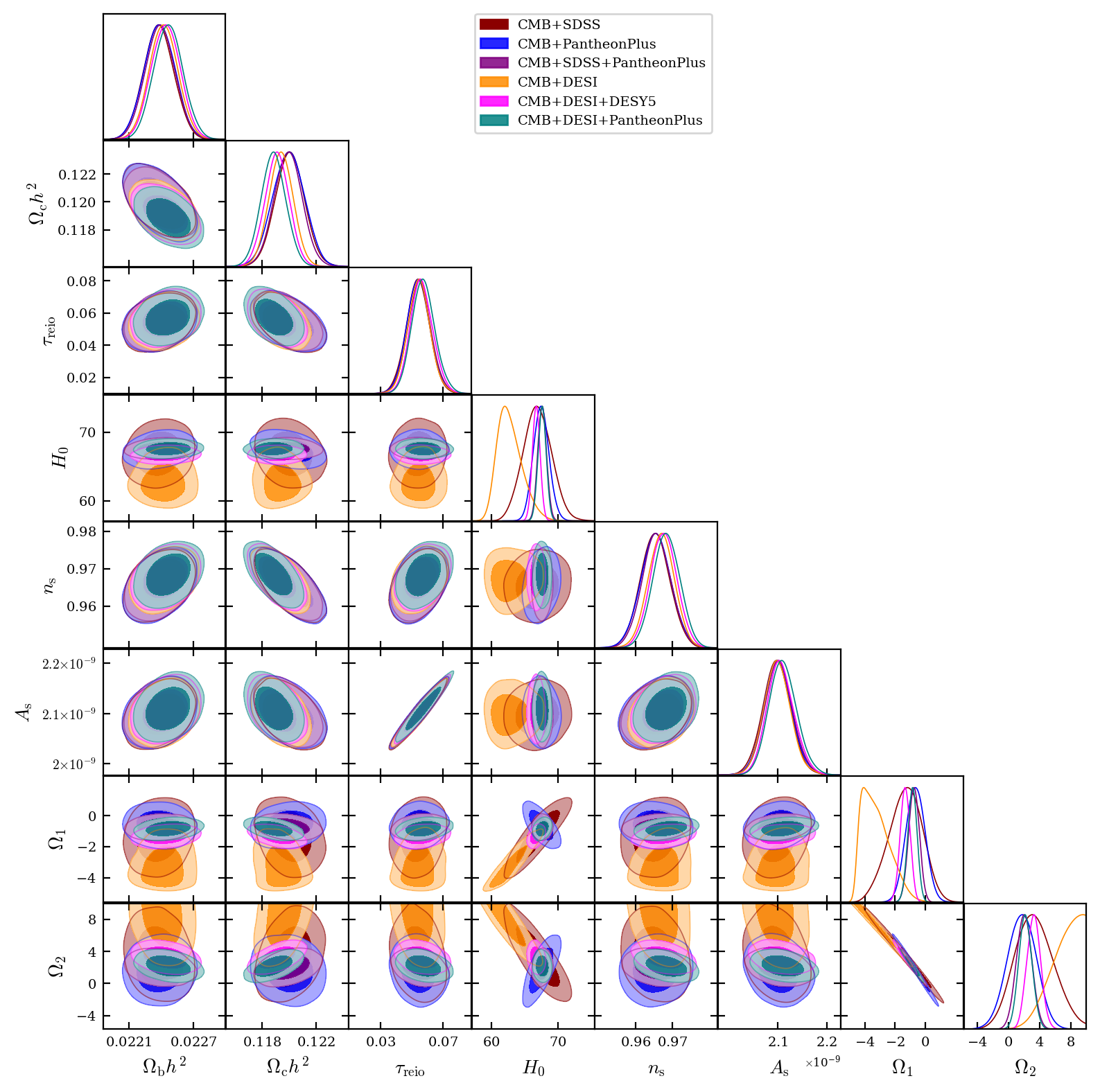}
    \caption{{\bf Expansion to second order.} One-dimensional posterior distributions and two-dimensional marginalized contours for the second-order expansion for the key parameters, as obtained from different dataset combinations: CMB+SDSS, CMB+PantheonPlus, CMB+SDSS+PantheonPlus, CMB+DESI, CMB+DESI+DESY5, and CMB+DESI+PantheonPlus.}
    \label{fig:Post_sec_10}
\end{figure}

For the second-order expansion, we allow both $\Omega_{\mathrm{1}}$ and $\Omega_{\mathrm{2}}$ to vary according to the priors given in table~\ref{tab:prior_table} (with $\Lambda$CDM corresponding to $\Omega_1 = \Omega_2 = 0$). The marginalized contours and posterior distributions are shown in figure~\ref{fig:Post_sec_10}, with parameter constraints at the 68\% confidence level summarized in table~\ref{tab:CosmoParams_sec}.

Our analysis reveals:
\begin{itemize}
\item With CMB data alone, both DE parameters remain poorly constrained, with $\Omega_1 = -0.6^{+1.7}_{-2.6}$ and only a lower limit on $\Omega_2 > 1.3$. Other cosmological parameters are also poorly constrained, with $H_0 = 79\pm10$\,km/s/Mpc and $\Omega_{\mathrm{m}} = 0.246^{+0.027}_{-0.099}$, with a shift of the mean values in the direction of alleviating the cosmological tensions.
\item Adding SDSS data significantly improves the constraints and shifts the preferred values, yielding $\Omega_1 = -1.3^{+1.2}_{-1.0}$ and $\Omega_2 = 3.3^{+2.3}_{-2.6}$. This indicates a mild preference for negative $\Omega_1$ and positive $\Omega_2$ at slightly more than $1\sigma$. The $\Delta \chi^2_{\mathrm{min},\Lambda \mathrm{CDM}}$ value of $-2.72$, corresponding to $1.1\sigma$ as calculated using eq.~\eqref{eq:Nsigma_conversion}, indicates that a mild preference for the DDE model than $\Lambda$CDM at slightly more than $1\sigma$ level. The Hubble constant becomes $H_0 = 66.9 \pm 2.1$\,km/s/Mpc, much closer to the value under $\Lambda$CDM.
\item The CMB+PantheonPlus combination yields $\Omega_1 = -0.64 \pm 0.63$ and $\Omega_2 = 1.7 \pm 1.9$, showing a similar preference for negative $\Omega_1$ and positive $\Omega_2$, though with substantial uncertainties. The $\Delta \chi^2_{\mathrm{min},\Lambda \mathrm{CDM}}$ value of $-1.49$ indicates that the DDE model is consistent with $\Lambda$CDM within the $1\sigma$ level. This combination provides $H_0 = 67.5 \pm 1.2$\,km/s/Mpc, in good agreement with the Planck 2018 $\Lambda$CDM results~\cite{Planck:2018vyg}.
\item More stringent constraints come from the CMB+SDSS+PantheonPlus combination, with $\Omega_1 = -0.74 \pm 0.36$ and $\Omega_2 = 2.01 \pm 0.99$. The $\Delta \chi^2_{\mathrm{min},\Lambda \mathrm{CDM}}$ value of $-3.35$ corresponds to a preference for the DDE model over $\Lambda$CDM at the $1.3\sigma$ level, as calculated using eq.~\eqref{eq:Nsigma_conversion}.
\item Replacing SDSS with DESI DR2 provides stronger evidence for deviations from $\Lambda$CDM. The CMB+DESI combination yields $\Omega_1 = -3.24^{+0.54}_{-1.2}$ and a lower limit of $\Omega_2 > 6.44$. The $\Delta \chi^2_{\mathrm{min},\Lambda \mathrm{CDM}}$ value of $-8.97$ corresponds to a preference for the DDE model over $\Lambda$CDM at the $2.5\sigma$ level, as calculated using eq.~\eqref{eq:Nsigma_conversion}. The Hubble constant becomes $H_0 = 62.8^{+1.4}_{-2.2}$\,km/s/Mpc, while the matter density rises to $\Omega_{\mathrm{m}} = 0.362^{+0.024}_{-0.017}$.
\item The most significant evidence for deviation comes from CMB+DESI+DESY5, which constrains $\Omega_1 = -1.33 \pm 0.33$ and $\Omega_2 = 3.25\pm0.88$. The $\Delta \chi^2_{\mathrm{min},\Lambda \mathrm{CDM}}$ value of $-18.41$ corresponds to a strong preference for the DDE model over $\Lambda$CDM at the $4\sigma$ level, as calculated using eq.~\eqref{eq:Nsigma_conversion}, providing strong evidence that DDE deviates substantially from a cosmological constant.
\item Finally, CMB+DESI+PantheonPlus yields $\Omega_1 = -0.85 \pm 0.31$ and $\Omega_2 = 2.17 \pm 0.85$. The $\Delta \chi^2_{\mathrm{min},\Lambda \mathrm{CDM}}$ value of $-6.53$ corresponds to a preference for the DDE model over $\Lambda$CDM at the $2\sigma$ level, as calculated using eq.~\eqref{eq:Nsigma_conversion}. This combination provides $H_0 = 67.62 \pm 0.60$\,km/s/Mpc and $\Omega_{\mathrm{m}} = 0.3106 \pm 0.0057$, values consistent with the Planck 2018 $\Lambda$CDM results~\cite{Planck:2018vyg}.
\end{itemize}

A notable feature in our results is the strong negative correlation between the $\Omega_1$ and $\Omega_2$ parameters, visible in figure~\ref{fig:Post_sec_10}. This correlation is consistent across all dataset combinations and indicates that models with more negative $\Omega_1$ values require more positive $\Omega_2$ values to maintain consistency with observational data.

We compare the goodness of fit and Bayesian evidence of the second-order expansion against the standard $\Lambda$CDM and CPL ($w_0$--$w_a$) models. When comparing the second-order expansion to $\Lambda$CDM, we find negative $\Delta \chi^2_{\mathrm{min},\Lambda \mathrm{CDM}}$ values for most dataset combinations, indicating a better fit of our model. The improvement is particularly dramatic for CMB+DESI+DESY5, with $\Delta \chi^2_{\mathrm{min},\Lambda \mathrm{CDM}} = -18.41$, suggesting a $4\sigma$ preference for DDE model.
When considering Bayesian evidence ($\Delta \ln \mathcal{Z}_{\Lambda \mathrm{CDM}}$), which penalizes additional model complexity, the results are more nuanced. For most dataset combinations, $\Lambda$CDM is favored due to its simplicity, with negative $\Delta \ln \mathcal{Z}_{\Lambda \mathrm{CDM}}$ values. However, for CMB+DESI+DESY5, we find $\Delta \ln \mathcal{Z}_{\Lambda \mathrm{CDM}} = 1.84$, indicating that even after accounting for the complexity penalty, the second-order expansion is moderately favored. According to the Jeffreys' scale~\cite{Kass:1995loi}, this represents ``moderate'' evidence in favor of our model.
Comparing the second-order expansion to the CPL parameterization, we find the absolute values of both $\Delta \chi^2_{\mathrm{min},\mathrm{CPL}}$ and $\Delta \ln \mathcal{Z}_{\mathrm{CPL}}$ are typically less than 1, implying that differences between the two models are minor in both likelihood and Bayesian terms. Thus, while our second-order model is marginally disfavored relative to CPL in terms of maximum-likelihood fit, the small values of $\Delta \ln \mathcal{Z}$ indicate that they remain nearly indistinguishable according to Bayesian evidence.


\begin{table}[htbp]
\scriptsize
\centering
\resizebox{\textwidth}{!}{%
\begin{tabular}{l|ccccccc}
\hline
\textbf{Parameters} 
& \textbf{CMB} 
& \textbf{CMB} 
& \textbf{CMB} 
& \textbf{CMB+SDSS}
& \textbf{CMB} 
& \textbf{CMB} 
& \textbf{CMB+DESI} \\
&
& \textbf{+SDSS} 
& \textbf{+PantheonPlus} 
& \textbf{+PantheonPlus}
& \textbf{+DESI} 
& \textbf{+DESI+DESY5} 
& \textbf{+PantheonPlus} \\
\hline
$\Omega_{\mathrm{b}}\,h^2$ 
& $0.02244 \pm 0.00016$ 
& $0.02238 \pm 0.00014$ 
& $0.02239 \pm 0.00014$ 
& $0.02239 \pm 0.00014$
& $0.02243 \pm 0.00013$ 
& $0.02244 \pm 0.00013$ 
& $0.02247 \pm 0.00013$ \\[6pt]

$\Omega_{\mathrm{c}}\,h^2$ 
& $0.1192 \pm 0.0013$ 
& $0.1201 \pm 0.0011$ 
& $0.1200 \pm 0.0012$ 
& $0.1199 \pm 0.0010$
& $0.11941 \pm 0.00090$ 
& $0.11913 \pm 0.00089$ 
& $0.11886 \pm 0.00091$ \\[6pt]

$100\,\theta_{\mathrm{MC}}$ 
& $1.04101 \pm 0.00032$ 
& $1.04092 \pm 0.00030$ 
& $1.04093 \pm 0.00030$ 
& $1.04094 \pm 0.00030$
& $1.04102 \pm 0.00029$ 
& $1.04106 \pm 0.00029$ 
& $1.04109 \pm 0.00029$ \\[6pt]

$\tau_{\mathrm{reio}}$ 
& $0.0522 \pm 0.0076$ 
& $0.0539 \pm 0.0075$ 
& $0.0543 \pm 0.0076$ 
& $0.0549 \pm 0.0075$
& $0.0546 \pm 0.0072$ 
& $0.0563^{+0.0067}_{-0.0076}$ 
& $0.0574^{+0.0069}_{-0.0077}$ \\[6pt]

$n_{\mathrm{s}}$ 
& $0.9675 \pm 0.0045$ 
& $0.9654 \pm 0.0040$ 
& $0.9656 \pm 0.0041$ 
& $0.9658 \pm 0.0039$
& $0.9670 \pm 0.0037$ 
& $0.9677 \pm 0.0036$ 
& $0.9684 \pm 0.0037$ \\[6pt]

$\ln\bigl(10^{10}\,A_{\mathrm{s}}\bigr)$ 
& $3.036 \pm 0.015$ 
& $3.043 \pm 0.014$ 
& $3.044 \pm 0.014$ 
& $3.045 \pm 0.014$
& $3.044 \pm 0.013$ 
& $3.047^{+0.012}_{-0.014}$ 
& $3.049 \pm 0.014$ \\[6pt]

$\Omega_{1}$ 
& $-0.6^{+1.7}_{-2.6}$ 
& $-1.3^{+1.2}_{-1.0}$ 
& $-0.64 \pm 0.63$ 
& $-0.74 \pm 0.36$
& $-3.24^{+0.54}_{-1.2}$ 
& $-1.33 \pm 0.33$ 
& $-0.85 \pm 0.31$ \\[6pt]

$\Omega_{2}$ 
& $> 1.30$ 
& $3.3^{+2.3}_{-2.6}$ 
& $1.7 \pm 1.9$ 
& $2.01 \pm 0.99$
& $> 6.44$ 
& $3.25 \pm 0.88$ 
& $2.17 \pm 0.85$ \\[6pt]
\midrule

$H_0$ [km/s/Mpc] 
& $79 \pm 10$ 
& $66.9 \pm 2.1$ 
& $67.5 \pm 1.2$ 
& $67.62 \pm 0.65$
& $62.8^{+1.4}_{-2.2}$ 
& $66.73 \pm 0.57$ 
& $67.62 \pm 0.60$ \\[6pt]

$\sigma_{8}$ 
& $0.904^{+0.11}_{-0.077}$ 
& $0.811 \pm 0.018$ 
& $0.814 \pm 0.012$ 
& $0.8163 \pm 0.0089$
& $0.772^{+0.015}_{-0.019}$ 
& $0.8047 \pm 0.0085$ 
& $0.8106 \pm 0.0087$ \\[6pt]

$S_{8}$ 
& $0.797^{+0.032}_{-0.045}$ 
& $0.838 \pm 0.014$ 
& $0.833 \pm 0.011$ 
& $0.8334 \pm 0.0098$
& $0.848^{+0.013}_{-0.0098}$ 
& $0.8304 \pm 0.0090$ 
& $0.8248 \pm 0.0090$ \\[6pt]

$\Omega_{\mathrm{m}}$ 
& $0.246^{+0.027}_{-0.099}$ 
& $0.321^{+0.019}_{-0.022}$ 
& $0.314 \pm 0.012$ 
& $0.3128 \pm 0.0065$
& $0.362^{+0.024}_{-0.017}$ 
& $0.3194 \pm 0.0058$ 
& $0.3106 \pm 0.0057$ \\[6pt]

\midrule

$\Delta \chi^2_{\mathrm{min},\Lambda \mathrm{CDM}}$ 
& $-6.31$ 
& $-2.72$ 
& $-1.49$ 
& $-3.35$
& $-8.97$ 
& $-18.41$ 
& $-6.53$ \\[6pt]

$\Delta \ln \mathcal{Z}_{\Lambda \mathrm{CDM}}$ 
& $-0.36$ 
& $-4.07$ 
& $-5.9$ 
& $-5.3$
& $-1.0$ 
& $1.84$ 
& $-3.52$ \\[6pt]

$\Delta \chi^2_{\mathrm{min},\mathrm{CPL}}$ 
& $-1.23$ 
& $-0.27$ 
& $0.01$ 
& $0.83$
& $-1.42$ 
& $1.04$ 
& $1.01$ \\[6pt]

$\Delta \ln \mathcal{Z}_{\mathrm{CPL}}$ 
& $-0.38$ 
& $-0.88$ 
& $-0.85$ 
& $-1.28$
& $-0.44$ 
& $-0.87$ 
& $-0.94$ \\
\hline
\end{tabular}}
\caption{{\bf Expansion to second order.} Parameter constraints at $68\%$ CL from different data sets for second-order expansion. $\Delta \chi^2_{\mathrm{min}}$ and $\Delta \ln \mathcal{Z}$ are defined as $\Delta \chi^2_{\mathrm{min},\Lambda \mathrm{CDM}/\mathrm{CPL}} = \chi^2_{\mathrm{min},2\mathrm{nd}} - \chi^2_{\mathrm{min},\Lambda \mathrm{CDM}/\mathrm{CPL}}$, and $\Delta \ln \mathcal{Z}_{\Lambda \mathrm{CDM}/\mathrm{CPL}} \equiv \ln \mathcal{Z}_{\mathrm{2nd}} - \ln \mathcal{Z}_{\mathrm{\Lambda CDM}/\mathrm{CPL}}$. Negative values of $\Delta \chi^2_{\mathrm{min},\Lambda \mathrm{CDM}/\mathrm{CPL}}$ favor the Pressure Parameterization DE model over the standard $\Lambda$CDM/CPL scenario, while positive values of $\Delta \ln \mathcal{Z}_{\Lambda \mathrm{CDM}/\mathrm{CPL}}$ indicate a preference for the DDE model.}
\label{tab:CosmoParams_sec}
\end{table}

\subsection{Dark Energy Evolution}

\begin{figure*}[t!]
    \centering
    \begin{subfigure}[t]{0.325\textwidth}
        \centering
        \includegraphics[width=\linewidth]{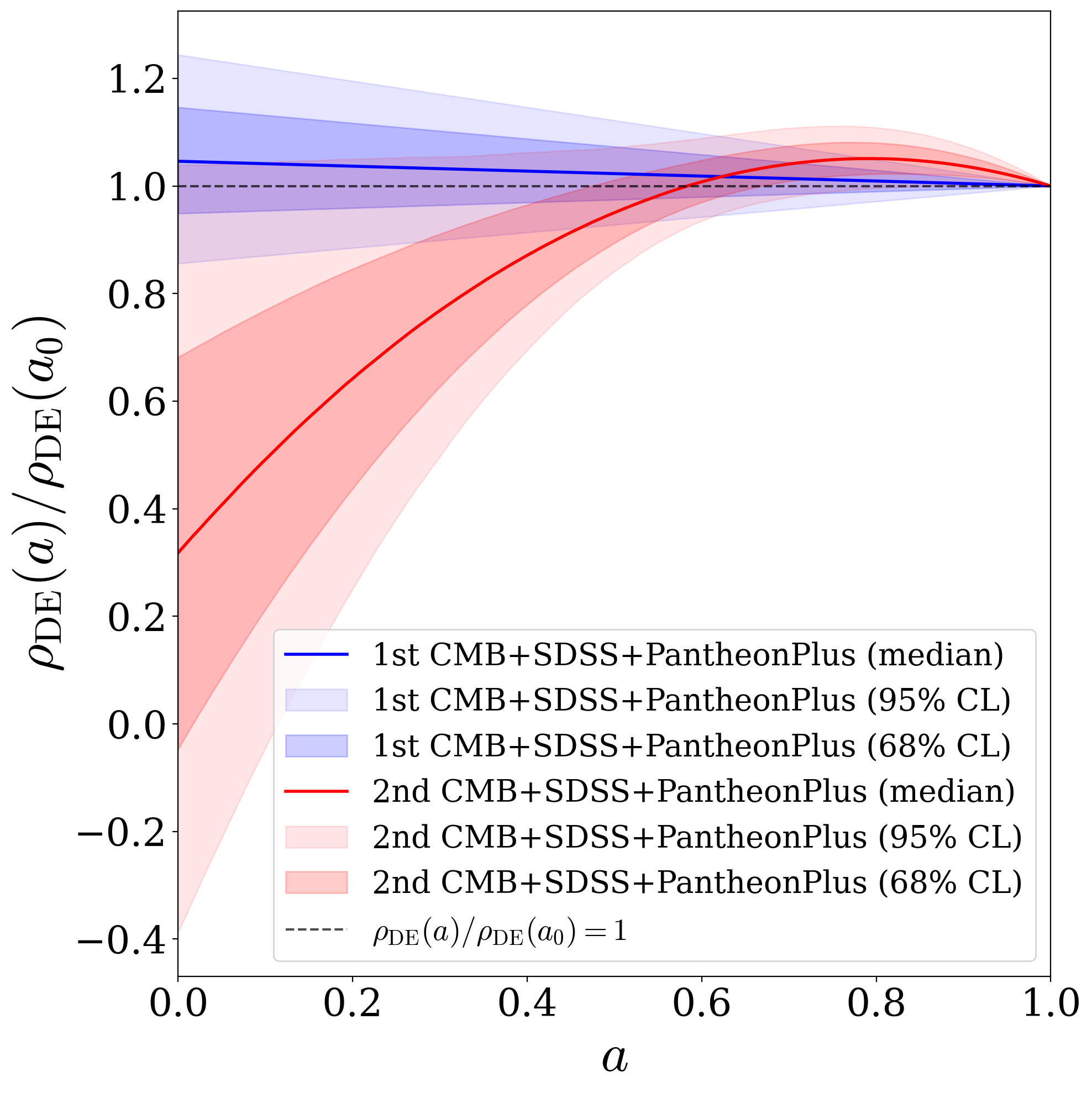}
    \end{subfigure}
    \begin{subfigure}[t]{0.325\textwidth}
        \centering
        \includegraphics[width=\linewidth]{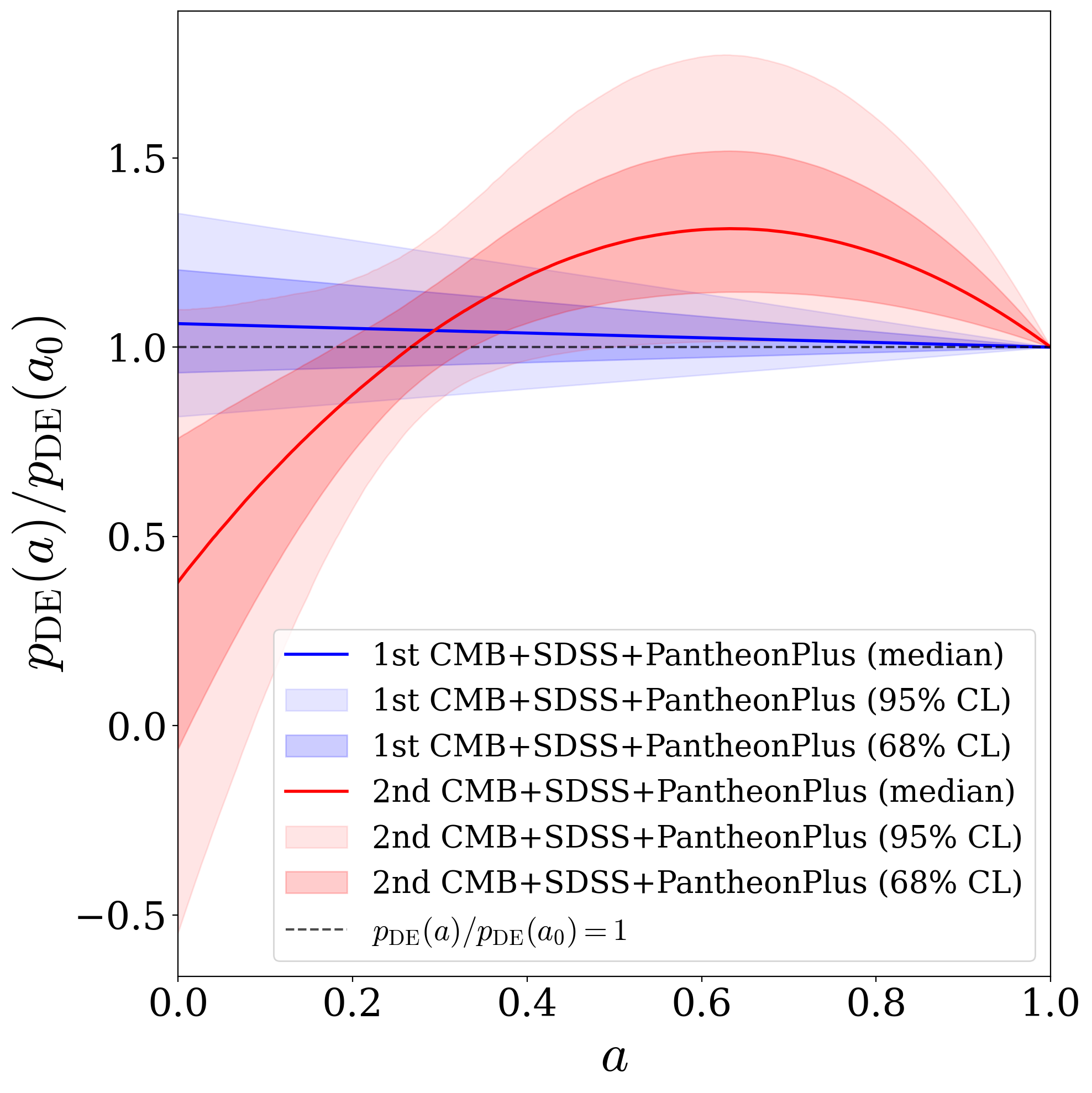}
    \end{subfigure}
    \begin{subfigure}[t]{0.325\textwidth}
        \centering
        \includegraphics[width=\linewidth]{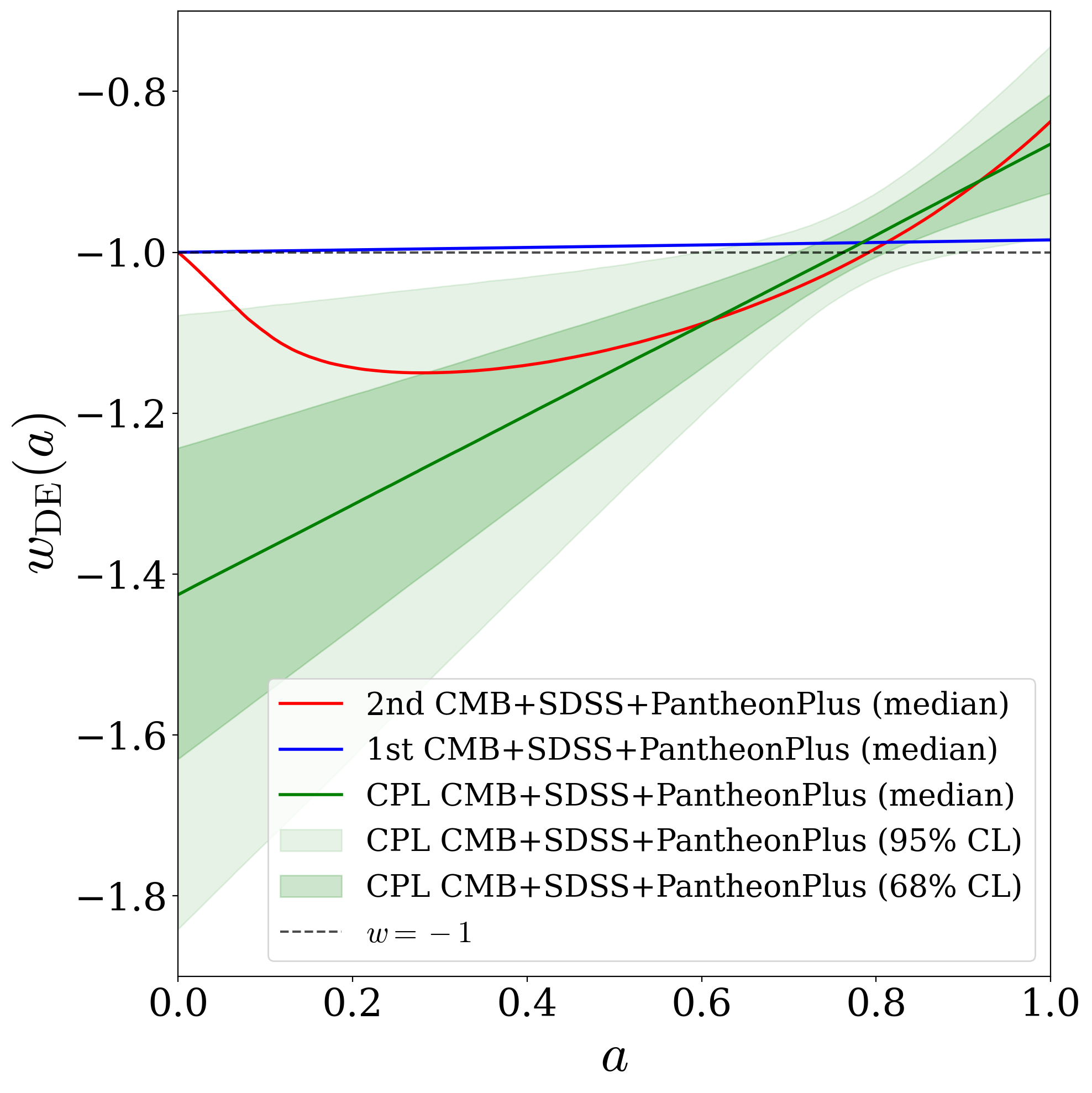}
    \end{subfigure}
    \caption{Reconstruction of the normalized DE density $\rho_{\mathrm{DE}}(a)/\rho_{\mathrm{DE}}(a_0)$ (left panel) and pressure $p_{\mathrm{DE}}(a)/p_{\mathrm{DE}}(a_0)$ (middle panel) as functions of the scale factor $a$. The solid red and blue lines represent the mean curves from the second-order and first-order parameterizations, respectively, based on MCMC samples using the combined datasets \textbf{CMB}+\textbf{SDSS}+\textbf{PantheonPlus}. Shaded regions denote the $68\%$ and $95\%$ CL for each parameterization. The horizontal dashed black line indicates the present-day value, normalized to unity at $a = a_0$, for reference. Also shown is the evolution of the dark-energy equation of state $w_{\rm DE}(a)$ (right panel) for the first-order (blue) and second-order (red) expansions, compared to the CPL parameterization (green), using the same data combination. The solid lines show the medians, while the dark (light) shaded bands represent the $68\%$ ($95\%$) CL for the CPL parameterization. The dashed horizontal line marks the cosmological constant value $w = -1$.}
    \label{fig:sdss_pp}
\end{figure*}

\begin{figure*}[t!]
    \centering
    \begin{subfigure}[t]{0.325\textwidth}
        \centering
        \includegraphics[width=\linewidth]{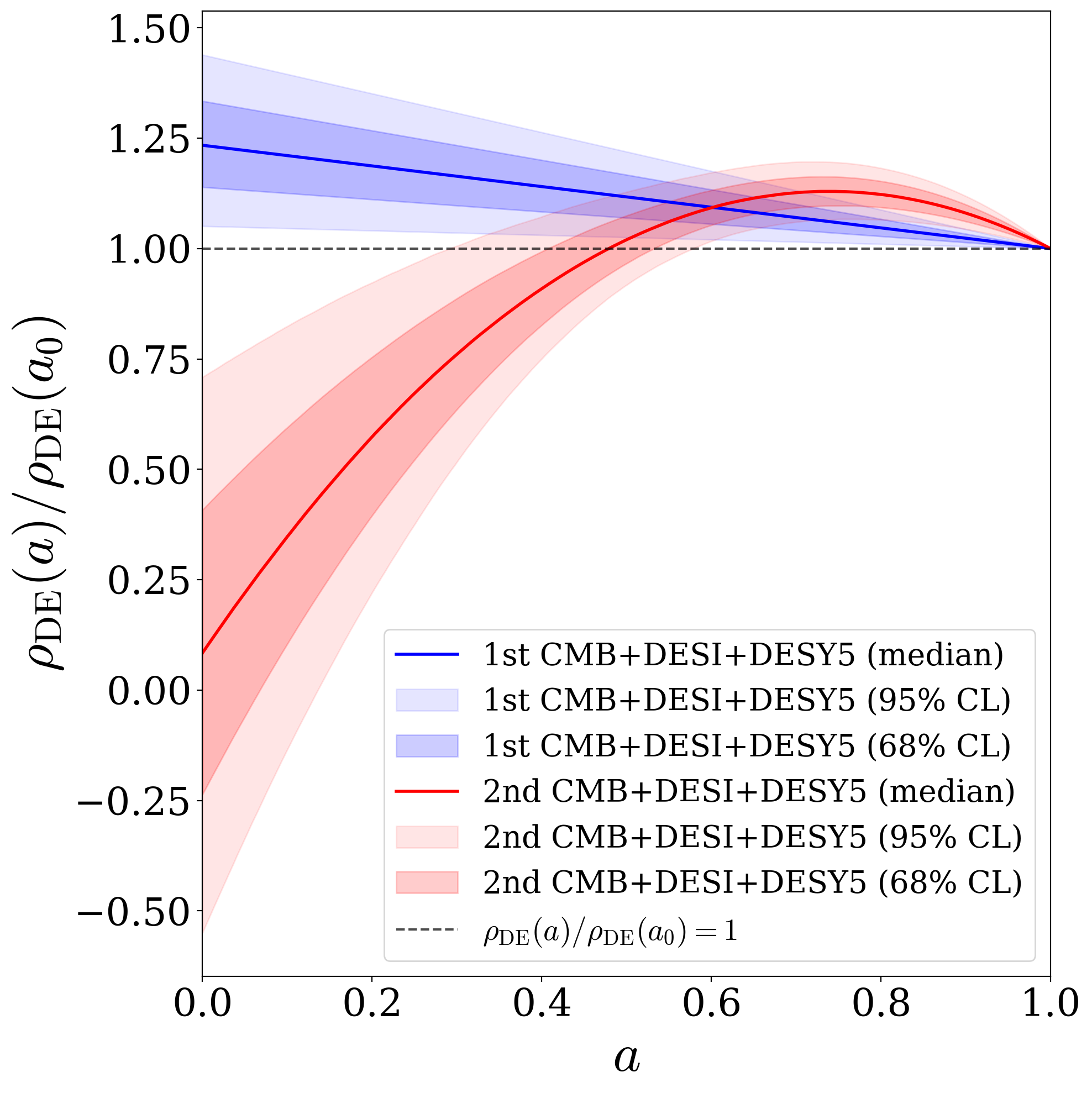}
    \end{subfigure}
    \begin{subfigure}[t]{0.325\textwidth}
        \centering
        \includegraphics[width=\linewidth]{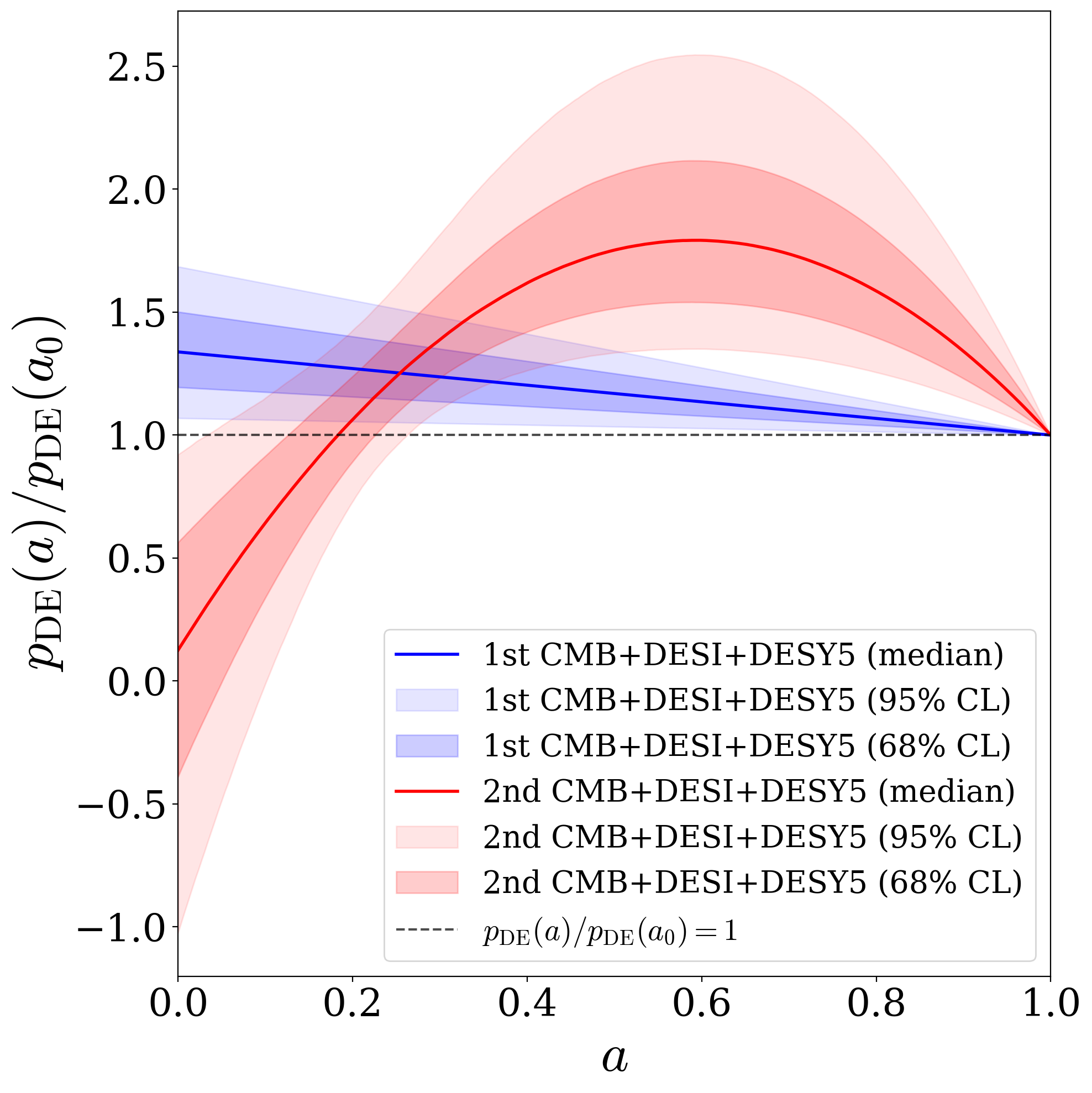}
    \end{subfigure}
    \begin{subfigure}[t]{0.325\textwidth}
        \centering
        \includegraphics[width=\linewidth]{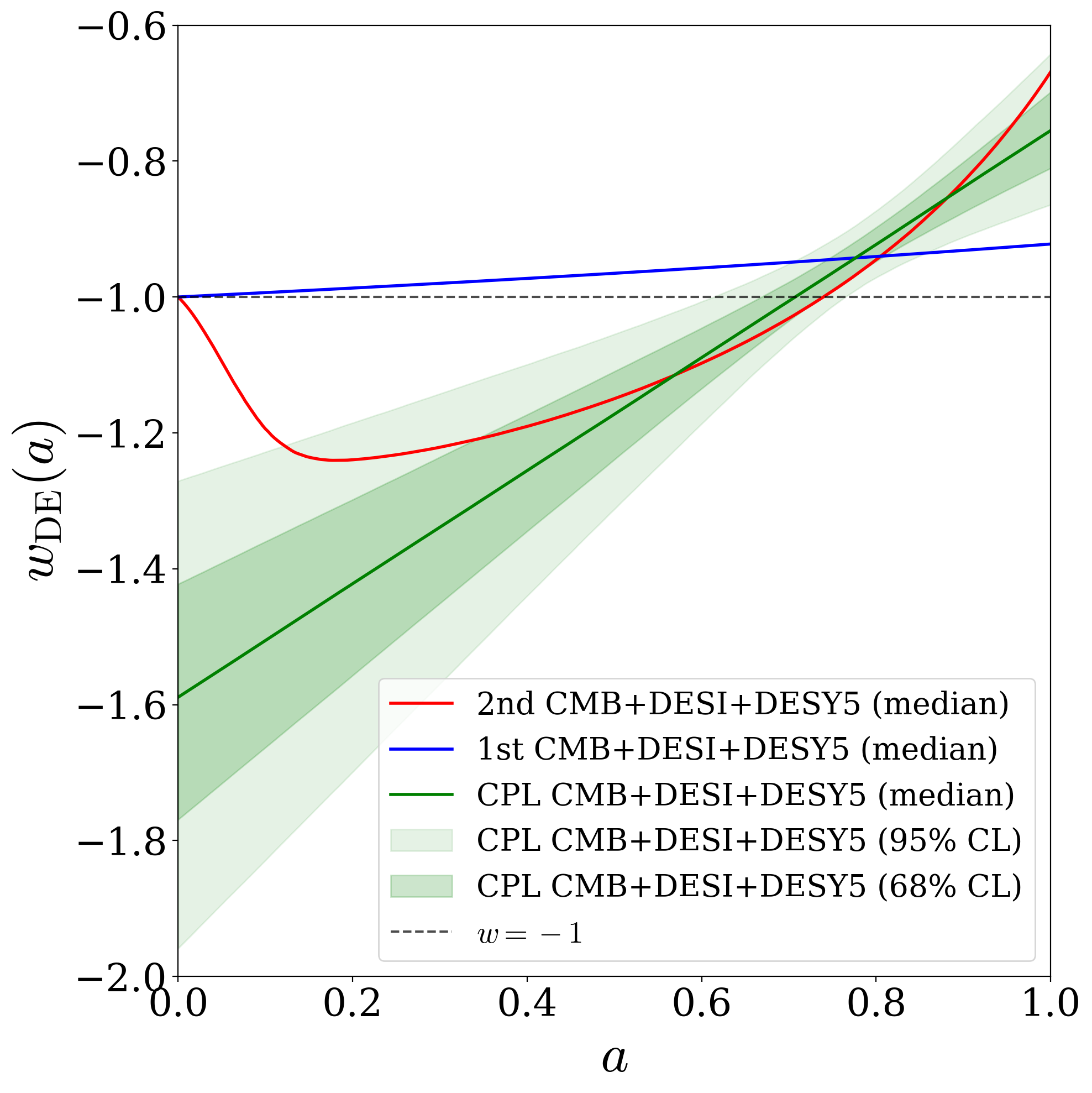}
    \end{subfigure}
    \caption{Same as figure~\ref{fig:sdss_pp}, for the combined datasets \textbf{CMB}+\textbf{DESI}+\textbf{DESY5}.}
    \label{fig:desi_desy5}
\end{figure*}

\begin{figure*}[t!]
    \centering
    \begin{subfigure}[t]{0.325\textwidth}
        \centering
        \includegraphics[width=\linewidth]{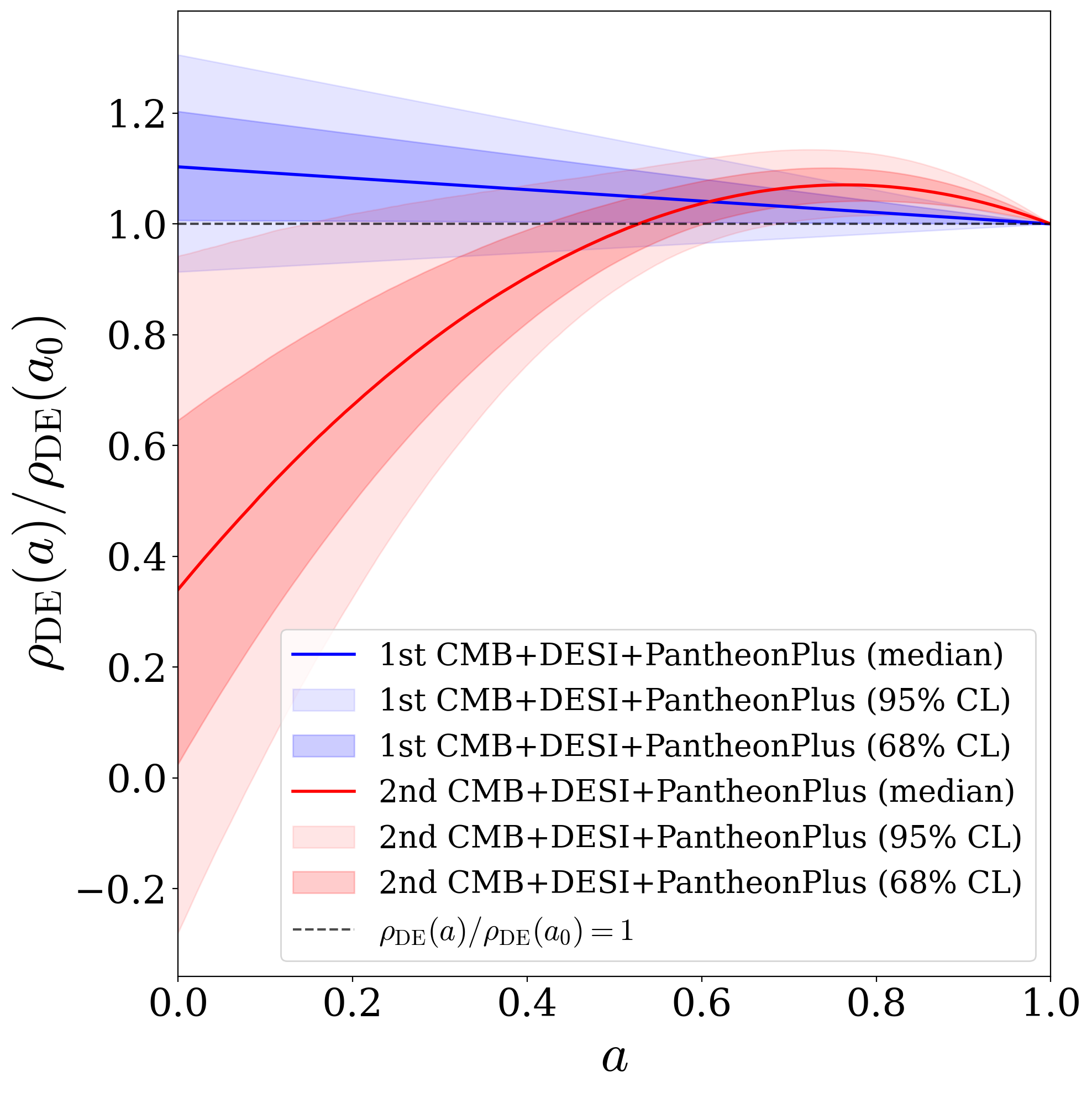}
    \end{subfigure}
    \begin{subfigure}[t]{0.325\textwidth}
        \centering
        \includegraphics[width=\linewidth]{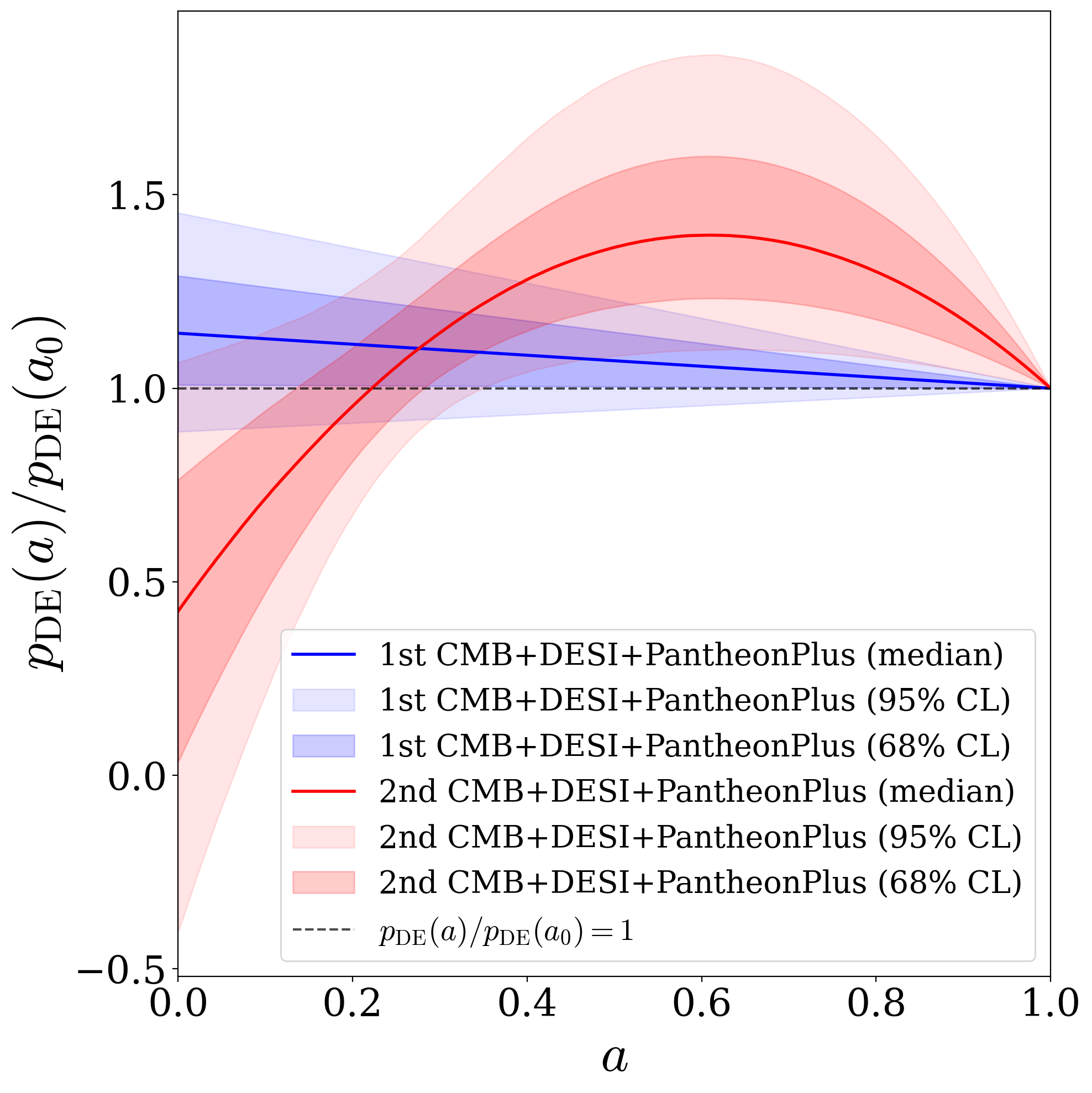}
    \end{subfigure}
    \begin{subfigure}[t]{0.325\textwidth}
        \centering
        \includegraphics[width=\linewidth]{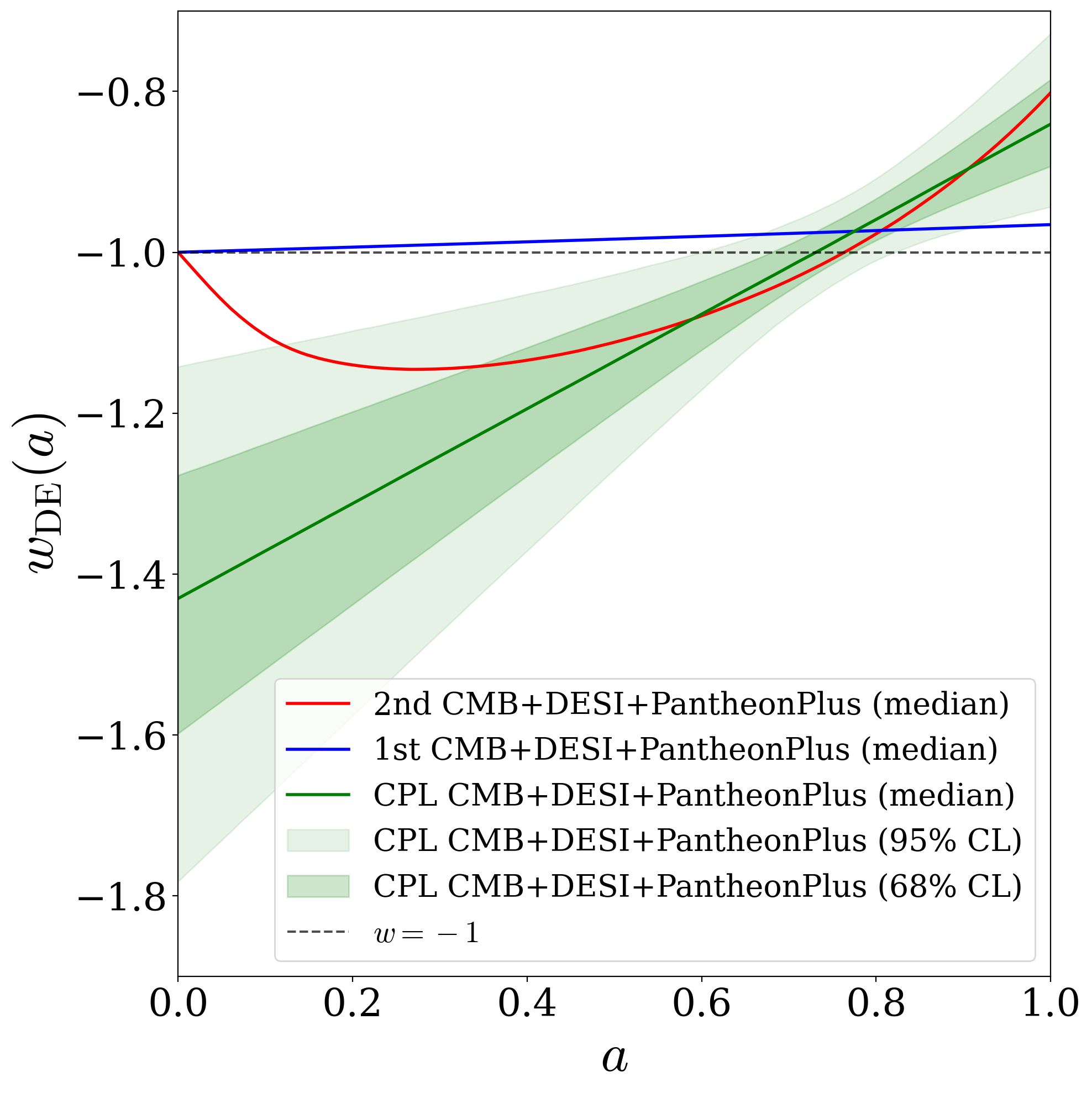}
    \end{subfigure}
    \caption{Same as figure~\ref{fig:sdss_pp}, for the combined datasets \textbf{CMB}+\textbf{DESI}+\textbf{PantheonPlus}.}
    \label{fig:desi_pp}
\end{figure*}

The left and middle panels in figures~\ref{fig:sdss_pp}, \ref{fig:desi_desy5}, and \ref{fig:desi_pp} compare the reconstruction of the normalized DE density $\rho_{\mathrm{DE}}(a)/\rho_{\mathrm{DE}}(a_0)$ (left) and pressure $p_{\mathrm{DE}}(a)/p_{\mathrm{DE}}(a_0)$ (middle) for the first-order and second-order expansions against a cosmological constant across multiple dataset combinations.
The first-order expansion model (blue curves) exhibits relatively mild evolution in the DE density across cosmic history. For all dataset combinations, this model shows:
\begin{itemize}
    \item The normalized density generally remains above unity, with a gradual decrease toward the present epoch ($a = 1$)
    \item Modest deviations from the $\Lambda$CDM prediction (dashed black line)
    \item Relatively narrow confidence intervals, indicating tighter constraints
\end{itemize}
The first-order model constrained by the CMB+DESI+DESY5 dataset combination (figure~\ref{fig:desi_desy5}) shows the strongest deviation from $\Lambda$CDM, exceeding $2\sigma$. By contrast, the CMB+DESI+PantheonPlus and CMB+SDSS+PantheonPlus combinations yield more modest deviations (nearly $1\sigma$ for the former and less than $1\sigma$ for the latter).

The second-order expansion model (red curves) reveals dramatically different behavior from both the first-order model and $\Lambda$CDM:
\begin{itemize}
    \item Significantly lower $\rho_{\rm DE}(a)$ values at early times ($a < 0.1$), with median values falling to $\sim (0.0 \textrm{--} 0.4)\,\rho_{\rm DE,0}$
    \item A non-monotonic evolution pattern, with the density first increasing with scale factor, crossing the $\Lambda$CDM prediction at $a \approx 0.4$–$0.6$, reaching a maximum at $a \approx 0.7$–$0.8$, and then decreasing to unity at the present day
    \item Much larger confidence intervals, particularly at early times, indicating the model's additional flexibility and the corresponding reduction in constraining power
\end{itemize}
The second-order model constrained by the CMB+DESI+DESY5 dataset (figure~\ref{fig:desi_desy5}) shows the strongest deviation from $\Lambda$CDM, well beyond $2\sigma$. The CMB+DESI+PantheonPlus and CMB+SDSS+PantheonPlus combinations yield more modest deviations (slightly more than $2\sigma$ for the former and nearly $2\sigma$ for the latter). This suggests that the DESY5 data may be particularly sensitive to DE dynamics, potentially due to its extended redshift coverage or different systematic uncertainties compared to PantheonPlus. Furthermore, DESI data appears to favor DDE more strongly than SDSS data, as already discussed in~\cite{DESI:2025zgx}.

The right panels of figures~\ref{fig:sdss_pp}, \ref{fig:desi_desy5}, and \ref{fig:desi_pp} show the reconstructed equation of state $w_{\rm DE}(a)$ under three different parameterizations. The first-order expansion (blue) median $w_{\rm DE}(a)$ remains very close to a cosmological constant throughout ($-1 \lesssim w \lesssim -0.9$) and never crosses into the phantom regime. In contrast, the second-order expansion (red) median $w_{\rm DE}(a)$ starts at $w = -1$ at $a = 0$, dips to $w \simeq -1.2$ to $-1.25$ at $a \sim 0.1$–$0.25$, before evolving upward and crossing the phantom divide ($w = -1$) at $a \approx 0.7$–$0.8$, remaining in the quintessence regime thereafter. The behavior of $w_{\rm DE}(a)$ after $a \sim 0.5$ (including its present value and the phantom crossing) is consistent across dataset combinations and agrees well with results from the CPL parameterization, demonstrating remarkable consistency in the qualitative behavior of late-time $w_{\rm DE}(a)$ across different models and dataset combinations. Unlike the CPL parametrization, the second-order pressure expansion exhibits a transient phantom phase, with $w_{\rm DE} < -1$ at intermediate redshifts followed by a return to $w_{\rm DE} > -1$ at later times. This crossing has important implications for model building; for instance, in multi-field scalar field models, the cosmological constant equation of state can be crossed without pathologies~\cite{Feng:2004ad, Vikman:2004dc}.

\begin{figure*}[t!]
    \centering
    \begin{subfigure}[t]{0.49\textwidth}
        \centering
        \includegraphics[width=0.99\linewidth]{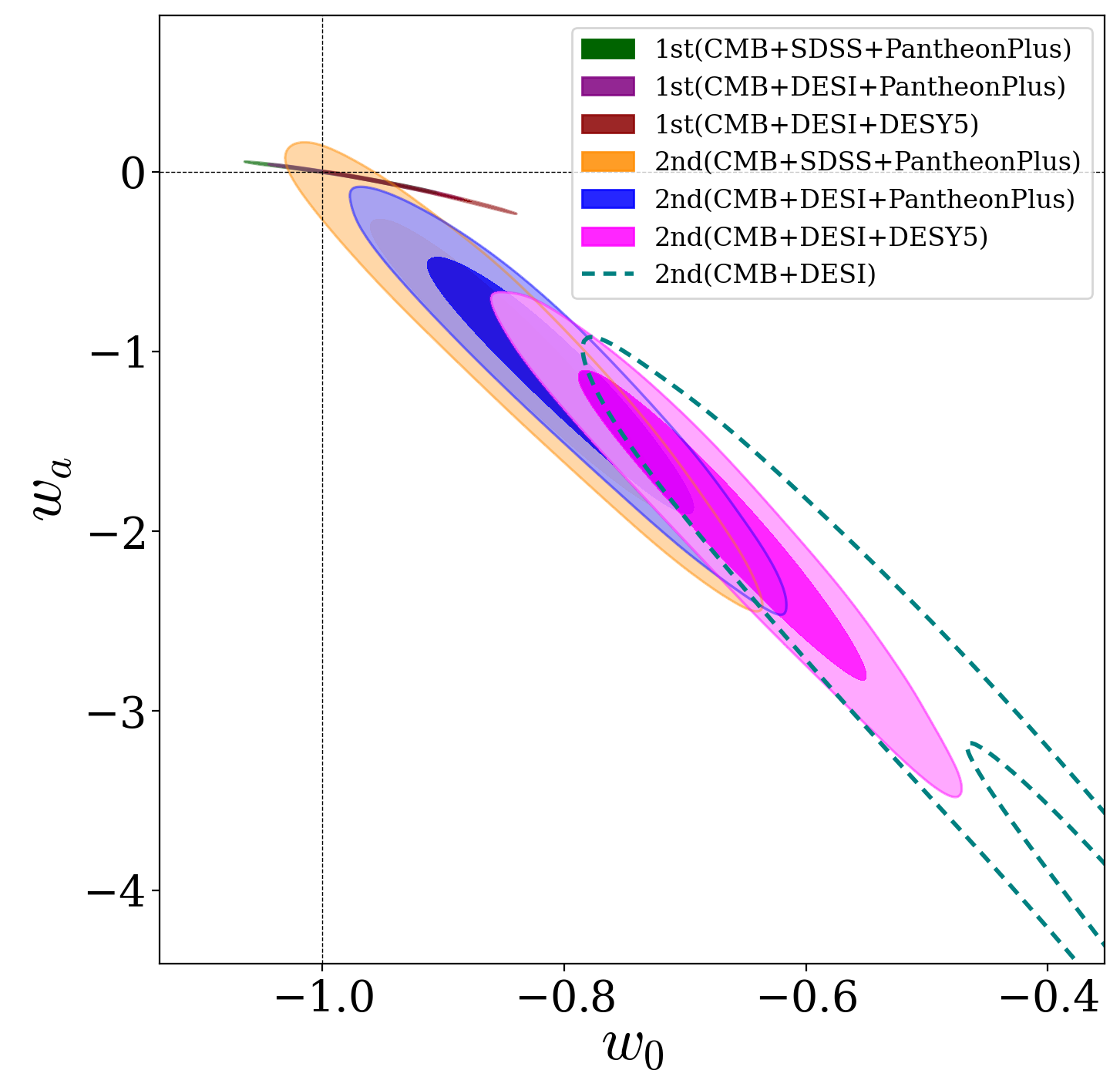}
    \end{subfigure}
    \begin{subfigure}[t]{0.49\textwidth}
        \centering
        \includegraphics[width=0.99\linewidth]{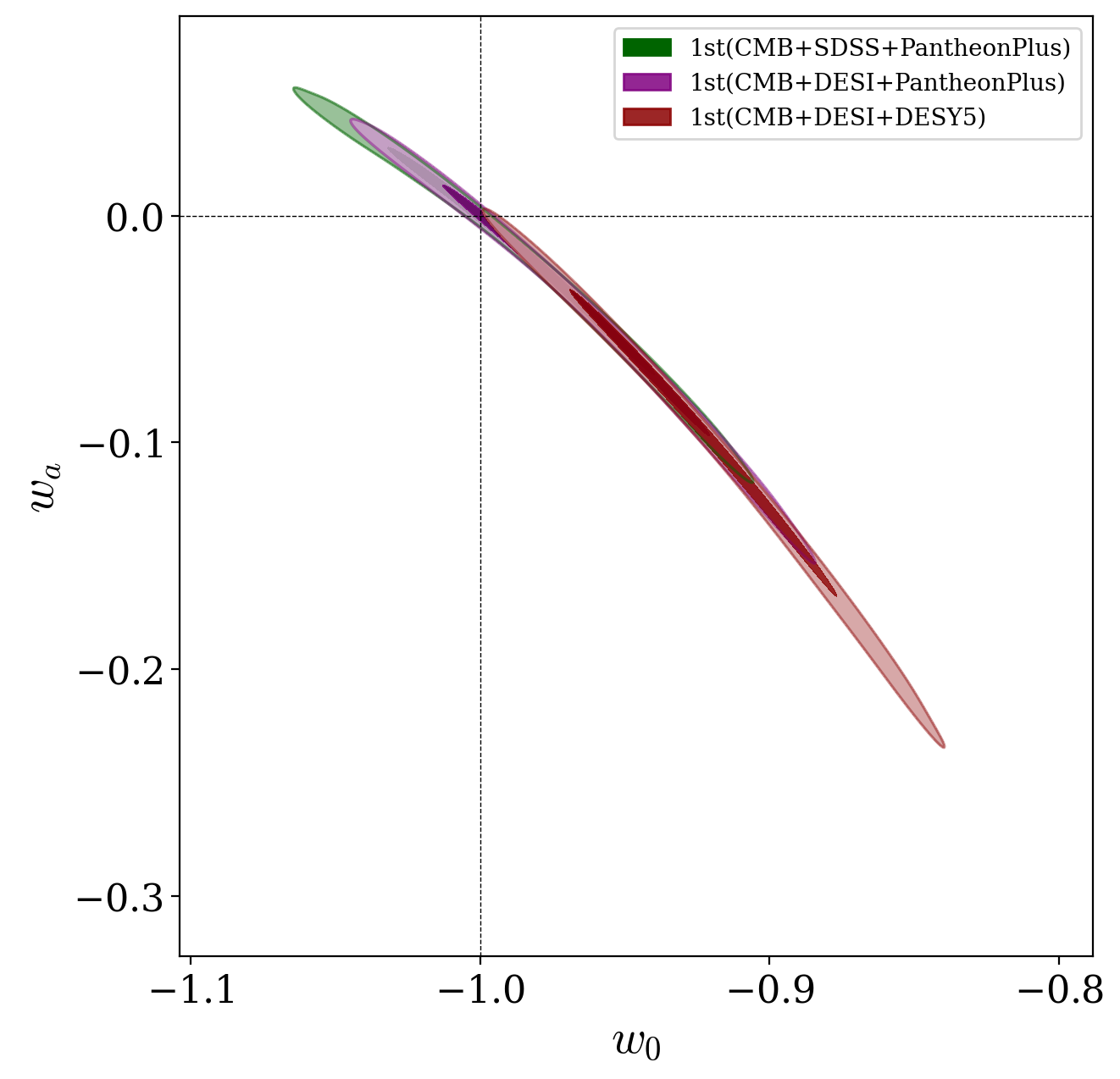}
    \end{subfigure}
    \caption{\textit{Left:} Confidence contours (68\% and 95\%) in the $w_0$–$w_a$ plane for various dataset combinations. \textit{Right:} Zoom-in on the $w_0$–$w_a$ plane, restricted to the first-order expansion model only.}
    \label{fig:w0wa_plot}
\end{figure*}

Figure~\ref{fig:w0wa_plot} summarizes the posterior constraints on the present-day equation-of-state parameter $w_0$ and its first derivative $w_a$, as defined in eqs.~\eqref{eq:w0_definition} and~\eqref{eq:wa_definition}. In the right panel, the contours derived from the first-order pressure expansion are tightly clustered around the $\Lambda$CDM point $(w_0, w_a) = (-1, 0)$. The dataset combinations CMB+SDSS+PantheonPlus (green) and CMB+DESI+PantheonPlus (purple) are consistent with a cosmological constant at the 68\% CL, while CMB+DESI+DESY5 (red) remains consistent at the 95\% CL. The first-order model yields relatively tight constraints, concentrated in a narrow diagonal band stretching from the upper left to the lower right of the parameter space. This structure indicates a strong degeneracy between the $w_0$ and $w_a$ parameters, with more negative values of $w_0$ corresponding to less negative values of $w_a$.
By contrast, the left panel shows the results from the second-order pressure expansion (orange, blue, magenta), which occupy a region of parameter space characterized by $w_0 > -1$ and $w_a < 0$. The increasingly negative values of $w_a$ indicate stronger evolution of $w(a)$ away from $-1$ at earlier times. CMB+DESI shows the strongest deviation from cosmological constant. A comparison between the results obtained from CMB+SDSS+PantheonPlus (orange) and CMB+DESI+PantheonPlus (blue) reveals that replacing SDSS with DESI data shifts the constraints further from the $\Lambda$CDM solution. A similar trend is observed when comparing CMB+DESI+PantheonPlus (blue) with CMB+DESI+DESY5 (magenta), indicating that substituting PantheonPlus with DESY5 data pushes the contours even further from the cosmological constant. The dataset combinations CMB+DESI+PantheonPlus and CMB+DESI+DESY5 show more than $2\sigma$ deviation from a cosmological constant, while CMB+SDSS+PantheonPlus remains consistent within the $2\sigma$ level, consistent with the findings of~\cite{DESI:2025zgx}.


\section{Discussion}
\label{sec:discussion}
In interpreting our model comparison, we emphasize that Bayesian evidence accounts for both the goodness of fit and model complexity, unlike the $\chi^2$ statistic, which reflects only the best-fit likelihood. For instance, in table~\ref{tab:CosmoParams_first}, while the CPL model yields a better fit compared to the first-order model, it also introduces one additional parameter. Bayesian evidence, defined as the integral of the likelihood weighted by the prior, inherently penalizes such increases in parameter space unless the improvement in fit is substantial. This embodies the principle of Occam’s razor: simpler models are favored unless additional complexity is strongly justified. Therefore, despite a worse $\chi^2$, the first-order model can still achieve higher Bayesian evidence due to its lower complexity. 
The choice of prior volumes plays a critical role in this comparison, as enforcing identical priors across models may obscure genuine differences in model behavior and penalization.

We previously discussed that a sign change in the dark energy density $\rho_{\mathrm{DE}}$ would lead to the presence of a pole in $w_{\rm DE}(a)$, as noted in Section~\ref{sec:DDE}. The normalized DE density curves $\rho_{\mathrm{DE}}(a)/\rho_{\mathrm{DE}}(a_0)$ shown in figures~\ref{fig:sdss_pp}, \ref{fig:desi_desy5}, and \ref{fig:desi_pp} demonstrate that while this pole is unlikely to appear within the 68\% CL region, it does emerge within the 95\% CL region. However, this pole has minimal impact on our findings. The primary reason is that it occurs only in the early universe—the sign change in $\rho_{\mathrm{DE}}(a)/\rho_{\mathrm{DE}}(a_0)$ happens at $a < 0.1$ in figures~\ref{fig:sdss_pp}, \ref{fig:desi_desy5}, and \ref{fig:desi_pp}—and thus affects only the shape of the confidence regions near $a = 0.1$.
Since dark energy becomes dominant in the late universe, our analysis focuses primarily on the behavior of the equation of state $w(a)$ at later times ($a > 0.2$), where the pole has negligible influence on our results.

\begin{figure}[htbp]
    \centering
    \includegraphics[width=0.425\linewidth]{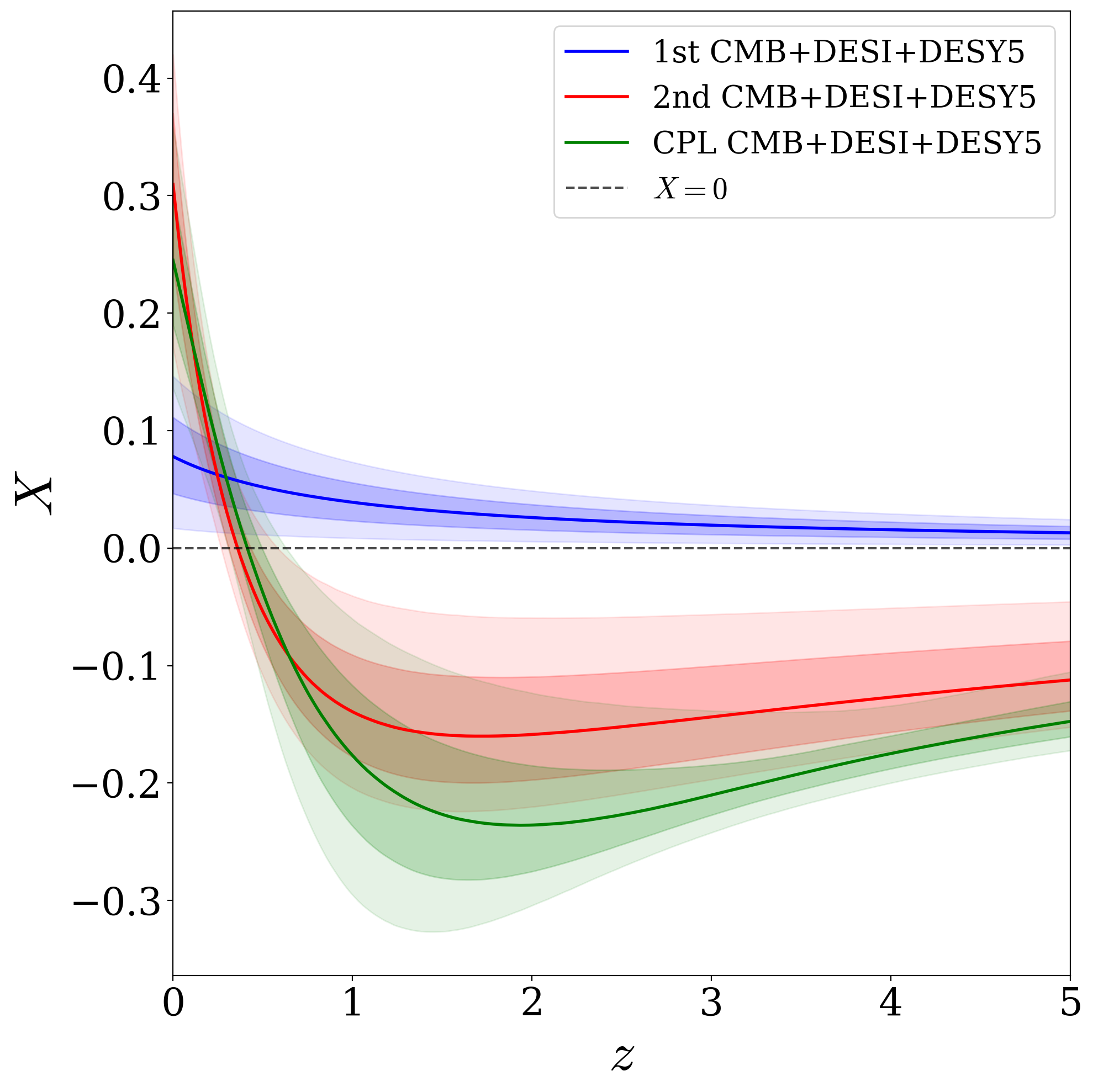}
    \caption{Reconstruction of $X$ shown in eq.~\eqref{eq:X} as a function of the redshift $z$ for the first-order, second-order, and CPL parameterizations. Shaded regions denote the $68\%$ and $95\%$ CL for each parameterization. Results are obtained using the CMB+DESI+DESY5 dataset combination.}
    \label{fig:X_Z}
\end{figure}

To test the sensitivity of the datasets to the signs of $\Omega_1$ and $\Omega_2$, we study the evolution of the quantity
\begin{equation}
X \equiv \frac{(1 + z)}{3 \rho_{\mathrm{DE,0}}} \frac{{\rm d}\rho_{\mathrm{DE}}}{{\rm d}z} = (1 + w_{\mathrm{DE}}) \frac{\rho_{\mathrm{DE}}}{\rho_{\mathrm{DE,0}}}\,,
\label{eq:X}
\end{equation}
where $w_{\mathrm{DE}}$ is the dark energy (DE) equation of state and $\rho_{\mathrm{DE}}$ its energy density. The behavior of $X$ can signal new physics in the DE sector. Within the DDE framework, as shown in figure~\ref{fig:X_Z}, analysis of the combined CMB+DESI+DESY5 datasets shows that $X = 0$ is excluded at the $2\sigma$ level, disfavoring the cosmological constant solution $w_{\mathrm{DE}} = -1$, in agreement with the 2D parameter contours shown in figures~\ref{fig:desi_desy5}, \ref{fig:w0wa_plot}. In the first-order scenario, $X$ stays positive over the redshift range $z \in [0,5]$, excluding both phantom behavior ($1 + w_{\mathrm{DE}} < 0$) and negative $\rho_{\mathrm{DE}}$. In the second-order case, $X$ becomes negative for $z \gtrsim 0.7$, indicating a single phantom crossing, similar to CPL.

\begin{table}[htbp]
  \centering\small
  \resizebox{\textwidth}{!}{%
  \begin{tabular}{lcccc}
    \toprule
    Parameter 
      & \multicolumn{2}{c}{First‐order expansion} 
      & \multicolumn{2}{c}{Second‐order expansion} \\
    \cmidrule(lr){2-3} \cmidrule(lr){4-5}
      & CMB+DESI 
      & CMB+DESI+PantheonPlus 
      & CMB+DESI 
      & CMB+DESI+PantheonPlus \\
    \midrule
    $\Omega_b h^2$ 
      & $0.02250\pm0.00013$ 
      & $0.02252\pm0.00013$ 
      & $0.02249\pm0.00013$ 
      & $0.02251\pm0.00013$ \\

    $\Omega_c h^2$ 
      & $0.11834\pm0.00077$ 
      & $0.11812\pm0.00066$ 
      & $0.11852\pm0.00075$ 
      & $0.11824\pm0.00066$ \\

    $100\,\theta_{\rm MC}$ 
      & $1.04115\pm0.00029$ 
      & $1.04119\pm0.00028$ 
      & $1.04114\pm0.00028$ 
      & $1.04118\pm0.00028$ \\

    $\tau_{\rm reio}$ 
      & $0.0596^{+0.0069}_{-0.0076}$ 
      & $0.0606^{+0.0068}_{-0.0077}$ 
      & $0.0591^{+0.0067}_{-0.0076}$ 
      & $0.0604\pm0.0071$ \\

    $n_s$ 
      & $0.9697\pm0.0035$ 
      & $0.9705\pm0.0033$ 
      & $0.9692\pm0.0035$ 
      & $0.9700\pm0.0034$ \\

    $\log(10^{10}A_s)$ 
      & $3.054\pm0.014$ 
      & $3.056^{+0.012}_{-0.014}$ 
      & $3.053\pm0.013$ 
      & $3.055\pm0.013$ \\

    $\Omega_1$ 
      & $0.137^{+0.048}_{-0.12}$ 
      & $<0.0401$ 
      & $0.253^{+0.091}_{-0.14}$ 
      & $0.069^{+0.023}_{-0.051}$ \\

    $\Omega_2$ 
      & --- 
      & --- 
      & $-0.150^{+0.15}_{-0.040}$ 
      & $-0.0398^{+0.040}_{-0.0087}$ \\
    \midrule
    $H_0$ [km/s/Mpc] 
      & $69.66^{+0.65}_{-1.1}$ 
      & $68.58^{+0.32}_{-0.39}$ 
      & $70.28^{+0.78}_{-1.1}$ 
      & $68.76^{+0.36}_{-0.42}$ \\

    $\sigma_8$ 
      & $0.8245^{+0.0087}_{-0.012}$ 
      & $0.8139^{+0.0057}_{-0.0065}$ 
      & $0.8308^{+0.0097}_{-0.012}$ 
      & $0.8162^{+0.0057}_{-0.0066}$ \\

    $S_8$ 
      & $0.8128\pm0.0078$ 
      & $0.8146\pm0.0077$ 
      & $0.8123\pm0.0077$ 
      & $0.8149\pm0.0076$ \\

    $\Omega_{\mathrm{m}}$
      & $0.2917^{+0.0079}_{-0.0061}$ 
      & $0.3005\pm0.0040$ 
      & $0.2869^{+0.0081}_{-0.0068}$ 
      & $0.2991\pm0.0042$ \\
    \midrule
    $\Delta\chi^2_{\min,\Lambda\mathrm{CDM}}$ 
      & $0.97$ 
      & $0.45$ 
      & $1.35$ 
      & $1.97$ \\
    \bottomrule
  \end{tabular}%
  }
\caption{Parameter constraints (68\% CL) for the first- and second-order expansions under the following prior conditions: $\Omega_1 > 0$ for the first-order expansion, and $\Omega_2 < 0$ with $\Omega_1 + \Omega_2 > 0$ for the second-order expansion. Results are shown for the CMB+DESI and CMB+DESI+PantheonPlus dataset combinations.}
\label{tab:with_prior}
\end{table}

\begin{figure}[htbp]
    \centering
    \includegraphics[width=0.9\linewidth]{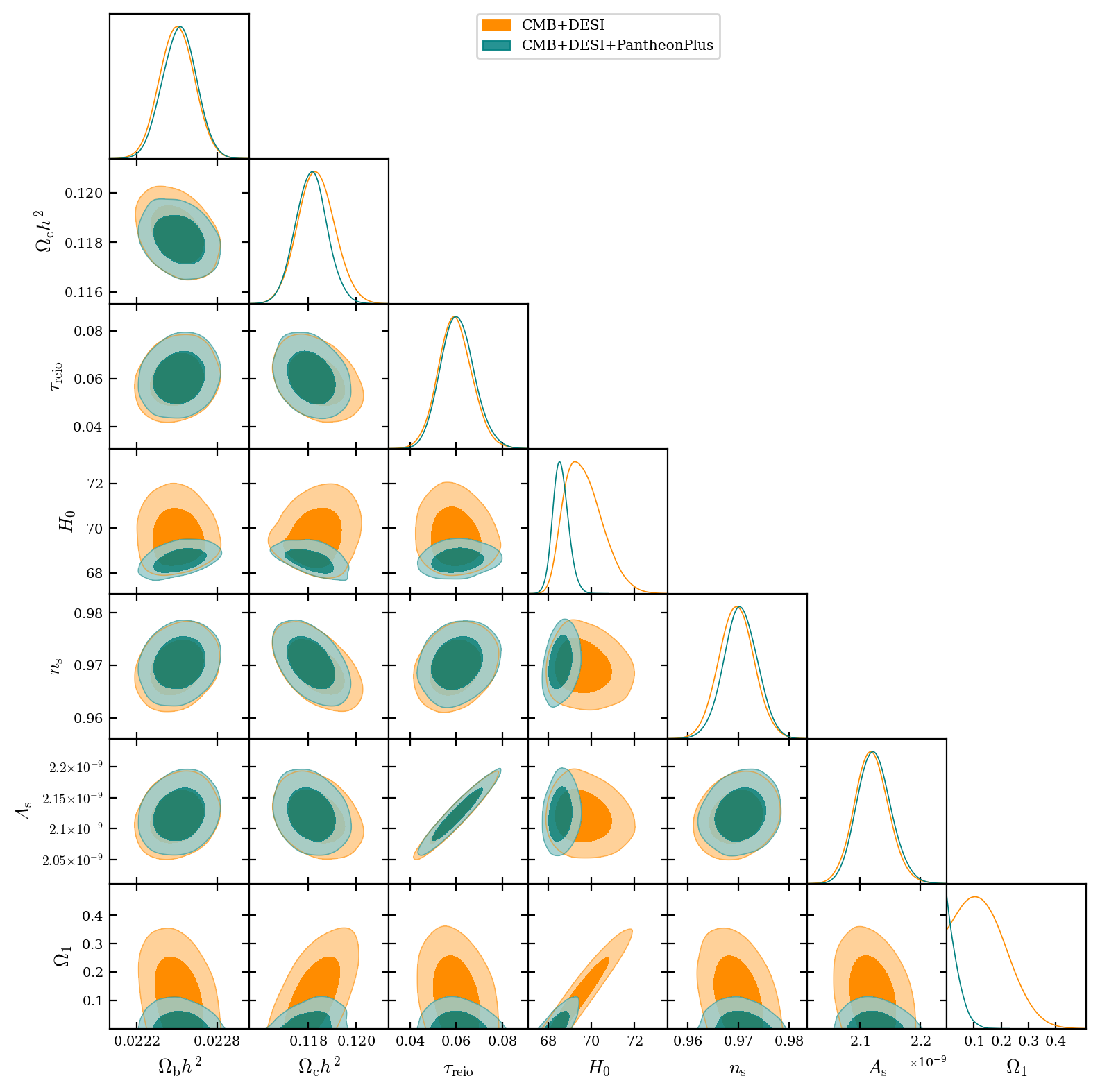}
    \caption{Expansion to first order with the prior condition $\Omega_1 > 0$. One-dimensional posterior distributions and two-dimensional marginalized contours for the key parameters of the first-order expansion. Results are obtained using the CMB+DESI and CMB+DESI+PantheonPlus dataset combinations.}
    \label{fig:anfst_post}
\end{figure}

\begin{figure}[htbp]
    \centering
    \includegraphics[width=0.9\linewidth]{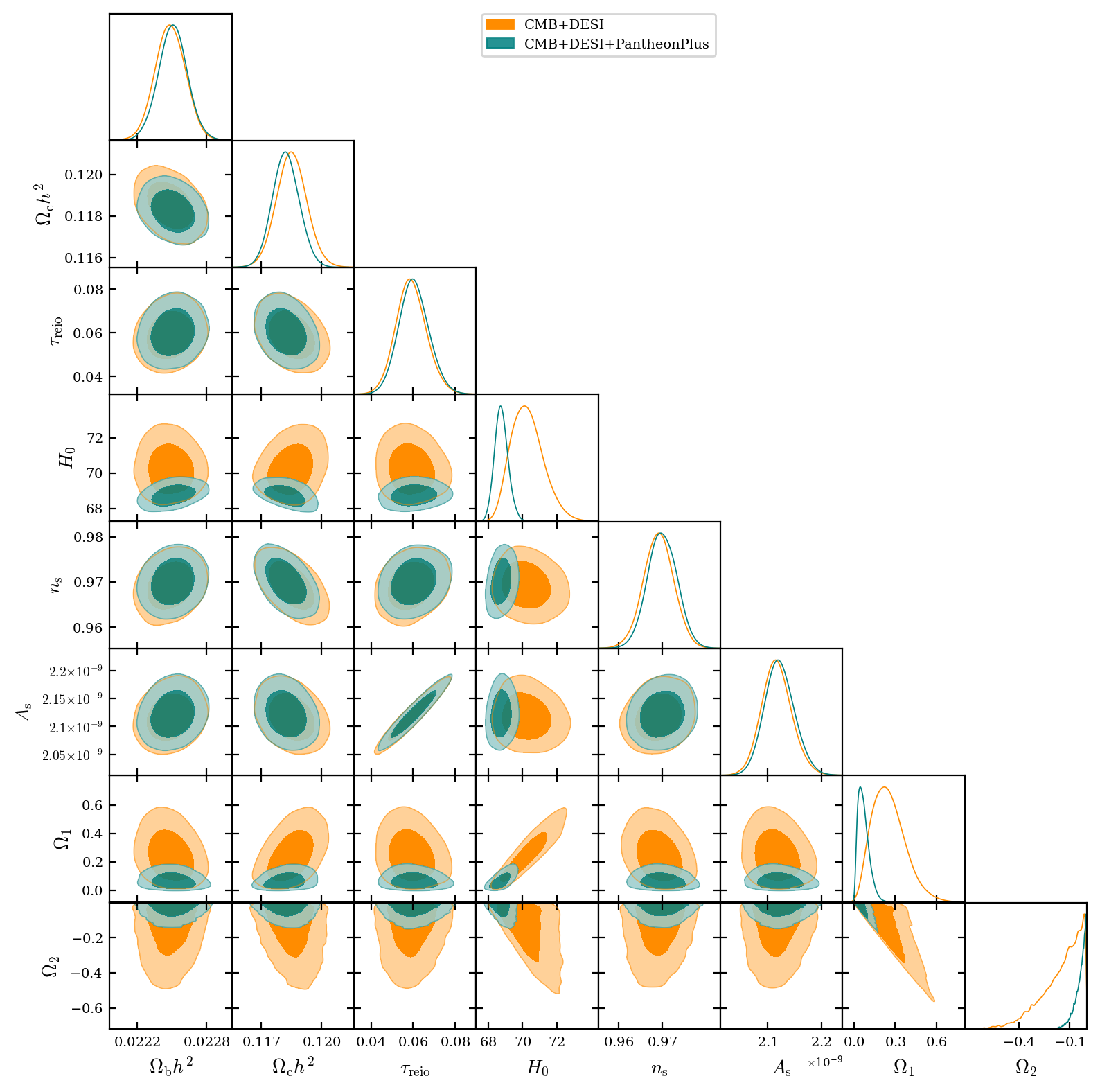}
    \caption{Expansion to second order with the prior conditions $\Omega_2 < 0$ and $\Omega_1 + \Omega_2 > 0$. One-dimensional posterior distributions and two-dimensional marginalized contours for the key parameters of the second-order expansion, obtained from the CMB+DESI and CMB+DESI+PantheonPlus dataset combinations.}
    \label{fig:ansec_post}
\end{figure}

The set of eqs.~\eqref{eq:rho_a} can be rewritten as follows for the first-order case ($\Omega_2 = 0$), showing that dark energy can be effectively decomposed into a cosmological constant plus an additional fluid characterized by an equation-of-state parameter $w = -4/3$, corresponding to a phantom fluid:

\begin{equation}
\rho = \rho_{\mathrm{DE},0}\left[ \left(1 - \frac{\Omega_1}{\Omega_{\mathrm{DE},0}}\right) + \frac{\Omega_1}{\Omega_{\mathrm{DE},0}} a \right].
\label{eq:rho_eff_first_order}
\end{equation}

Similarly, for the second-order expansion, equations~\eqref{eq:rho_a} can be rewritten as follows, indicating that dark energy can be effectively decomposed into a cosmological constant accompanied by two additional fluids with equation-of-state parameters $w = -4/3$ and $w = -5/3$, respectively:

\begin{equation}
\rho = \rho_{\mathrm{DE},0}\left[\left(1-\frac{\Omega_1}{\Omega_{\mathrm{DE},0}}-\frac{3}{5}\frac{\Omega_2}{\Omega_{\mathrm{DE},0}}\right) + \left(\frac{\Omega_1}{\Omega_{\mathrm{DE},0}} + \frac{\Omega_2}{\Omega_{\mathrm{DE},0}}\right)a - \frac{2}{5}\frac{\Omega_2}{\Omega_{\mathrm{DE},0}}a^2\right].
\label{eq:rho_eff_second_order}
\end{equation}

Ref.~\cite{Sen:2007gk} proposed a Lagrangian that can model such a system, consisting of a cosmological constant plus $n$ additional phantom components, corresponding to an $n$-th order Taylor expansion around the cosmological constant value in the pressure parameterization. While we do not delve into the details of this Lagrangian, it is important to note that if the dark energy behavior arises from this framework, it imposes certain prior conditions: $\Omega_1 > 0$ for the first-order expansion, and $\Omega_2 < 0$ with $\Omega_1 + \Omega_2 > 0$ for the second-order expansion. These constraints arise because the energy density of each phantom component, derived from a dynamical field, must remain positive.

We present the posterior distributions and parameter constraints (68\% CL) for the first- and second-order expansion models in figure~\ref{fig:anfst_post}, figure~\ref{fig:ansec_post}, and table~\ref{tab:with_prior}. Our analysis demonstrates that the results obtained under these imposed priors remain consistent with the previous results shown in tables~\ref{tab:CosmoParams_first}, \ref{tab:CosmoParams_sec} and figures~\ref{fig:Post_distribution_O1}, \ref{fig:Post_sec_10}, which used the more general prior ranges defined in table~\ref{tab:prior_table}.
Moreover, comparing the values of $\Delta\chi^2_{\min,\Lambda\mathrm{CDM}}$ in table~\ref{tab:with_prior} with those in tables~\ref{tab:CosmoParams_first} and~\ref{tab:CosmoParams_sec} reveals that models constrained by these priors provide a poorer fit to the CMB+DESI and CMB+DESI+PantheonPlus datasets, relative to the more general prior scenario. This suggests that a purely phantom behavior of DDE is not favored by the data. Instead, a larger parameter space (as shown in table~\ref{tab:prior_table}) is required to allow greater flexibility in DDE behavior and to produce more compelling results, as discussed in Section~\ref{sec:results}.

\section{Conclusions}
\label{sec:conclusions}
In this work, we have examined constraints on DDE models expanded to first and second order, comparing their performance against the $\Lambda$CDM and CPL parameterizations using various observational datasets. Our analysis has provided valuable insights into the nature of dark energy and its evolution throughout cosmic history.

The first-order expansion model remains close to a cosmological constant across cosmic history ($-1 \lesssim w \lesssim -0.95$) and never enters the phantom regime. The parameter $\Omega_1$, which characterizes deviations from $\Lambda$CDM, typically remains consistent with zero at approximately the $1\sigma$ level for all dataset combinations, except for the CMB-only and CMB+DESI+DESY5 cases. Notably, the CMB+DESI+DESY5 combination indicates the strongest deviation from $\Lambda$CDM, at the $2.7\sigma$ level. However, Bayesian evidence analysis reveals no clear preference for the first-order model over the simpler $\Lambda$CDM scenario. Interestingly, the Bayesian evidence shows that the first-order expansion model consistently outperforms the CPL parameterization for dataset combinations that do not include DESI, whereas the inclusion of DESI data reverses this preference, favoring the CPL parameterization despite its higher complexity (i.e., greater number of free parameters).

The results from the second-order expansion, which introduces an additional parameter $\Omega_2$, present significantly stronger indications of DDE. Particularly striking is the dataset combination CMB+DESI+DESY5, which shows a deviation from $\Lambda$CDM exceeding $4\sigma$. Bayesian evidence further supports this model for this specific combination, representing ``moderate'' statistical preference according to the Jeffreys' scale. Other dataset combinations, such as CMB+SDSS+PantheonPlus and CMB+DESI+PantheonPlus, also show deviations from $\Lambda$CDM at the $>2\sigma$ level and improved maximum-likelihood fits. However, the increased complexity of the second-order model often leads to weaker Bayesian support compared to $\Lambda$CDM. In terms of comparison with the CPL parameterization, our second-order expansion model performs comparably in both maximum-likelihood fit and Bayesian evidence across most dataset combinations. This concern has been studied in detail, for example in Ref.~\cite{Nesseris:2025lke}, where Taylor expansions of $w_{\rm DE}(a)$ can suffer from prior-volume effects in Bayesian model comparison.

When examining the reconstructed DE evolution, we observe that the second-order model reveals a distinctive non-monotonic behavior in the dark energy density, with significantly lower values at early times compared to $\Lambda$CDM, followed by an increase that peaks around $a \approx 0.7$–$0.8$, before decreasing to its present value. The corresponding equation of state exhibits analogous non-monotonic features and phantom-crossing behavior. Notably, the evolution of $w_{\rm DE}(a)$ after $a \sim 0.5$, including its present value and the phantom crossing, remains consistent across various dataset combinations and shows strong agreement with results obtained from the CPL parameterization. This demonstrates remarkable consistency in the qualitative behavior of late-time $w_{\rm DE}(a)$ across different models and datasets. 
The phantom-crossing behavior has important theoretical implications, as it cannot be explained by standard quintessence or phantom models. It suggests that DDE may require more exotic frameworks, such as scalar-tensor theories or modified gravity.

Our results highlight the importance of dataset selection in constraining DE models. The combination of CMB, DESI, and DESY5 data consistently shows the strongest preference for DDE. The DESY5 data appears particularly sensitive to DE dynamics, potentially due to its extended redshift coverage or distinct systematic uncertainties compared to PantheonPlus. Similarly, DESI data shows a stronger preference for DDE than SDSS data.

Overall, while the evidence for DDE remains tentative for most dataset combinations, the strong preference shown by certain datasets, particularly when using our second-order expansion model, suggests that the cosmological constant may not provide a complete description of DE. Our multi-model analysis across various dataset combinations offers compelling evidence for a DDE component exhibiting phantom-crossing behavior. Future surveys with improved precision and extended redshift coverage will be crucial for definitively resolving this question and potentially uncovering the true, possibly dynamical, nature of dark energy.

\section*{Acknowledgements}
HC acknowledges the support of the China Scholarship Council (CSC) program (Project ID No.\ 202406230341). HC and LV acknowledge support by the National Natural Science Foundation of China (NSFC) through the grant No.\ 12350610240 ``Astrophysical Axion Laboratories''. EDV is supported by a Royal Society Dorothy Hodgkin Research Fellowship. AAS acknowledges the funding from ANRF, Govt.\ of India, under the research grant No.\ CRG/2023/003984. LV also thanks Istituto Nazionale di Fisica Nucleare (INFN) through the ``QGSKY'' Iniziativa Specifica project. This article is based upon work from COST Action CA21136 ``Addressing observational tensions in cosmology with systematics and fundamental physics'' (CosmoVerse), supported by COST (European Cooperation in Science and Technology). The authors acknowledge the use of High-Performance Computing resources from the IT Services at the University of Sheffield.
\bibliographystyle{JHEP}
\bibliography{biblio}

\providecommand{\href}[2]{#2}\begingroup\raggedright\begin{thebibliography}{100}

\bibitem{SupernovaSearchTeam:1998fmf}
{\scshape Supernova Search Team} collaboration, \emph{{Observational evidence from supernovae for an accelerating universe and a cosmological constant}}, \href{https://doi.org/10.1086/300499}{\emph{Astron. J.} {\bfseries 116} (1998) 1009} [\href{https://arxiv.org/abs/astro-ph/9805201}{{\ttfamily astro-ph/9805201}}].

\bibitem{SupernovaCosmologyProject:1998vns}
{\scshape Supernova Cosmology Project} collaboration, \emph{{Measurements of $\Omega$ and $\Lambda$ from 42 High Redshift Supernovae}}, \href{https://doi.org/10.1086/307221}{\emph{Astrophys. J.} {\bfseries 517} (1999) 565} [\href{https://arxiv.org/abs/astro-ph/9812133}{{\ttfamily astro-ph/9812133}}].

\bibitem{WMAP:2012fli}
{\scshape WMAP} collaboration, \emph{{Nine-Year Wilkinson Microwave Anisotropy Probe (WMAP) Observations: Final Maps and Results}}, \href{https://doi.org/10.1088/0067-0049/208/2/20}{\emph{Astrophys. J. Suppl.} {\bfseries 208} (2013) 20} [\href{https://arxiv.org/abs/1212.5225}{{\ttfamily 1212.5225}}].

\bibitem{Planck:2018nkj}
{\scshape Planck} collaboration, \emph{{Planck 2018 results. I. Overview and the cosmological legacy of Planck}}, \href{https://doi.org/10.1051/0004-6361/201833880}{\emph{Astron. Astrophys.} {\bfseries 641} (2020) A1} [\href{https://arxiv.org/abs/1807.06205}{{\ttfamily 1807.06205}}].

\bibitem{SDSS:2003eyi}
{\scshape SDSS} collaboration, \emph{{Cosmological parameters from SDSS and WMAP}}, \href{https://doi.org/10.1103/PhysRevD.69.103501}{\emph{Phys. Rev. D} {\bfseries 69} (2004) 103501} [\href{https://arxiv.org/abs/astro-ph/0310723}{{\ttfamily astro-ph/0310723}}].

\bibitem{SDSS:2003lnz}
{\scshape SDSS} collaboration, \emph{{Physical Evidence for Dark Energy}},  \href{https://arxiv.org/abs/astro-ph/0307335}{{\ttfamily astro-ph/0307335}}.

\bibitem{SDSS:2005xqv}
{\scshape SDSS} collaboration, \emph{{Detection of the Baryon Acoustic Peak in the Large-Scale Correlation Function of SDSS Luminous Red Galaxies}}, \href{https://doi.org/10.1086/466512}{\emph{Astrophys. J.} {\bfseries 633} (2005) 560} [\href{https://arxiv.org/abs/astro-ph/0501171}{{\ttfamily astro-ph/0501171}}].

\bibitem{2dFGRSTeam:2002tzq}
{\scshape 2dFGRS Team} collaboration, \emph{{Parameter constraints for flat cosmologies from CMB and 2dFGRS power spectra}}, \href{https://doi.org/10.1046/j.1365-8711.2002.06001.x}{\emph{Mon. Not. Roy. Astron. Soc.} {\bfseries 337} (2002) 1068} [\href{https://arxiv.org/abs/astro-ph/0206256}{{\ttfamily astro-ph/0206256}}].

\bibitem{Efstathiou:2021ocp}
G.~Efstathiou, \emph{{To H0 or not to H0?}}, \href{https://doi.org/10.1093/mnras/stab1588}{\emph{Mon. Not. Roy. Astron. Soc.} {\bfseries 505} (2021) 3866} [\href{https://arxiv.org/abs/2103.08723}{{\ttfamily 2103.08723}}].

\bibitem{Krishnan:2021dyb}
C.~Krishnan, R.~Mohayaee, E.{\'O}.~Colg{\'a}in, M.M.~Sheikh-Jabbari and L.~Yin, \emph{{Does Hubble tension signal a breakdown in FLRW cosmology?}}, \href{https://doi.org/10.1088/1361-6382/ac1a81}{\emph{Class. Quant. Grav.} {\bfseries 38} (2021) 184001} [\href{https://arxiv.org/abs/2105.09790}{{\ttfamily 2105.09790}}].

\bibitem{Keeley:2022ojz}
R.E.~Keeley and A.~Shafieloo, \emph{{Ruling Out New Physics at Low Redshift as a Solution to the H0 Tension}}, \href{https://doi.org/10.1103/PhysRevLett.131.111002}{\emph{Phys. Rev. Lett.} {\bfseries 131} (2023) 111002} [\href{https://arxiv.org/abs/2206.08440}{{\ttfamily 2206.08440}}].

\bibitem{Gariazzo:2024sil}
S.~Gariazzo, W.~Giar{\`e}, O.~Mena and E.~Di~Valentino, \emph{{How robust are the parameter constraints extending the {\ensuremath{\Lambda}}CDM model?}}, \href{https://doi.org/10.1103/PhysRevD.111.023540}{\emph{Phys. Rev. D} {\bfseries 111} (2025) 023540} [\href{https://arxiv.org/abs/2404.11182}{{\ttfamily 2404.11182}}].

\bibitem{DESI:2024mwx}
{\scshape DESI} collaboration, \emph{{DESI 2024 VI: cosmological constraints from the measurements of baryon acoustic oscillations}}, \href{https://doi.org/10.1088/1475-7516/2025/02/021}{\emph{JCAP} {\bfseries 02} (2025) 021} [\href{https://arxiv.org/abs/2404.03002}{{\ttfamily 2404.03002}}].

\bibitem{DESI:2024aqx}
{\scshape DESI} collaboration, \emph{{DESI 2024: reconstructing dark energy using crossing statistics with DESI DR1 BAO data}}, \href{https://doi.org/10.1088/1475-7516/2024/10/048}{\emph{JCAP} {\bfseries 10} (2024) 048} [\href{https://arxiv.org/abs/2405.04216}{{\ttfamily 2405.04216}}].

\bibitem{DESI:2024jis}
{\scshape DESI} collaboration, \emph{{DESI 2024 V: Full-Shape Galaxy Clustering from Galaxies and Quasars}},  \href{https://arxiv.org/abs/2411.12021}{{\ttfamily 2411.12021}}.

\bibitem{DESI:2025zgx}
{\scshape DESI} collaboration, \emph{{DESI DR2 Results II: Measurements of Baryon Acoustic Oscillations and Cosmological Constraints}},  \href{https://arxiv.org/abs/2503.14738}{{\ttfamily 2503.14738}}.

\bibitem{DESI:2025fii}
{\scshape DESI} collaboration, \emph{{Extended Dark Energy analysis using DESI DR2 BAO measurements}},  \href{https://arxiv.org/abs/2503.14743}{{\ttfamily 2503.14743}}.

\bibitem{DES:2024jxu}
{\scshape DES} collaboration, \emph{{The Dark Energy Survey: Cosmology Results with {\ensuremath{\sim}}1500 New High-redshift Type Ia Supernovae Using the Full 5 yr Data Set}}, \href{https://doi.org/10.3847/2041-8213/ad6f9f}{\emph{Astrophys. J. Lett.} {\bfseries 973} (2024) L14} [\href{https://arxiv.org/abs/2401.02929}{{\ttfamily 2401.02929}}].

\bibitem{Giare:2024smz}
W.~Giar{\`e}, M.A.~Sabogal, R.C.~Nunes and E.~Di~Valentino, \emph{{Interacting Dark Energy after DESI Baryon Acoustic Oscillation Measurements}}, \href{https://doi.org/10.1103/PhysRevLett.133.251003}{\emph{Phys. Rev. Lett.} {\bfseries 133} (2024) 251003} [\href{https://arxiv.org/abs/2404.15232}{{\ttfamily 2404.15232}}].

\bibitem{Wang:2024hks}
D.~Wang, \emph{{Constraining Cosmological Physics with DESI BAO Observations}},  \href{https://arxiv.org/abs/2404.06796}{{\ttfamily 2404.06796}}.

\bibitem{Cortes:2024lgw}
M.~Cort{\^e}s and A.R.~Liddle, \emph{{Interpreting DESI's evidence for evolving dark energy}}, \href{https://doi.org/10.1088/1475-7516/2024/12/007}{\emph{JCAP} {\bfseries 12} (2024) 007} [\href{https://arxiv.org/abs/2404.08056}{{\ttfamily 2404.08056}}].

\bibitem{Colgain:2024xqj}
E.{\'O}.~Colg{\'a}in, M.G.~Dainotti, S.~Capozziello, S.~Pourojaghi, M.M.~Sheikh-Jabbari and D.~Stojkovic, \emph{{Does DESI 2024 confirm {\ensuremath{\Lambda}}CDM?}}, \href{https://doi.org/10.1016/j.jheap.2025.100428}{\emph{JHEAp} {\bfseries 49} (2026) 100428} [\href{https://arxiv.org/abs/2404.08633}{{\ttfamily 2404.08633}}].

\bibitem{Yin:2024hba}
W.~Yin, \emph{{Cosmic clues: DESI, dark energy, and the cosmological constant problem}}, \href{https://doi.org/10.1007/JHEP05(2024)327}{\emph{JHEP} {\bfseries 05} (2024) 327} [\href{https://arxiv.org/abs/2404.06444}{{\ttfamily 2404.06444}}].

\bibitem{Sabogal:2025mkp}
M.A.~Sabogal, E.~Silva, R.C.~Nunes, S.~Kumar and E.~Di~Valentino, \emph{{Sign switching in dark sector coupling interactions as a candidate for resolving cosmological tensions}}, \href{https://doi.org/10.1103/PhysRevD.111.043531}{\emph{Phys. Rev. D} {\bfseries 111} (2025) 043531} [\href{https://arxiv.org/abs/2501.10323}{{\ttfamily 2501.10323}}].

\bibitem{Abdalla:2022yfr}
E.~Abdalla et~al., \emph{{Cosmology intertwined: A review of the particle physics, astrophysics, and cosmology associated with the cosmological tensions and anomalies}}, \href{https://doi.org/10.1016/j.jheap.2022.04.002}{\emph{JHEAp} {\bfseries 34} (2022) 49} [\href{https://arxiv.org/abs/2203.06142}{{\ttfamily 2203.06142}}].

\bibitem{DiValentino:2020vvd}
E.~Di~Valentino et~al., \emph{{Cosmology Intertwined III: $f \sigma_8$ and $S_8$}}, \href{https://doi.org/10.1016/j.astropartphys.2021.102604}{\emph{Astropart. Phys.} {\bfseries 131} (2021) 102604} [\href{https://arxiv.org/abs/2008.11285}{{\ttfamily 2008.11285}}].

\bibitem{DiValentino:2020srs}
E.~Di~Valentino et~al., \emph{{Snowmass2021 - Letter of interest cosmology intertwined IV: The age of the universe and its curvature}}, \href{https://doi.org/10.1016/j.astropartphys.2021.102607}{\emph{Astropart. Phys.} {\bfseries 131} (2021) 102607} [\href{https://arxiv.org/abs/2008.11286}{{\ttfamily 2008.11286}}].

\bibitem{DiValentino:2020zio}
E.~Di~Valentino et~al., \emph{{Snowmass2021 - Letter of interest cosmology intertwined II: The hubble constant tension}}, \href{https://doi.org/10.1016/j.astropartphys.2021.102605}{\emph{Astropart. Phys.} {\bfseries 131} (2021) 102605} [\href{https://arxiv.org/abs/2008.11284}{{\ttfamily 2008.11284}}].

\bibitem{Perivolaropoulos:2021jda}
L.~Perivolaropoulos and F.~Skara, \emph{{Challenges for {\ensuremath{\Lambda}}CDM: An update}}, \href{https://doi.org/10.1016/j.newar.2022.101659}{\emph{New Astron. Rev.} {\bfseries 95} (2022) 101659} [\href{https://arxiv.org/abs/2105.05208}{{\ttfamily 2105.05208}}].

\bibitem{DiValentino:2025sru}
{\scshape CosmoVerse} collaboration, \emph{{The CosmoVerse White Paper: Addressing observational tensions in cosmology with systematics and fundamental physics}},  \href{https://arxiv.org/abs/2504.01669}{{\ttfamily 2504.01669}}.

\bibitem{Verde:2019ivm}
L.~Verde, T.~Treu and A.G.~Riess, \emph{{Tensions between the Early and the Late Universe}}, \href{https://doi.org/10.1038/s41550-019-0902-0}{\emph{Nature Astron.} {\bfseries 3} (2019) 891} [\href{https://arxiv.org/abs/1907.10625}{{\ttfamily 1907.10625}}].

\bibitem{DiValentino:2021izs}
E.~Di~Valentino, O.~Mena, S.~Pan, L.~Visinelli, W.~Yang, A.~Melchiorri et~al., \emph{{In the realm of the Hubble tension{\textemdash}a review of solutions}}, \href{https://doi.org/10.1088/1361-6382/ac086d}{\emph{Class. Quant. Grav.} {\bfseries 38} (2021) 153001} [\href{https://arxiv.org/abs/2103.01183}{{\ttfamily 2103.01183}}].

\bibitem{Schoneberg:2021qvd}
N.~Sch{\"o}neberg, G.~Franco~Abell{\'a}n, A.~P{\'e}rez~S{\'a}nchez, S.J.~Witte, V.~Poulin and J.~Lesgourgues, \emph{{The H0 Olympics: A fair ranking of proposed models}}, \href{https://doi.org/10.1016/j.physrep.2022.07.001}{\emph{Phys. Rept.} {\bfseries 984} (2022) 1} [\href{https://arxiv.org/abs/2107.10291}{{\ttfamily 2107.10291}}].

\bibitem{Shah:2021onj}
P.~Shah, P.~Lemos and O.~Lahav, \emph{{A buyer{\textquoteright}s guide to the Hubble constant}}, \href{https://doi.org/10.1007/s00159-021-00137-4}{\emph{Astron. Astrophys. Rev.} {\bfseries 29} (2021) 9} [\href{https://arxiv.org/abs/2109.01161}{{\ttfamily 2109.01161}}].

\bibitem{DiValentino:2022fjm}
E.~Di~Valentino, \emph{{Challenges of the Standard Cosmological Model}}, \href{https://doi.org/10.3390/universe8080399}{\emph{Universe} {\bfseries 8} (2022) 399}.

\bibitem{Kamionkowski:2022pkx}
M.~Kamionkowski and A.G.~Riess, \emph{{The Hubble Tension and Early Dark Energy}}, \href{https://doi.org/10.1146/annurev-nucl-111422-024107}{\emph{Ann. Rev. Nucl. Part. Sci.} {\bfseries 73} (2023) 153} [\href{https://arxiv.org/abs/2211.04492}{{\ttfamily 2211.04492}}].

\bibitem{Giare:2023xoc}
W.~Giar{\`e}, \emph{{CMB Anomalies and the Hubble Tension}},  \href{https://arxiv.org/abs/2305.16919}{{\ttfamily 2305.16919}}.

\bibitem{Hu:2023jqc}
J.-P.~Hu and F.-Y.~Wang, \emph{{Hubble Tension: The Evidence of New Physics}}, \href{https://doi.org/10.3390/universe9020094}{\emph{Universe} {\bfseries 9} (2023) 94} [\href{https://arxiv.org/abs/2302.05709}{{\ttfamily 2302.05709}}].

\bibitem{Verde:2023lmm}
L.~Verde, N.~Sch{\"o}neberg and H.~Gil-Mar{\'\i}n, \emph{{A Tale of Many H0}}, \href{https://doi.org/10.1146/annurev-astro-052622-033813}{\emph{Ann. Rev. Astron. Astrophys.} {\bfseries 62} (2024) 287} [\href{https://arxiv.org/abs/2311.13305}{{\ttfamily 2311.13305}}].

\bibitem{DiValentino:2024yew}
E.~Di~Valentino and D.~Brout, eds., \emph{{The Hubble Constant Tension}}, Springer Series in Astrophysics and Cosmology, Springer (2024), \href{https://doi.org/10.1007/978-981-99-0177-7}{10.1007/978-981-99-0177-7}.

\bibitem{Perivolaropoulos:2024yxv}
L.~Perivolaropoulos, \emph{{Hubble tension or distance ladder crisis?}}, \href{https://doi.org/10.1103/PhysRevD.110.123518}{\emph{Phys. Rev. D} {\bfseries 110} (2024) 123518} [\href{https://arxiv.org/abs/2408.11031}{{\ttfamily 2408.11031}}].

\bibitem{Planck:2018vyg}
{\scshape Planck} collaboration, \emph{{Planck 2018 results. VI. Cosmological parameters}}, \href{https://doi.org/10.1051/0004-6361/201833910}{\emph{Astron. Astrophys.} {\bfseries 641} (2020) A6} [\href{https://arxiv.org/abs/1807.06209}{{\ttfamily 1807.06209}}].

\bibitem{SPT-3G:2022hvq}
{\scshape SPT-3G} collaboration, \emph{{Measurement of the CMB temperature power spectrum and constraints on cosmology from the SPT-3G 2018 TT, TE, and EE dataset}}, \href{https://doi.org/10.1103/PhysRevD.108.023510}{\emph{Phys. Rev. D} {\bfseries 108} (2023) 023510} [\href{https://arxiv.org/abs/2212.05642}{{\ttfamily 2212.05642}}].

\bibitem{ACT:2025fju}
{\scshape ACT} collaboration, \emph{{The Atacama Cosmology Telescope: DR6 Power Spectra, Likelihoods and $\Lambda$CDM Parameters}},  \href{https://arxiv.org/abs/2503.14452}{{\ttfamily 2503.14452}}.

\bibitem{Freedman:2020dne}
W.L.~Freedman, B.F.~Madore, T.~Hoyt, I.S.~Jang, R.~Beaton, M.G.~Lee et~al., \emph{{Calibration of the Tip of the Red Giant Branch (TRGB)}},  \href{https://arxiv.org/abs/2002.01550}{{\ttfamily 2002.01550}}.

\bibitem{Birrer:2020tax}
S.~Birrer et~al., \emph{{TDCOSMO - IV. Hierarchical time-delay cosmography {\textendash} joint inference of the Hubble constant and galaxy density profiles}}, \href{https://doi.org/10.1051/0004-6361/202038861}{\emph{Astron. Astrophys.} {\bfseries 643} (2020) A165} [\href{https://arxiv.org/abs/2007.02941}{{\ttfamily 2007.02941}}].

\bibitem{Wu:2021jyk}
Q.~Wu, G.-Q.~Zhang and F.-Y.~Wang, \emph{{An 8~per{\,}cent determination of the Hubble constant from localized fast radio bursts}}, \href{https://doi.org/10.1093/mnrasl/slac022}{\emph{Mon. Not. Roy. Astron. Soc.} {\bfseries 515} (2022) L1} [\href{https://arxiv.org/abs/2108.00581}{{\ttfamily 2108.00581}}].

\bibitem{Anderson:2023aga}
R.I.~Anderson, N.W.~Koblischke and L.~Eyer, \emph{{Small-amplitude Red Giants Elucidate the Nature of the Tip of the Red Giant Branch as a Standard Candle}}, \href{https://doi.org/10.3847/2041-8213/ad284d}{\emph{Astrophys. J. Lett.} {\bfseries 963} (2024) L43} [\href{https://arxiv.org/abs/2303.04790}{{\ttfamily 2303.04790}}].

\bibitem{Scolnic:2023mrv}
D.~Scolnic, A.G.~Riess, J.~Wu, S.~Li, G.S.~Anand, R.~Beaton et~al., \emph{{CATS: The Hubble Constant from Standardized TRGB and Type Ia Supernova Measurements}}, \href{https://doi.org/10.3847/2041-8213/ace978}{\emph{Astrophys. J. Lett.} {\bfseries 954} (2023) L31} [\href{https://arxiv.org/abs/2304.06693}{{\ttfamily 2304.06693}}].

\bibitem{Jones:2022mvo}
D.O.~Jones et~al., \emph{{Cosmological Results from the RAISIN Survey: Using Type Ia Supernovae in the Near Infrared as a Novel Path to Measure the Dark Energy Equation of State}}, \href{https://doi.org/10.3847/1538-4357/ac755b}{\emph{Astrophys. J.} {\bfseries 933} (2022) 172} [\href{https://arxiv.org/abs/2201.07801}{{\ttfamily 2201.07801}}].

\bibitem{Anand:2021sum}
G.S.~Anand, R.B.~Tully, L.~Rizzi, A.G.~Riess and W.~Yuan, \emph{{Comparing Tip of the Red Giant Branch Distance Scales: An Independent Reduction of the Carnegie-Chicago Hubble Program and the Value of the Hubble Constant}}, \href{https://doi.org/10.3847/1538-4357/ac68df}{\emph{Astrophys. J.} {\bfseries 932} (2022) 15} [\href{https://arxiv.org/abs/2108.00007}{{\ttfamily 2108.00007}}].

\bibitem{Freedman:2021ahq}
W.L.~Freedman, \emph{{Measurements of the Hubble Constant: Tensions in Perspective}}, \href{https://doi.org/10.3847/1538-4357/ac0e95}{\emph{Astrophys. J.} {\bfseries 919} (2021) 16} [\href{https://arxiv.org/abs/2106.15656}{{\ttfamily 2106.15656}}].

\bibitem{Uddin:2023iob}
S.A.~Uddin et~al., \emph{{Carnegie Supernova Project I and II: Measurements of H $_{0}$ Using Cepheid, Tip of the Red Giant Branch, and Surface Brightness Fluctuation Distance Calibration to Type Ia Supernovae*}}, \href{https://doi.org/10.3847/1538-4357/ad3e63}{\emph{Astrophys. J.} {\bfseries 970} (2024) 72} [\href{https://arxiv.org/abs/2308.01875}{{\ttfamily 2308.01875}}].

\bibitem{Huang:2023frr}
C.D.~Huang et~al., \emph{{The Mira Distance to M101 and a 4{\%} Measurement of H $_{0}$}}, \href{https://doi.org/10.3847/1538-4357/ad1ff8}{\emph{Astrophys. J.} {\bfseries 963} (2024) 83} [\href{https://arxiv.org/abs/2312.08423}{{\ttfamily 2312.08423}}].

\bibitem{Pesce:2020xfe}
D.W.~Pesce et~al., \emph{{The Megamaser Cosmology Project. XIII. Combined Hubble constant constraints}}, \href{https://doi.org/10.3847/2041-8213/ab75f0}{\emph{Astrophys. J. Lett.} {\bfseries 891} (2020) L1} [\href{https://arxiv.org/abs/2001.09213}{{\ttfamily 2001.09213}}].

\bibitem{Kourkchi:2020iyz}
E.~Kourkchi, R.B.~Tully, G.S.~Anand, H.M.~Courtois, A.~Dupuy, J.D.~Neill et~al., \emph{{Cosmicflows-4: The Calibration of Optical and Infrared Tully{\textendash}Fisher Relations}}, \href{https://doi.org/10.3847/1538-4357/ab901c}{\emph{Astrophys. J.} {\bfseries 896} (2020) 3} [\href{https://arxiv.org/abs/2004.14499}{{\ttfamily 2004.14499}}].

\bibitem{Schombert:2020pxm}
J.~Schombert, S.~McGaugh and F.~Lelli, \emph{{Using the Baryonic Tully{\textendash}Fisher Relation to Measure H o}}, \href{https://doi.org/10.3847/1538-3881/ab9d88}{\emph{Astron. J.} {\bfseries 160} (2020) 71} [\href{https://arxiv.org/abs/2006.08615}{{\ttfamily 2006.08615}}].

\bibitem{Blakeslee:2021rqi}
J.P.~Blakeslee, J.B.~Jensen, C.-P.~Ma, P.A.~Milne and J.E.~Greene, \emph{{The Hubble Constant from Infrared Surface Brightness Fluctuation Distances}}, \href{https://doi.org/10.3847/1538-4357/abe86a}{\emph{Astrophys. J.} {\bfseries 911} (2021) 65} [\href{https://arxiv.org/abs/2101.02221}{{\ttfamily 2101.02221}}].

\bibitem{deJaeger:2022lit}
T.~de~Jaeger, L.~Galbany, A.G.~Riess, B.E.~Stahl, B.J.~Shappee, A.V.~Filippenko et~al., \emph{{A 5~per{\,}cent measurement of the Hubble{\textendash}Lema{\^\i}tre constant from Type II supernovae}}, \href{https://doi.org/10.1093/mnras/stac1661}{\emph{Mon. Not. Roy. Astron. Soc.} {\bfseries 514} (2022) 4620} [\href{https://arxiv.org/abs/2203.08974}{{\ttfamily 2203.08974}}].

\bibitem{Murakami:2023xuy}
Y.S.~Murakami, A.G.~Riess, B.E.~Stahl, W.D.~Kenworthy, D.-M.A.~Pluck, A.~Macoretta et~al., \emph{{Leveraging SN Ia spectroscopic similarity to improve the measurement of H $_{0}$}}, \href{https://doi.org/10.1088/1475-7516/2023/11/046}{\emph{JCAP} {\bfseries 11} (2023) 046} [\href{https://arxiv.org/abs/2306.00070}{{\ttfamily 2306.00070}}].

\bibitem{Breuval:2024lsv}
L.~Breuval, A.G.~Riess, S.~Casertano, W.~Yuan, L.M.~Macri, M.~Romaniello et~al., \emph{{Small Magellanic Cloud Cepheids Observed with the Hubble Space Telescope Provide a New Anchor for the SH0ES Distance Ladder}}, \href{https://doi.org/10.3847/1538-4357/ad630e}{\emph{Astrophys. J.} {\bfseries 973} (2024) 30} [\href{https://arxiv.org/abs/2404.08038}{{\ttfamily 2404.08038}}].

\bibitem{Freedman:2024eph}
W.L.~Freedman, B.F.~Madore, T.J.~Hoyt, I.S.~Jang, A.J.~Lee and K.A.~Owens, \emph{{Status Report on the Chicago-Carnegie Hubble Program (CCHP): Measurement of the Hubble Constant Using the Hubble and James Webb Space Telescopes}}, \href{https://doi.org/10.3847/1538-4357/adce78}{\emph{Astrophys. J.} {\bfseries 985} (2025) 203} [\href{https://arxiv.org/abs/2408.06153}{{\ttfamily 2408.06153}}].

\bibitem{Riess:2024vfa}
A.G.~Riess et~al., \emph{{JWST Validates HST Distance Measurements: Selection of Supernova Subsample Explains Differences in JWST Estimates of Local H $_{0}$}}, \href{https://doi.org/10.3847/1538-4357/ad8c21}{\emph{Astrophys. J.} {\bfseries 977} (2024) 120} [\href{https://arxiv.org/abs/2408.11770}{{\ttfamily 2408.11770}}].

\bibitem{Vogl:2024bum}
C.~Vogl et~al., \emph{{No rungs attached: A distance-ladder free determination of the Hubble constant through type II supernova spectral modelling}},  \href{https://arxiv.org/abs/2411.04968}{{\ttfamily 2411.04968}}.

\bibitem{Gao:2024kkx}
D.H.~Gao, Q.~Wu, J.P.~Hu, S.X.~Yi, X.~Zhou, F.Y.~Wang et~al., \emph{{Measuring the Hubble constant using localized and nonlocalized fast radio bursts}}, \href{https://doi.org/10.1051/0004-6361/202453006}{\emph{Astron. Astrophys.} {\bfseries 698} (2025) A215} [\href{https://arxiv.org/abs/2410.03994}{{\ttfamily 2410.03994}}].

\bibitem{Scolnic:2024hbh}
D.~Scolnic et~al., \emph{{The Hubble Tension in Our Own Backyard: DESI and the Nearness of the Coma Cluster}}, \href{https://doi.org/10.3847/2041-8213/ada0bd}{\emph{Astrophys. J. Lett.} {\bfseries 979} (2025) L9} [\href{https://arxiv.org/abs/2409.14546}{{\ttfamily 2409.14546}}].

\bibitem{Said:2024pwm}
K.~Said et~al., \emph{{DESI Peculiar Velocity Survey -- Fundamental Plane}}, \href{https://doi.org/10.1093/mnras/staf700}{\emph{Mon. Not. Roy. Astron. Soc.} {\bfseries 539} (2025) 3627} [\href{https://arxiv.org/abs/2408.13842}{{\ttfamily 2408.13842}}].

\bibitem{Boubel:2024cqw}
P.~Boubel, M.~Colless, K.~Said and L.~Staveley-Smith, \emph{{An improved Tully{\textendash}Fisher estimate of H0}}, \href{https://doi.org/10.1093/mnras/stae1925}{\emph{Mon. Not. Roy. Astron. Soc.} {\bfseries 533} (2024) 1550} [\href{https://arxiv.org/abs/2408.03660}{{\ttfamily 2408.03660}}].

\bibitem{Scolnic:2024oth}
D.~Scolnic, P.~Boubel, J.~Byrne, A.G.~Riess and G.S.~Anand, \emph{{Calibrating the Tully-Fisher Relation to Measure the Hubble Constant}},  \href{https://arxiv.org/abs/2412.08449}{{\ttfamily 2412.08449}}.

\bibitem{Li:2025ife}
S.~Li, A.G.~Riess, D.~Scolnic, S.~Casertano and G.S.~Anand, \emph{{JAGB 2.0: Improved Constraints on the J-region Asymptotic Giant Branch{\textendash}based Hubble Constant from an Expanded Sample of JWST Observations}}, \href{https://doi.org/10.3847/1538-4357/addd0c}{\emph{Astrophys. J.} {\bfseries 988} (2025) 97} [\href{https://arxiv.org/abs/2502.05259}{{\ttfamily 2502.05259}}].

\bibitem{Jensen:2025aai}
J.B.~Jensen, J.P.~Blakeslee, M.~Cantiello, M.~Cowles, G.S.~Anand, R.B.~Tully et~al., \emph{{The TRGB-SBF Project. III. Refining the HST Surface Brightness Fluctuation Distance Scale Calibration with JWST}},  \href{https://arxiv.org/abs/2502.15935}{{\ttfamily 2502.15935}}.

\bibitem{Chevallier:2000qy}
M.~Chevallier and D.~Polarski, \emph{{Accelerating universes with scaling dark matter}}, \href{https://doi.org/10.1142/S0218271801000822}{\emph{Int. J. Mod. Phys. D} {\bfseries 10} (2001) 213} [\href{https://arxiv.org/abs/gr-qc/0009008}{{\ttfamily gr-qc/0009008}}].

\bibitem{Linder:2002et}
E.V.~Linder, \emph{{Exploring the expansion history of the universe}}, \href{https://doi.org/10.1103/PhysRevLett.90.091301}{\emph{Phys. Rev. Lett.} {\bfseries 90} (2003) 091301} [\href{https://arxiv.org/abs/astro-ph/0208512}{{\ttfamily astro-ph/0208512}}].

\bibitem{Shlivko:2024llw}
D.~Shlivko and P.J.~Steinhardt, \emph{{Assessing observational constraints on dark energy}}, \href{https://doi.org/10.1016/j.physletb.2024.138826}{\emph{Phys. Lett. B} {\bfseries 855} (2024) 138826} [\href{https://arxiv.org/abs/2405.03933}{{\ttfamily 2405.03933}}].

\bibitem{Luongo:2024fww}
O.~Luongo and M.~Muccino, \emph{{Model-independent cosmographic constraints from DESI 2024}}, \href{https://doi.org/10.1051/0004-6361/202450512}{\emph{Astron. Astrophys.} {\bfseries 690} (2024) A40} [\href{https://arxiv.org/abs/2404.07070}{{\ttfamily 2404.07070}}].

\bibitem{Gialamas:2024lyw}
I.D.~Gialamas, G.~H{\"u}tsi, K.~Kannike, A.~Racioppi, M.~Raidal, M.~Vasar et~al., \emph{{Interpreting DESI 2024 BAO: Late-time dynamical dark energy or a local effect?}}, \href{https://doi.org/10.1103/PhysRevD.111.043540}{\emph{Phys. Rev. D} {\bfseries 111} (2025) 043540} [\href{https://arxiv.org/abs/2406.07533}{{\ttfamily 2406.07533}}].

\bibitem{Dinda:2024kjf}
B.R.~Dinda, \emph{{A new diagnostic for the null test of dynamical dark energy in light of DESI 2024 and other BAO data}}, \href{https://doi.org/10.1088/1475-7516/2024/09/062}{\emph{JCAP} {\bfseries 09} (2024) 062} [\href{https://arxiv.org/abs/2405.06618}{{\ttfamily 2405.06618}}].

\bibitem{Najafi:2024qzm}
M.~Najafi, S.~Pan, E.~Di~Valentino and J.T.~Firouzjaee, \emph{{Dynamical dark energy confronted with multiple CMB missions}}, \href{https://doi.org/10.1016/j.dark.2024.101539}{\emph{Phys. Dark Univ.} {\bfseries 45} (2024) 101539} [\href{https://arxiv.org/abs/2407.14939}{{\ttfamily 2407.14939}}].

\bibitem{Wang:2024dka}
H.~Wang and Y.-S.~Piao, \emph{{Dark energy in light of recent DESI BAO and Hubble tension}},  \href{https://arxiv.org/abs/2404.18579}{{\ttfamily 2404.18579}}.

\bibitem{Tada:2024znt}
Y.~Tada and T.~Terada, \emph{{Quintessential interpretation of the evolving dark energy in light of DESI observations}}, \href{https://doi.org/10.1103/PhysRevD.109.L121305}{\emph{Phys. Rev. D} {\bfseries 109} (2024) L121305} [\href{https://arxiv.org/abs/2404.05722}{{\ttfamily 2404.05722}}].

\bibitem{Carloni:2024zpl}
Y.~Carloni, O.~Luongo and M.~Muccino, \emph{{Does dark energy really revive using DESI 2024 data?}}, \href{https://doi.org/10.1103/PhysRevD.111.023512}{\emph{Phys. Rev. D} {\bfseries 111} (2025) 023512} [\href{https://arxiv.org/abs/2404.12068}{{\ttfamily 2404.12068}}].

\bibitem{Chan-GyungPark:2024mlx}
C.-G.~Park, J.~de~Cruz~P{\'e}rez and B.~Ratra, \emph{{Using non-DESI data to confirm and strengthen the DESI 2024 spatially flat w0waCDM cosmological parametrization result}}, \href{https://doi.org/10.1103/PhysRevD.110.123533}{\emph{Phys. Rev. D} {\bfseries 110} (2024) 123533} [\href{https://arxiv.org/abs/2405.00502}{{\ttfamily 2405.00502}}].

\bibitem{DESI:2024kob}
{\scshape DESI} collaboration, \emph{{DESI 2024: Constraints on physics-focused aspects of dark energy using DESI DR1 BAO data}}, \href{https://doi.org/10.1103/PhysRevD.111.023532}{\emph{Phys. Rev. D} {\bfseries 111} (2025) 023532} [\href{https://arxiv.org/abs/2405.13588}{{\ttfamily 2405.13588}}].

\bibitem{Ramadan:2024kmn}
O.F.~Ramadan, J.~Sakstein and D.~Rubin, \emph{{DESI constraints on exponential quintessence}}, \href{https://doi.org/10.1103/PhysRevD.110.L041303}{\emph{Phys. Rev. D} {\bfseries 110} (2024) L041303} [\href{https://arxiv.org/abs/2405.18747}{{\ttfamily 2405.18747}}].

\bibitem{Notari:2024rti}
A.~Notari, M.~Redi and A.~Tesi, \emph{{Consistent theories for the DESI dark energy fit}}, \href{https://doi.org/10.1088/1475-7516/2024/11/025}{\emph{JCAP} {\bfseries 11} (2024) 025} [\href{https://arxiv.org/abs/2406.08459}{{\ttfamily 2406.08459}}].

\bibitem{Orchard:2024bve}
L.~Orchard and V.H.~C{\'a}rdenas, \emph{{Probing dark energy evolution post-DESI 2024}}, \href{https://doi.org/10.1016/j.dark.2024.101678}{\emph{Phys. Dark Univ.} {\bfseries 46} (2024) 101678} [\href{https://arxiv.org/abs/2407.05579}{{\ttfamily 2407.05579}}].

\bibitem{Hernandez-Almada:2024ost}
A.~Hern{\'a}ndez-Almada, M.L.~Mendoza-Mart{\'\i}nez, M.A.~Garc{\'\i}a-Aspeitia and V.~Motta, \emph{{Phenomenological emergent dark energy in the light of DESI Data Release 1}}, \href{https://doi.org/10.1016/j.dark.2024.101668}{\emph{Phys. Dark Univ.} {\bfseries 46} (2024) 101668} [\href{https://arxiv.org/abs/2407.09430}{{\ttfamily 2407.09430}}].

\bibitem{Malekjani:2024bgi}
{\scshape DESI} collaboration, \emph{{Cosmological constraints on dark energy parametrizations after DESI 2024: Persistent deviation from standard {\ensuremath{\Lambda}}CDM cosmology}}, \href{https://doi.org/10.1103/PhysRevD.111.083547}{\emph{Phys. Rev. D} {\bfseries 111} (2025) 083547} [\href{https://arxiv.org/abs/2407.09767}{{\ttfamily 2407.09767}}].

\bibitem{Giare:2024gpk}
W.~Giar{\`e}, M.~Najafi, S.~Pan, E.~Di~Valentino and J.T.~Firouzjaee, \emph{{Robust preference for Dynamical Dark Energy in DESI BAO and SN measurements}}, \href{https://doi.org/10.1088/1475-7516/2024/10/035}{\emph{JCAP} {\bfseries 10} (2024) 035} [\href{https://arxiv.org/abs/2407.16689}{{\ttfamily 2407.16689}}].

\bibitem{Reboucas:2024smm}
J.~Rebou{\c{c}}as, D.H.F.~de~Souza, K.~Zhong, V.~Miranda and R.~Rosenfeld, \emph{{Investigating late-time dark energy and massive neutrinos in light of DESI Y1 BAO}}, \href{https://doi.org/10.1088/1475-7516/2025/02/024}{\emph{JCAP} {\bfseries 02} (2025) 024} [\href{https://arxiv.org/abs/2408.14628}{{\ttfamily 2408.14628}}].

\bibitem{Giare:2024ocw}
W.~Giar{\`e}, \emph{{Dynamical dark energy beyond Planck? Constraints from multiple CMB probes, DESI BAO, and type-Ia supernovae}}, \href{https://doi.org/10.1103/ss37-cxhn}{\emph{Phys. Rev. D} {\bfseries 112} (2025) 023508} [\href{https://arxiv.org/abs/2409.17074}{{\ttfamily 2409.17074}}].

\bibitem{Chan-GyungPark:2024brx}
C.-G.~Park, J.~de~Cruz~P{\'e}rez and B.~Ratra, \emph{{Is the $w_0w_a$CDM cosmological parameterization evidence for dark energy dynamics partially caused by the excess smoothing of Planck CMB anisotropy data?}},  \href{https://arxiv.org/abs/2410.13627}{{\ttfamily 2410.13627}}.

\bibitem{Menci:2024hop}
N.~Menci, A.A.~Sen and M.~Castellano, \emph{{The Excess of JWST Bright Galaxies: A Possible Origin in the Ground State of Dynamical Dark Energy in the Light of DESI 2024 Data}}, \href{https://doi.org/10.3847/1538-4357/ad8d5b}{\emph{Astrophys. J.} {\bfseries 976} (2024) 227} [\href{https://arxiv.org/abs/2410.22940}{{\ttfamily 2410.22940}}].

\bibitem{Li:2024qus}
T.-N.~Li, Y.-H.~Li, G.-H.~Du, P.-J.~Wu, L.~Feng, J.-F.~Zhang et~al., \emph{{Revisiting holographic dark energy after DESI 2024}}, \href{https://doi.org/10.1140/epjc/s10052-025-14279-7}{\emph{Eur. Phys. J. C} {\bfseries 85} (2025) 608} [\href{https://arxiv.org/abs/2411.08639}{{\ttfamily 2411.08639}}].

\bibitem{Li:2024hrv}
J.-X.~Li and S.~Wang, \emph{{A comprehensive numerical study on four categories of holographic dark energy models}}, \href{https://doi.org/10.1088/1475-7516/2025/07/047}{\emph{JCAP} {\bfseries 07} (2025) 047} [\href{https://arxiv.org/abs/2412.09064}{{\ttfamily 2412.09064}}].

\bibitem{Notari:2024zmi}
A.~Notari, M.~Redi and A.~Tesi, \emph{{BAO vs. SN evidence for evolving dark energy}}, \href{https://doi.org/10.1088/1475-7516/2025/04/048}{\emph{JCAP} {\bfseries 04} (2025) 048} [\href{https://arxiv.org/abs/2411.11685}{{\ttfamily 2411.11685}}].

\bibitem{Gao:2024ily}
Q.~Gao, Z.~Peng, S.~Gao and Y.~Gong, \emph{{On the Evidence of Dynamical Dark Energy}}, \href{https://doi.org/10.3390/universe11010010}{\emph{Universe} {\bfseries 11} (2025) 10} [\href{https://arxiv.org/abs/2411.16046}{{\ttfamily 2411.16046}}].

\bibitem{Fikri:2024klc}
R.~Fikri, E.~ElKhateeb, E.S.~Lashin and W.~El~Hanafy, \emph{{A preference for dynamical phantom dark energy using one-parameter model with Planck, DESI DR1 BAO and SN data}},  \href{https://arxiv.org/abs/2411.19362}{{\ttfamily 2411.19362}}.

\bibitem{Jiang:2024xnu}
J.-Q.~Jiang, D.~Pedrotti, S.S.~da~Costa and S.~Vagnozzi, \emph{{Nonparametric late-time expansion history reconstruction and implications for the Hubble tension in light of recent DESI and type Ia supernovae data}}, \href{https://doi.org/10.1103/PhysRevD.110.123519}{\emph{Phys. Rev. D} {\bfseries 110} (2024) 123519} [\href{https://arxiv.org/abs/2408.02365}{{\ttfamily 2408.02365}}].

\bibitem{Zheng:2024qzi}
J.~Zheng, D.-C.~Qiang and Z.-Q.~You, \emph{{Cosmological constraints on dark energy models using DESI BAO 2024}},  \href{https://arxiv.org/abs/2412.04830}{{\ttfamily 2412.04830}}.

\bibitem{Gomez-Valent:2024ejh}
A.~G{\'o}mez-Valent and J.~Sol{\`a}~Peracaula, \emph{{Composite dark energy and the cosmological tensions}}, \href{https://doi.org/10.1016/j.physletb.2025.139391}{\emph{Phys. Lett. B} {\bfseries 864} (2025) 139391} [\href{https://arxiv.org/abs/2412.15124}{{\ttfamily 2412.15124}}].

\bibitem{RoyChoudhury:2024wri}
S.~Roy~Choudhury and T.~Okumura, \emph{{Updated Cosmological Constraints in Extended Parameter Space with Planck PR4, DESI Baryon Acoustic Oscillations, and Supernovae: Dynamical Dark Energy, Neutrino Masses, Lensing Anomaly, and the Hubble Tension}}, \href{https://doi.org/10.3847/2041-8213/ad8c26}{\emph{Astrophys. J. Lett.} {\bfseries 976} (2024) L11} [\href{https://arxiv.org/abs/2409.13022}{{\ttfamily 2409.13022}}].

\bibitem{Lewis:2024cqj}
A.~Lewis and E.~Chamberlain, \emph{{Understanding acoustic scale observations: the one-sided fight against {\ensuremath{\Lambda}}}}, \href{https://doi.org/10.1088/1475-7516/2025/05/065}{\emph{JCAP} {\bfseries 05} (2025) 065} [\href{https://arxiv.org/abs/2412.13894}{{\ttfamily 2412.13894}}].

\bibitem{Wolf:2025jlc}
W.J.~Wolf, C.~Garc{\'\i}a-Garc{\'\i}a and P.G.~Ferreira, \emph{{Robustness of dark energy phenomenology across different parameterizations}}, \href{https://doi.org/10.1088/1475-7516/2025/05/034}{\emph{JCAP} {\bfseries 05} (2025) 034} [\href{https://arxiv.org/abs/2502.04929}{{\ttfamily 2502.04929}}].

\bibitem{Shajib:2025tpd}
A.J.~Shajib and J.A.~Frieman, \emph{{Scalar field dark energy models: Current and forecast constraints}},  \href{https://arxiv.org/abs/2502.06929}{{\ttfamily 2502.06929}}.

\bibitem{Giare:2025pzu}
W.~Giar{\`e}, T.~Mahassen, E.~Di~Valentino and S.~Pan, \emph{{An overview of what current data can (and cannot yet) say about evolving dark energy}}, \href{https://doi.org/10.1016/j.dark.2025.101906}{\emph{Phys. Dark Univ.} {\bfseries 48} (2025) 101906} [\href{https://arxiv.org/abs/2502.10264}{{\ttfamily 2502.10264}}].

\bibitem{Chaussidon:2025npr}
E.~Chaussidon et~al., \emph{{Early time solution as an alternative to the late time evolving dark energy with DESI DR2 BAO}},  \href{https://arxiv.org/abs/2503.24343}{{\ttfamily 2503.24343}}.

\bibitem{Kessler:2025kju}
D.A.~Kessler, L.A.~Escamilla, S.~Pan and E.~Di~Valentino, \emph{{One-parameter dynamical dark energy: Hints for oscillations}},  \href{https://arxiv.org/abs/2504.00776}{{\ttfamily 2504.00776}}.

\bibitem{Pang:2025lvh}
Y.-H.~Pang, X.~Zhang and Q.-G.~Huang, \emph{{The impact of the Hubble tension on the evidence for dynamical dark energy}}, \href{https://doi.org/10.1007/s11433-025-2713-8}{\emph{Sci. China Phys. Mech. Astron.} {\bfseries 68} (2025) 280410} [\href{https://arxiv.org/abs/2503.21600}{{\ttfamily 2503.21600}}].

\bibitem{RoyChoudhury:2025dhe}
S.~Roy~Choudhury, \emph{{Cosmology in Extended Parameter Space with DESI Data Release 2 Baryon Acoustic Oscillations: A 2{\ensuremath{\sigma}}+ Detection of Nonzero Neutrino Masses with an Update on Dynamical Dark Energy and Lensing Anomaly}}, \href{https://doi.org/10.3847/2041-8213/ade1cc}{\emph{Astrophys. J. Lett.} {\bfseries 986} (2025) L31} [\href{https://arxiv.org/abs/2504.15340}{{\ttfamily 2504.15340}}].

\bibitem{Scherer:2025esj}
M.~Scherer, M.A.~Sabogal, R.C.~Nunes and A.~De~Felice, \emph{{Challenging $\Lambda$CDM: 5$\sigma$ Evidence for a Dynamical Dark Energy Late-Time Transition}},  \href{https://arxiv.org/abs/2504.20664}{{\ttfamily 2504.20664}}.

\bibitem{Li:2025cxn}
C.~Li, J.~Wang, D.~Zhang, E.N.~Saridakis and Y.-F.~Cai, \emph{{Quantum Gravity Meets DESI: Dynamical Dark Energy in Light of the Trans-Planckian Censorship Conjecture}},  \href{https://arxiv.org/abs/2504.07791}{{\ttfamily 2504.07791}}.

\bibitem{Yang:2025mws}
Y.~Yang, Q.~Wang, X.~Ren, E.N.~Saridakis and Y.-F.~Cai, \emph{{Modified Gravity Realizations of Quintom Dark Energy after DESI DR2}}, \href{https://doi.org/10.3847/1538-4357/ade43f}{\emph{Astrophys. J.} {\bfseries 988} (2025) 123} [\href{https://arxiv.org/abs/2504.06784}{{\ttfamily 2504.06784}}].

\bibitem{Lin:2025gne}
W.~Lin, L.~Visinelli and T.T.~Yanagida, \emph{{Testing Quintessence Axion Dark Energy with Recent Cosmological Results}},  \href{https://arxiv.org/abs/2504.17638}{{\ttfamily 2504.17638}}.

\bibitem{Cheng:2025hug}
H.~Cheng, E.~Di~Valentino and L.~Visinelli, \emph{{Cosmic Strings as Dynamical Dark Energy: Novel Constraints}},  \href{https://arxiv.org/abs/2505.22066}{{\ttfamily 2505.22066}}.

\bibitem{An:2025vfz}
H.~An, C.~Han and B.~Zhang, \emph{{Topological defects as effective dynamical dark energy}},  \href{https://arxiv.org/abs/2506.10075}{{\ttfamily 2506.10075}}.

\bibitem{Sen:2007gk}
A.A.~Sen, \emph{{Deviation From LambdaCDM: Pressure Parametrization}}, \href{https://doi.org/10.1103/PhysRevD.77.043508}{\emph{Phys. Rev. D} {\bfseries 77} (2008) 043508} [\href{https://arxiv.org/abs/0708.1072}{{\ttfamily 0708.1072}}].

\bibitem{Ozulker:2022slu}
E.~Ozulker, \emph{{Is the dark energy equation of state parameter singular?}}, \href{https://doi.org/10.1103/PhysRevD.106.063509}{\emph{Phys. Rev. D} {\bfseries 106} (2022) 063509} [\href{https://arxiv.org/abs/2203.04167}{{\ttfamily 2203.04167}}].

\bibitem{Torrado:2020dgo}
J.~Torrado and A.~Lewis, \emph{{Cobaya: Code for Bayesian Analysis of hierarchical physical models}}, \href{https://doi.org/10.1088/1475-7516/2021/05/057}{\emph{JCAP} {\bfseries 05} (2021) 057} [\href{https://arxiv.org/abs/2005.05290}{{\ttfamily 2005.05290}}].

\bibitem{Lewis:1999bs}
A.~Lewis, A.~Challinor and A.~Lasenby, \emph{{Efficient computation of CMB anisotropies in closed FRW models}}, \href{https://doi.org/10.1086/309179}{\emph{Astrophys. J.} {\bfseries 538} (2000) 473} [\href{https://arxiv.org/abs/astro-ph/9911177}{{\ttfamily astro-ph/9911177}}].

\bibitem{Hu:2007pj}
W.~Hu and I.~Sawicki, \emph{{A Parameterized Post-Friedmann Framework for Modified Gravity}}, \href{https://doi.org/10.1103/PhysRevD.76.104043}{\emph{Phys. Rev. D} {\bfseries 76} (2007) 104043} [\href{https://arxiv.org/abs/0708.1190}{{\ttfamily 0708.1190}}].

\bibitem{Fang:2008sn}
W.~Fang, W.~Hu and A.~Lewis, \emph{{Crossing the Phantom Divide with Parameterized Post-Friedmann Dark Energy}}, \href{https://doi.org/10.1103/PhysRevD.78.087303}{\emph{Phys. Rev. D} {\bfseries 78} (2008) 087303} [\href{https://arxiv.org/abs/0808.3125}{{\ttfamily 0808.3125}}].

\bibitem{gelman_inference_1992}
A.~Gelman and D.~Rubin, \emph{Inference from iterative simulation using multiple sequences}, \href{https://doi.org/10.1214/ss/1177011136}{\emph{Statistical Science} {\bfseries 7} (1992) 457}.

\bibitem{Lewis:2019xzd}
A.~Lewis, \emph{{GetDist: a Python package for analysing Monte Carlo samples}},  \href{https://arxiv.org/abs/1910.13970}{{\ttfamily 1910.13970}}.

\bibitem{Planck:2019nip}
{\scshape Planck} collaboration, \emph{{Planck 2018 results. V. CMB power spectra and likelihoods}}, \href{https://doi.org/10.1051/0004-6361/201936386}{\emph{Astron. Astrophys.} {\bfseries 641} (2020) A5} [\href{https://arxiv.org/abs/1907.12875}{{\ttfamily 1907.12875}}].

\bibitem{Planck:2018lbu}
{\scshape Planck} collaboration, \emph{{Planck 2018 results. VIII. Gravitational lensing}}, \href{https://doi.org/10.1051/0004-6361/201833886}{\emph{Astron. Astrophys.} {\bfseries 641} (2020) A8} [\href{https://arxiv.org/abs/1807.06210}{{\ttfamily 1807.06210}}].

\bibitem{ACT:2023kun}
{\scshape ACT} collaboration, \emph{{The Atacama Cosmology Telescope: DR6 Gravitational Lensing Map and Cosmological Parameters}}, \href{https://doi.org/10.3847/1538-4357/acff5f}{\emph{Astrophys. J.} {\bfseries 962} (2024) 113} [\href{https://arxiv.org/abs/2304.05203}{{\ttfamily 2304.05203}}].

\bibitem{ACT:2023dou}
{\scshape ACT} collaboration, \emph{{The Atacama Cosmology Telescope: A Measurement of the DR6 CMB Lensing Power Spectrum and Its Implications for Structure Growth}}, \href{https://doi.org/10.3847/1538-4357/acfe06}{\emph{Astrophys. J.} {\bfseries 962} (2024) 112} [\href{https://arxiv.org/abs/2304.05202}{{\ttfamily 2304.05202}}].

\bibitem{eBOSS:2020yzd}
{\scshape eBOSS} collaboration, \emph{{Completed SDSS-IV extended Baryon Oscillation Spectroscopic Survey: Cosmological implications from two decades of spectroscopic surveys at the Apache Point Observatory}}, \href{https://doi.org/10.1103/PhysRevD.103.083533}{\emph{Phys. Rev. D} {\bfseries 103} (2021) 083533} [\href{https://arxiv.org/abs/2007.08991}{{\ttfamily 2007.08991}}].

\bibitem{DESI:2025qqy}
{\scshape DESI} collaboration, \emph{{Validation of the DESI DR2 Measurements of Baryon Acoustic Oscillations from Galaxies and Quasars}},  \href{https://arxiv.org/abs/2503.14742}{{\ttfamily 2503.14742}}.

\bibitem{Scolnic:2021amr}
D.~Scolnic et~al., \emph{{The Pantheon+ Analysis: The Full Data Set and Light-curve Release}}, \href{https://doi.org/10.3847/1538-4357/ac8b7a}{\emph{Astrophys. J.} {\bfseries 938} (2022) 113} [\href{https://arxiv.org/abs/2112.03863}{{\ttfamily 2112.03863}}].

\bibitem{Brout:2022vxf}
D.~Brout et~al., \emph{{The Pantheon+ Analysis: Cosmological Constraints}}, \href{https://doi.org/10.3847/1538-4357/ac8e04}{\emph{Astrophys. J.} {\bfseries 938} (2022) 110} [\href{https://arxiv.org/abs/2202.04077}{{\ttfamily 2202.04077}}].

\bibitem{DES:2024hip}
{\scshape DES} collaboration, \emph{{The Dark Energy Survey Supernova Program: Cosmological Analysis and Systematic Uncertainties}}, \href{https://doi.org/10.3847/1538-4357/ad5e6c}{\emph{Astrophys. J.} {\bfseries 975} (2024) 86} [\href{https://arxiv.org/abs/2401.02945}{{\ttfamily 2401.02945}}].

\bibitem{DES:2024upw}
{\scshape DES} collaboration, \emph{{The Dark Energy Survey Supernova Program: Light Curves and 5 Yr Data Release}}, \href{https://doi.org/10.3847/1538-4357/ad739a}{\emph{Astrophys. J.} {\bfseries 975} (2024) 5} [\href{https://arxiv.org/abs/2406.05046}{{\ttfamily 2406.05046}}].

\bibitem{Wilks:1938dza}
S.S.~Wilks, \emph{{The Large-Sample Distribution of the Likelihood Ratio for Testing Composite Hypotheses}}, \href{https://doi.org/10.1214/aoms/1177732360}{\emph{Annals Math. Statist.} {\bfseries 9} (1938) 60}.

\bibitem{Heavens:2017afc}
A.~Heavens, Y.~Fantaye, A.~Mootoovaloo, H.~Eggers, Z.~Hosenie, S.~Kroon et~al., \emph{{Marginal Likelihoods from Monte Carlo Markov Chains}},  \href{https://arxiv.org/abs/1704.03472}{{\ttfamily 1704.03472}}.

\bibitem{giare2025wgcosmo}
W.~Giare, ``wgcosmo: Cosmological analysis tools.'' \url{https://github.com/williamgiare/wgcosmo}, 2025.

\bibitem{Kass:1995loi}
R.E.~Kass and A.E.~Raftery, \emph{{Bayes Factors}}, \href{https://doi.org/10.1080/01621459.1995.10476572}{\emph{J. Am. Statist. Assoc.} {\bfseries 90} (1995) 773}.

\bibitem{Feng:2004ad}
B.~Feng, X.-L.~Wang and X.-M.~Zhang, \emph{{Dark energy constraints from the cosmic age and supernova}}, \href{https://doi.org/10.1016/j.physletb.2004.12.071}{\emph{Phys. Lett. B} {\bfseries 607} (2005) 35} [\href{https://arxiv.org/abs/astro-ph/0404224}{{\ttfamily astro-ph/0404224}}].

\bibitem{Vikman:2004dc}
A.~Vikman, \emph{{Can dark energy evolve to the phantom?}}, \href{https://doi.org/10.1103/PhysRevD.71.023515}{\emph{Phys. Rev. D} {\bfseries 71} (2005) 023515} [\href{https://arxiv.org/abs/astro-ph/0407107}{{\ttfamily astro-ph/0407107}}].

\bibitem{Nesseris:2025lke}
S.~Nesseris, Y.~Akrami and G.D.~Starkman, \emph{{To CPL, or not to CPL? What we have not learned about the dark energy equation of state}},  \href{https://arxiv.org/abs/2503.22529}{{\ttfamily 2503.22529}}.

\end{thebibliography}\endgroup
\end{document}